\documentclass{egpubl}
\usepackage{egsr2026_DLOnlyTrack}
\usepackage{amsmath} 

\pdfpagewidth=\paperwidth
\pdfpageheight=\paperheight

\WsPaper           

\usepackage[T1]{fontenc}
\usepackage{sty/dfadobe}

\usepackage{cite}  
\BibtexOrBiblatex
\electronicVersion
\PrintedOrElectronic
\ifpdf \usepackage[pdftex]{graphicx} \pdfcompresslevel=9
\else \usepackage[dvips]{graphicx} \fi

\usepackage{amsmath}
\usepackage{amssymb}
\usepackage{sty/egweblnk} 
\usepackage{tikz}
\usetikzlibrary{arrows.meta,positioning,fit,calc,decorations.pathreplacing,spy,shapes.geometric}
\usepackage{pgfplots}
\usepgfplotslibrary{statistics} 
\pgfplotsset{compat=1.18}

\DeclareMathOperator{\Var}{{Var}}
\DeclareMathOperator{\Mean}{{E}}
\newcommand{\mlp}{\mbox{mlp}}
\newcommand{\net}{\mbox{net}}

\newcommand{\wb}{\mathbf{wb}}
\newcommand{\lat}[1]{\pmb{l}_{#1}}
\newcommand{\tokent}{\pmb{t}}
\newcommand{\tokenx}{\pmb{x}}
\newcommand{\ts}{\mathcal{T}}

\title[Hypernetworks for NTC]%
{Neural Texture Compression using Hypernetworks}

\author[L. Belcour]{
    {\parbox{\textwidth}{\centering
        L. Belcour~\orcid{0000-0002-1982-0717}
    }}
    \\
    {\parbox{\textwidth}{\centering
        Intel Corporation
    }}
}

\begin{document}

\teaser{
    \vspace{-0.2cm}
    \begin{tikzpicture}[ 
        arrow/.style={-Latex, thick},
        spy using outlines={rectangle, magnification=2.5, size=2.2cm}
    ]
        \node[anchor=south west,inner sep=0] (image) at (0,0) {\includegraphics[height=5cm]{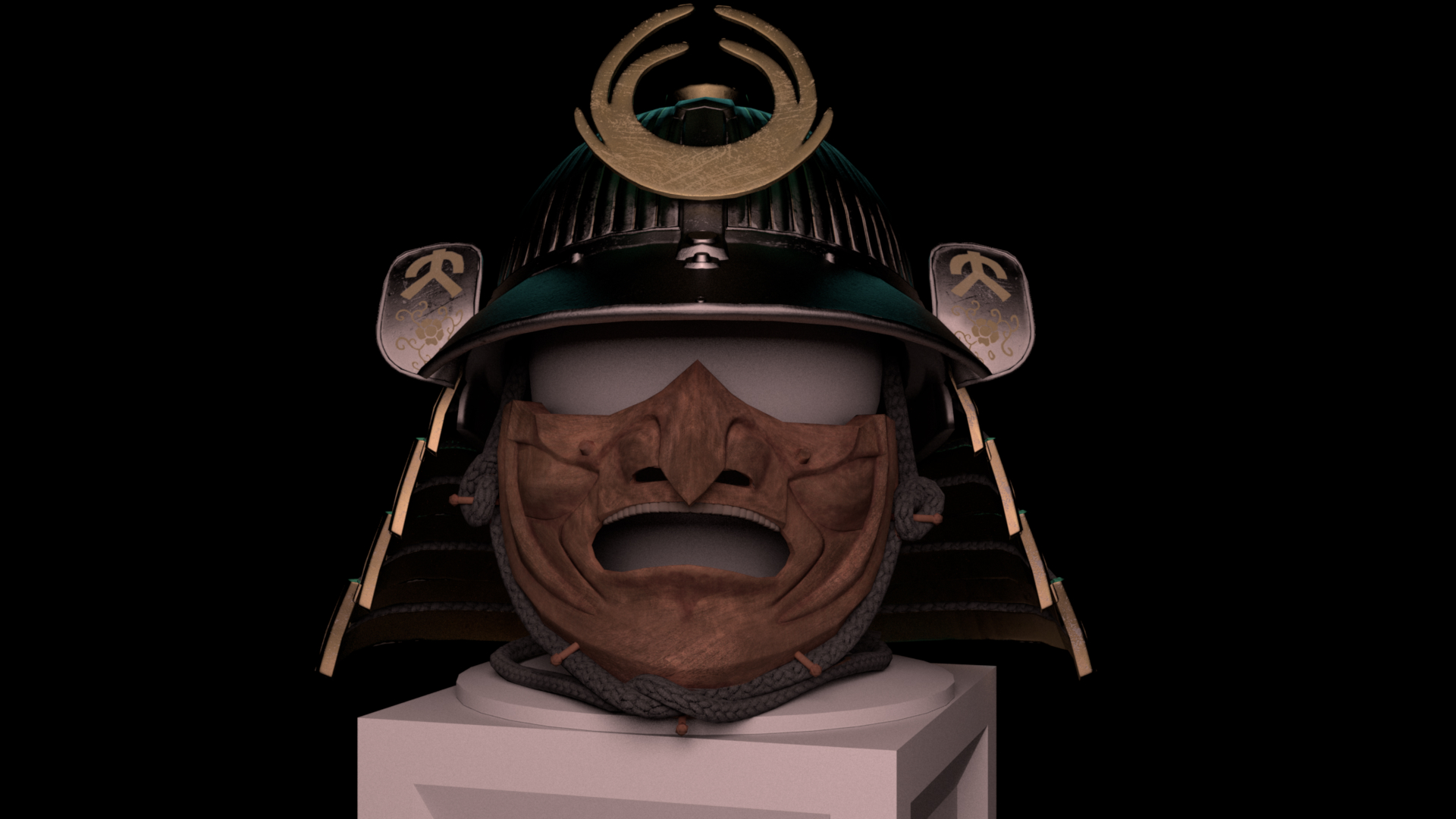}};
        \begin{scope}[x={(image.south east)},y={(image.north west)}]
            \draw[red,thick] (0.44,0.725) rectangle ++(0.07425,0.131);
            \draw[olive,thick] (0.43,0.12) rectangle ++(0.07425,0.131)  node[, inner sep=0pt] (crop0) {};
        \end{scope}
        \begin{scope}[xshift=1.6cm, yshift=3.9cm]
            \node[draw,thick,inner sep=0pt] (input_arm) { \includegraphics[width=2cm]{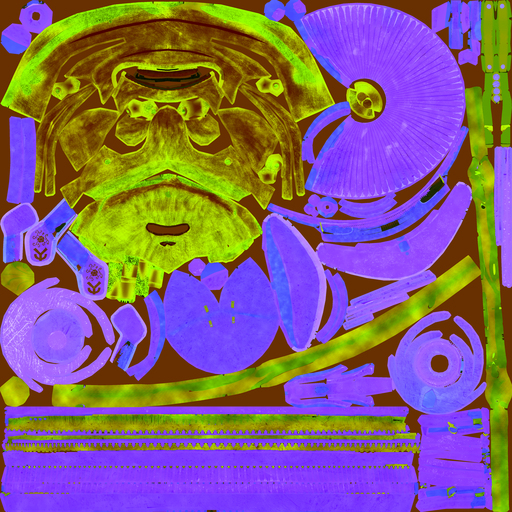} };
            \node[draw,thick,inner sep=0pt] (input_nor) at ($(input_arm) + (-0.2,-0.2)$) { \includegraphics[width=2cm]{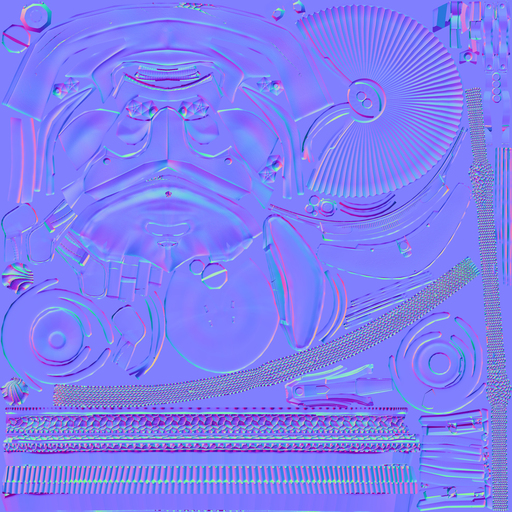} };
            \node[draw,thick,inner sep=0pt] (input_diff) at ($(input_nor) + (-0.2,-0.2)$) { \includegraphics[width=2cm]{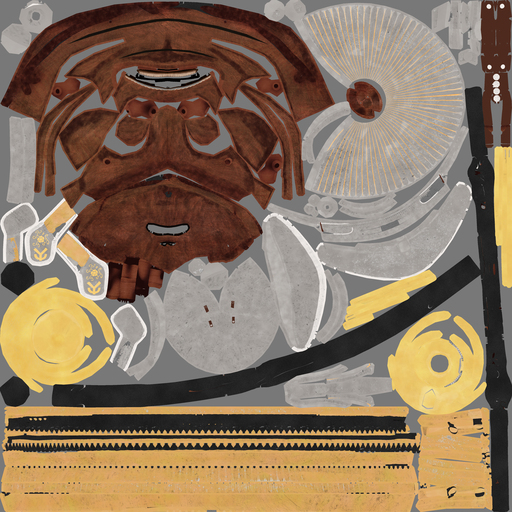} };
            \draw[red,thick] (1.0, 0) node[circle, minimum size=3pt, inner sep=0pt, draw, thick, fill=white]{} -- (2, 0) node[circle, minimum size=3pt, inner sep=0pt, draw, thick, fill=white]{};
            \draw[red,thick,arrow] (-1.38, 0) node[circle, minimum size=3pt, inner sep=0pt, draw, thick, fill=white]{} -| (-3, -0.35) node (A) {};
            \node at ($(input_nor.south) + (0.0,-0.4)$) {\tiny\color{white}{input texture set}};
        \end{scope}

        \node[draw, thick, rounded corners=2pt, fill=white, below=0pt of A, minimum width=1.0cm, minimum height=0.8cm] (enc_1) {\textbf{Hypernetwork}}; 
        \draw[arrow] ($(enc_1.south) + (-0.0,0)$) -- ++(0,-0.5) node (B) {};
        \node[draw,thick,inner sep=0pt, anchor=north west] (out_lat0)  at ($(B.south) + (-0.3,+0.0)$) { \includegraphics[width=1cm]{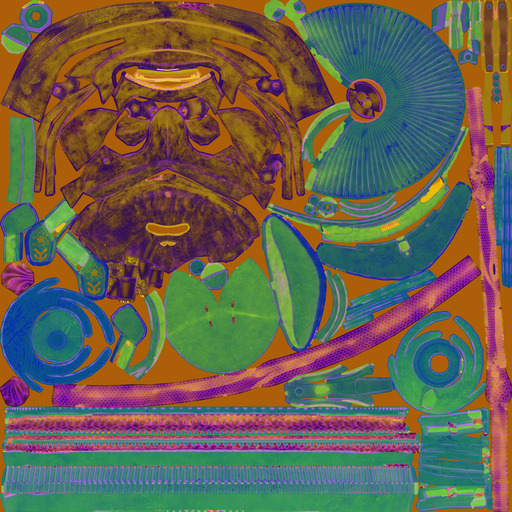} };
        \node[draw,thick,inner sep=0pt] (out_lat1) at ($(out_lat0) + (-0.15,-0.15)$) { \includegraphics[width=1cm]{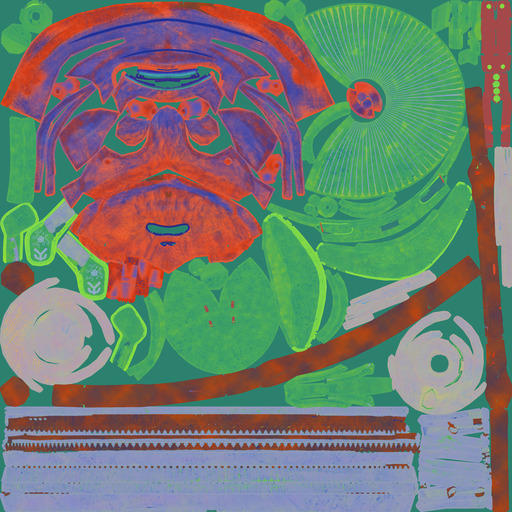} };
        \node[draw,thick,inner sep=0pt] (out_lat2) at ($(out_lat0) + (-0.3,-0.3)$) { \includegraphics[width=1cm]{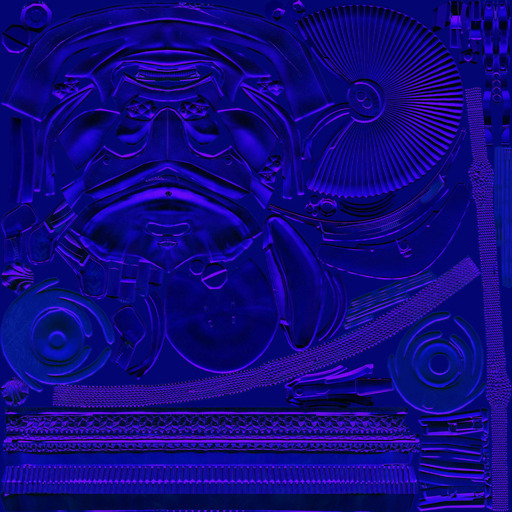} };
        \node[draw,thick,inner sep=0pt] (out_lat3) at ($(out_lat0) + (-0.45,-0.45)$) { \includegraphics[width=1cm]{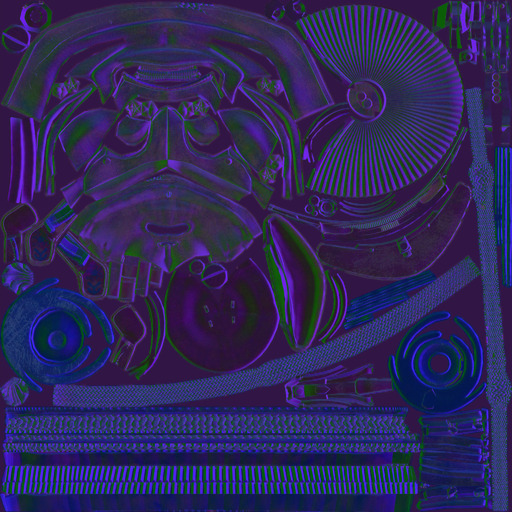} };
        \node[fit={(out_lat0.south west) (out_lat3.north east)}] (input) {};
        \node[inner sep=0pt, anchor=north west] (out_lat0)  at ($(B.south) + (-0.6,-0.3)$) { \includegraphics[width=1cm]{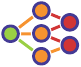} };
        \node at ($(out_lat0) + (0.0,-1.0)$) {\footnotesize\textbf{BCF1}};

        \node[red,draw,thick,inner sep=0pt, right=-2cm of image.north east, anchor=north west, yshift=-0.25cm] (closeup_0) { \includegraphics[width=2.2cm]{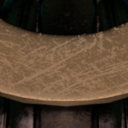} };
        \node[below=-12pt of closeup_0,white!90!black] {\tiny\textbf{Reference}};
        \node[olive,draw,thick,inner sep=0pt, below=2pt of closeup_0] (closeup_1) { \includegraphics[width=2.2cm]{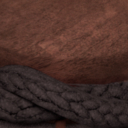} };
        \node[below=-12pt of closeup_1,white!90!black] {\tiny\textbf{Reference}};

        \node[red,draw,thick,inner sep=0pt, right=2pt of closeup_0] (closeup_0_b) { \includegraphics[width=2.2cm]{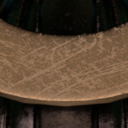} };
        \node[below=-12pt of closeup_0_b,white!90!black] {\tiny\textbf{Ours}};
        \node[olive,draw,thick,inner sep=0pt, right=2pt of closeup_1] (closeup_1_b) { \includegraphics[width=2.2cm]{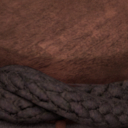} };
        \node[below=-12pt of closeup_1_b,white!90!black] {\tiny\textbf{Ours}};
    \end{tikzpicture}
    \vspace{-.2cm}
 \centering
  \caption{We train a single hypernetwork to learn to compress any texture set into a neural compressed representation (BCF1~\cite{belcour2025tsnc}) and still preserve high-frequency details such as scratches or wood grain once decoded (see insets).}
\label{fig:teaser}
}

\maketitle
\begin{abstract}
Recent work on neural texture compression has demonstrated that it is possible to learn small, per-material texture representations (composed of latent textures and a small Multi-Layer Perceptron decoder) that can be decoded in real-time during shading to reproduce the input to a physically based shading model. However, existing methods require performing gradient-descent optimization per material for a given MLP and latent configuration. In this work, we train a single hypernetwork that outputs both the latent features and the MLP's weights and biases. Though the solution space is high-dimensional, this approach produces results comparable in quality to the current reference neural texture compressors. We further extend this approach to infer multiple decoders at once or even produce decoders that learn super-resolution. 

\ccsdesc[300]{Computing methodologies~Neural networks}
\ccsdesc[300]{Computing methodologies~Machine learning}
\ccsdesc[200]{Computing methodologies~Computer graphics}

\end{abstract}

\section{Introduction}
\label{sec:introduction}
Physically based rendering (PBR) has become the standard for generating photorealistic images in computer graphics~\cite{pharr2023pbrt}. A key component of PBR is the use of material models that describe how light interacts with surfaces. These material models often require multiple texture maps (e.g., diffuse, normal, roughness, metallic) to capture the complex appearance of real-world materials (see OpenPBR specification~\cite{andersson2024openpbr} as an example). However, these texture maps can be large and require significant storage and bandwidth, especially when used in real-time applications such as video games. As an example, developers of \textit{Call of Duty} could not even store the whole texture atlas on console storage~\cite{fowler202extending}.

To mitigate this issue, texture compression algorithms have been developed to reduce the storage and bandwidth requirements of texture maps~\cite{khronos2025image}. Recently, neural material compression methods~\cite{vaidyanathan2023ntc,weinreich2024nbc,farhadzadeh2024neural,belcour2025tsnc} have shown promising results by leveraging the correlation between different texture maps of the same material. These methods use a small neural network (e.g., a Multi-Layer Perceptron or MLP) to decode a compressed representation, achieving higher compression ratios than hardware compression methods.

However, existing neural material compression methods require a separate optimization process for each material, which can be time-consuming and computationally expensive. In this work, we propose a novel approach that trains a single hypernetwork to output both the compressed representation and the decoder's weights and biases for any given material. This allows us to achieve good compression ratios while significantly reducing the time required for optimization. Furthermore, we show that our hypernetwork can be repurposed for other tasks such as super-resolution, demonstrating the versatility of our approach.

\begin{figure*}[ht!]
    \centering
    \begin{tikzpicture}[
    font=\small,
    arrow/.style={-Latex, thick},
    dashedArrow/.style={-Latex, very thick, dashed},
    blk/.style={draw, very thick, minimum height=10mm, minimum width=22mm, align=center, fill=black!2},
    sblk/.style={draw, thick, minimum height=8mm, minimum width=18mm, align=center, fill=black!1},
    tinylabel/.style={font=\scriptsize, align=center},
    ]
        \node[draw,thick,inner sep=0pt, anchor=west] (input_arm) { \includegraphics[width=2cm]{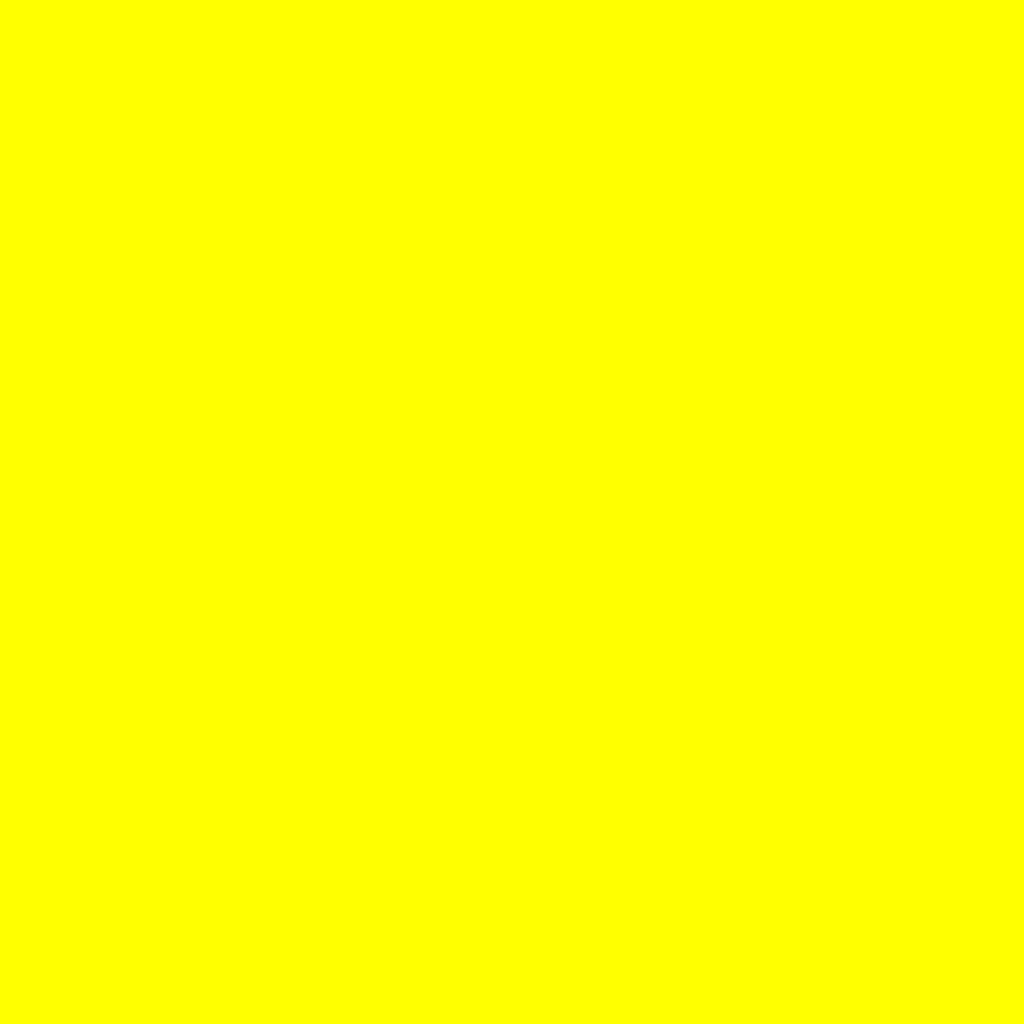} };
        \node[draw,thick,inner sep=0pt] (input_nor) at ($(input_arm) + (-0.2,-0.2)$) { \includegraphics[width=2cm]{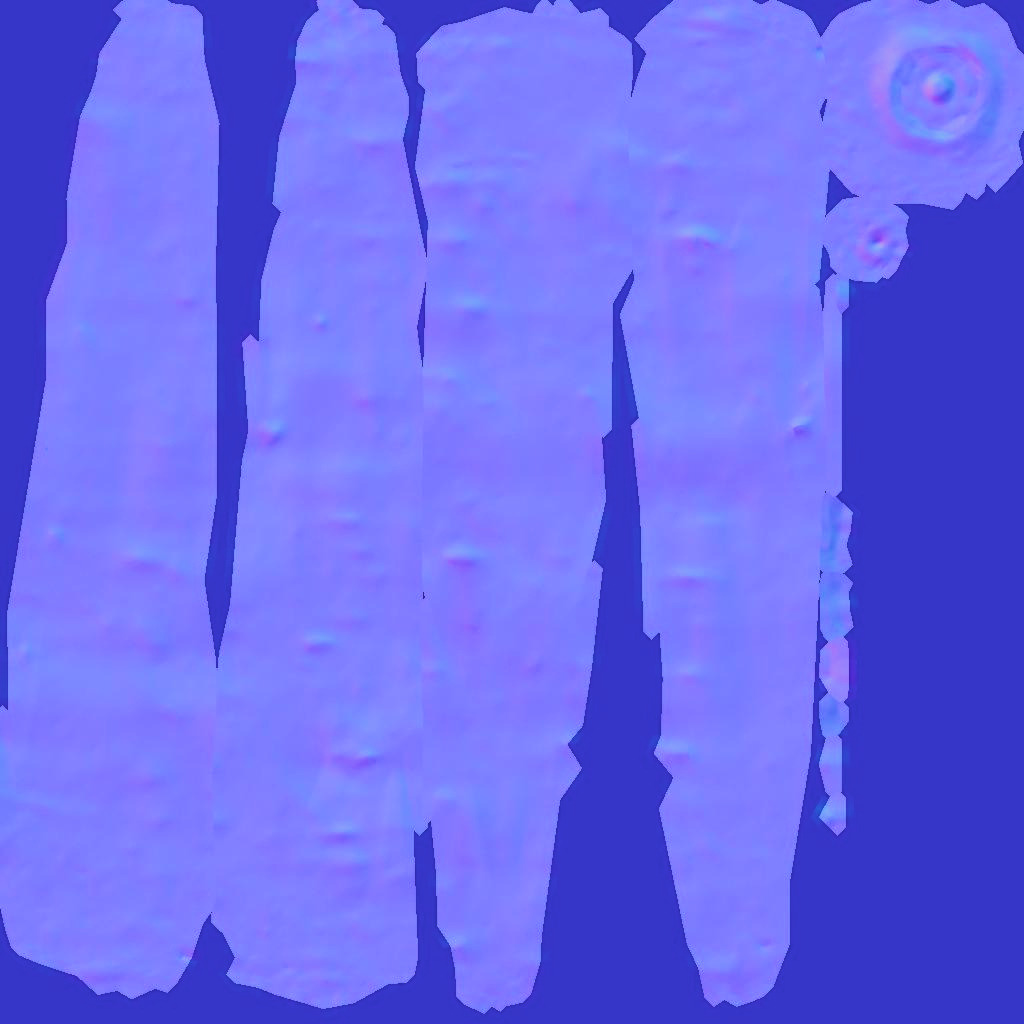} };
        \node[draw,thick,inner sep=0pt] (input_diff) at ($(input_nor) + (-0.2,-0.2)$) { \includegraphics[width=2cm]{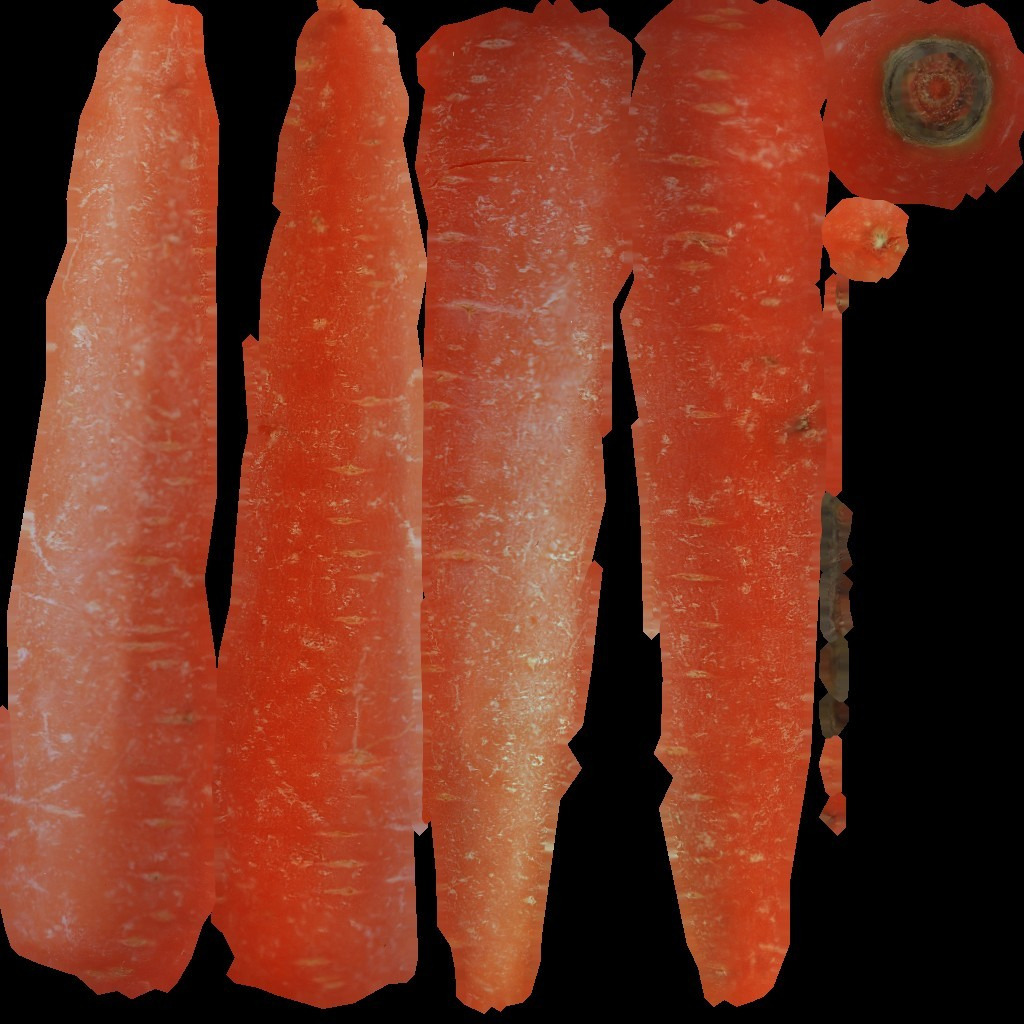} };

        \node[fit={(input_diff.south west) (input_arm.north east)}] (input) {};
        \node[below=0pt of input] {input textures};

        \node[draw, thick, rounded corners=2pt, fill=gray, right=17pt of input, minimum width=3.22cm, minimum height=2cm] (encoder) {};
        \node[below=0pt of encoder] {\textbf{Hypernetwork}};

        \draw[arrow] (input.east) -- ($(encoder.west)+(-2pt,0pt)$);

        \node[draw,thick,inner sep=0pt, anchor=north west] (out_lat0)  at ($(encoder.north east) + (+1.3,+0.2)$) { \includegraphics[width=1cm]{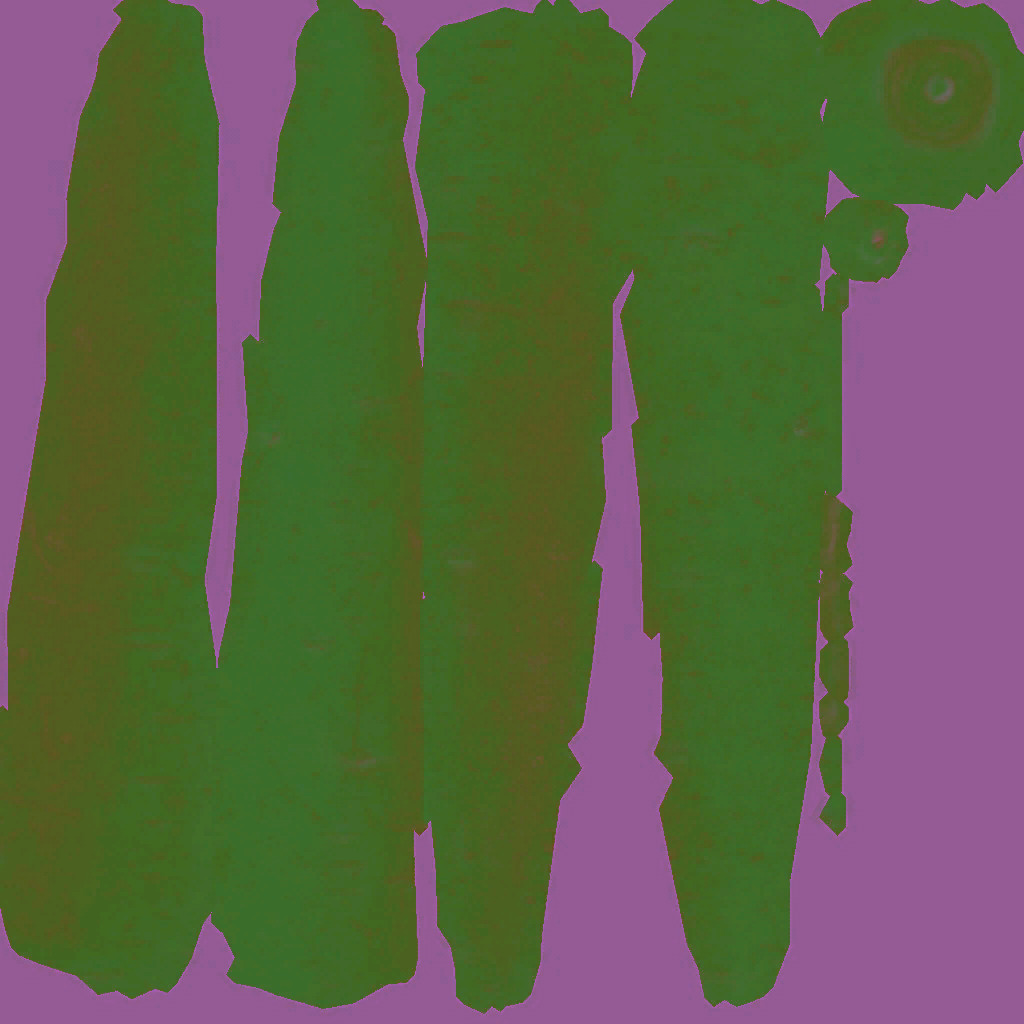} };
        \node[draw,thick,inner sep=0pt] (out_lat1) at ($(out_lat0) + (-0.15,-0.15)$) { \includegraphics[width=1cm]{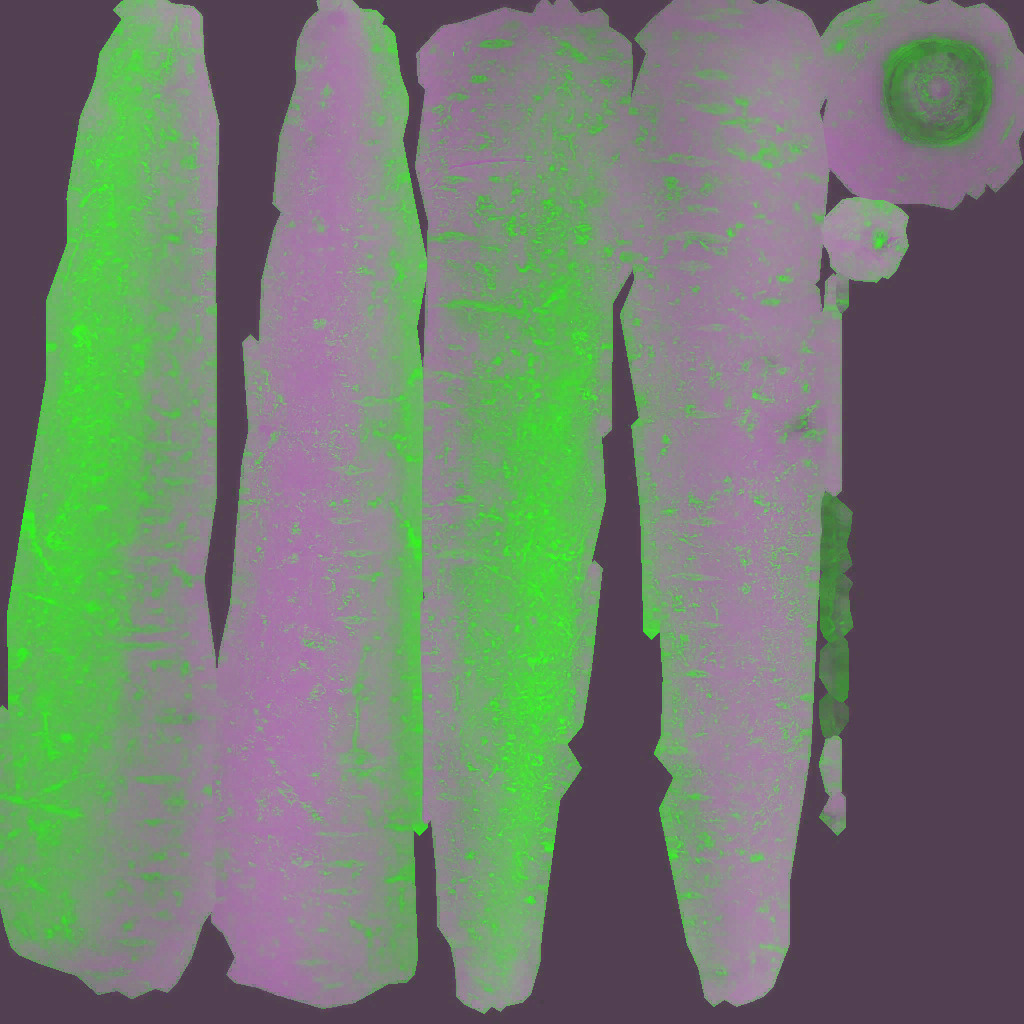} };
        \node[draw,thick,inner sep=0pt] (out_lat2) at ($(out_lat0) + (-0.3,-0.3)$) { \includegraphics[width=1cm]{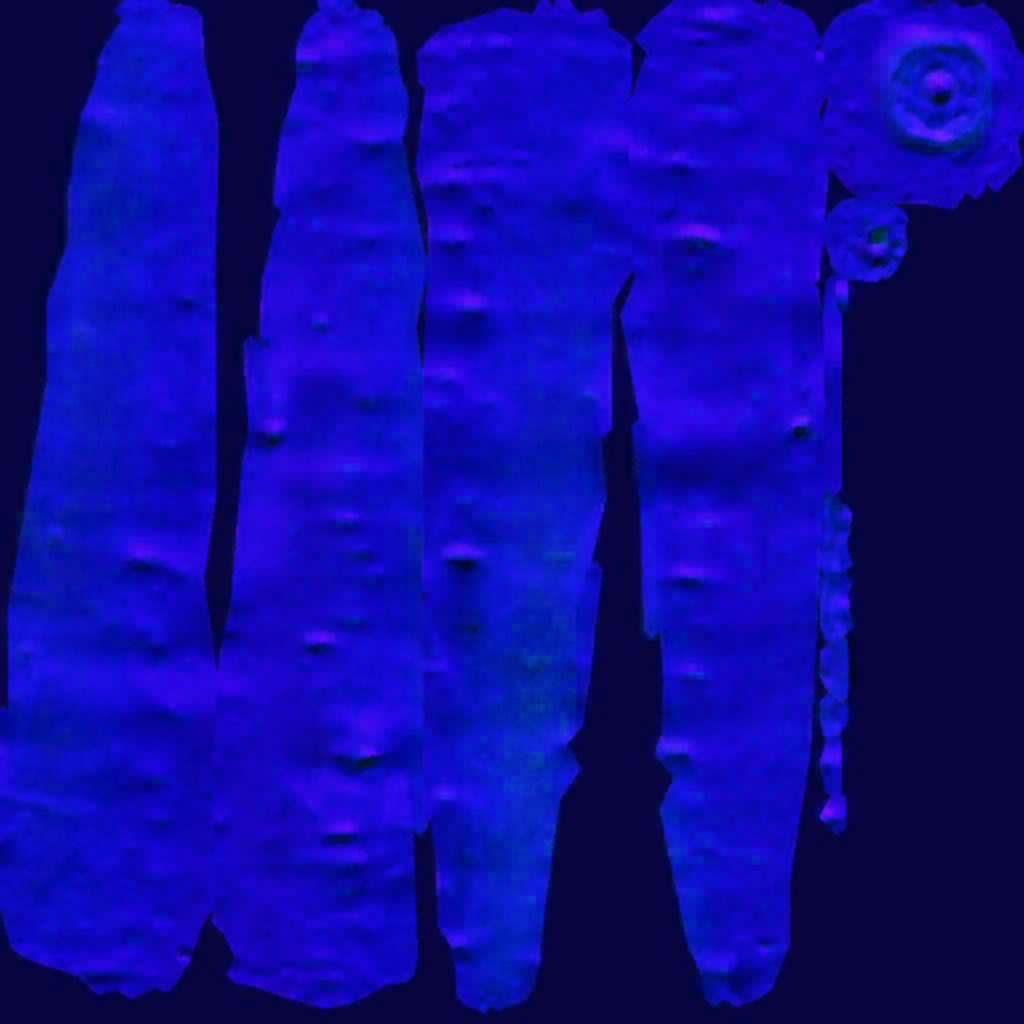} };
        \node[draw,thick,inner sep=0pt] (out_lat3) at ($(out_lat0) + (-0.45,-0.45)$) { \includegraphics[width=1cm]{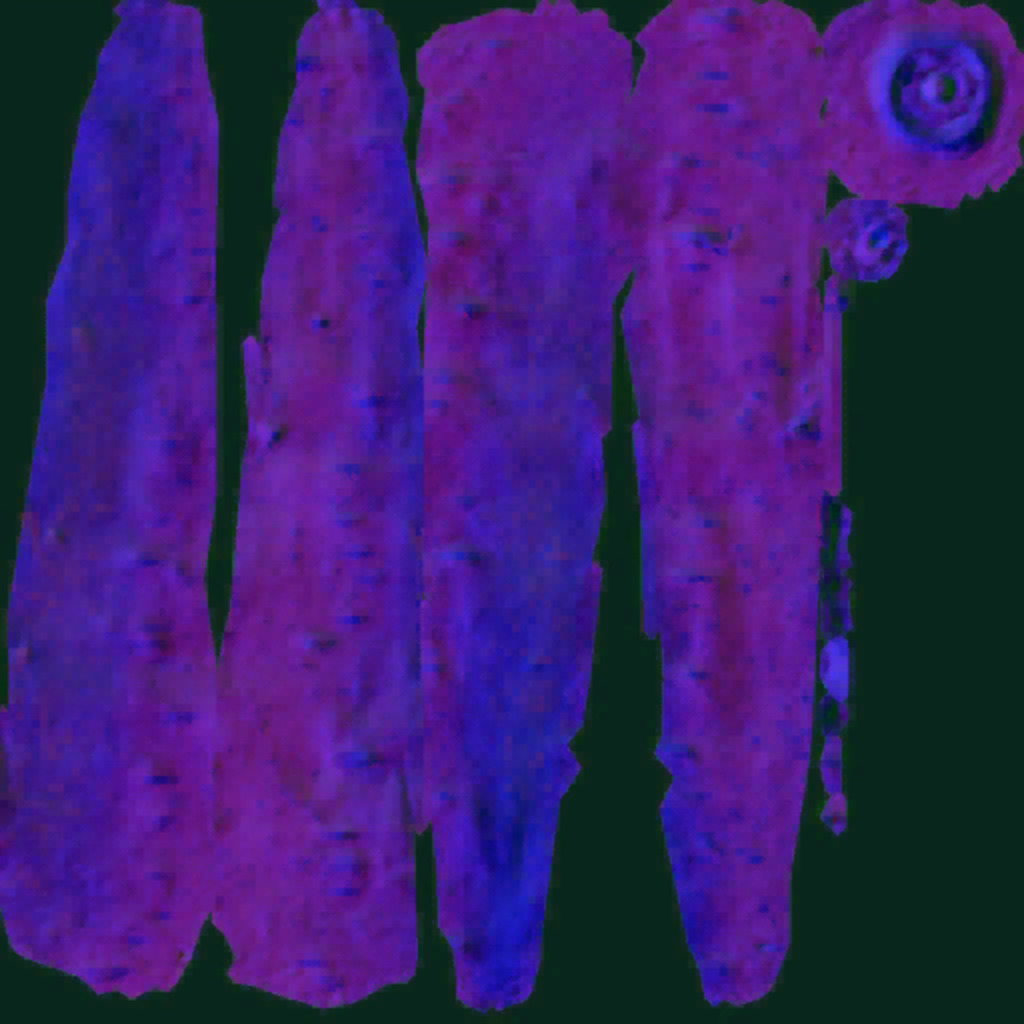} };
        \node[fit={(out_lat0.south west) (out_lat3.north east)}] (input) {};

        \node[draw, thick, rounded corners=2pt, fill=white, right=1cm of input_nor.east, minimum width=20pt, minimum height=1.8cm] (enc_0) {\rotatebox{90}{tokenization}}; 
        
        \node[draw, thick, rounded corners=2pt, fill=white, right=10pt of enc_0.east, minimum width=20pt, minimum height=1.8cm] (enc_1) {\rotatebox{90}{\textbf{$N\times$ DiT}}}; 
        \draw[arrow] (enc_0.east) -- (enc_1.west);

        \node[draw, thick, rounded corners=2pt, fill=white, right=15pt of enc_1.north east, anchor=north west, minimum width=20pt, minimum height=1.3cm] (enc_2) {\rotatebox{90}{$4 \times $\textbf{BC1}}}; 
        \draw[arrow] ($(enc_1.east) + (0,0.25)$) -- (enc_2.west);

        \draw[arrow] (enc_2.east) -- ($(out_lat3.west) + (-0.05,0.0)$);
        \draw[arrow] ($(enc_1.east) + (0,-0.6cm)$) -- ++(1.9,0.0) node (wb_in) {};
        
        \draw[arrow] ($(out_lat3.west) + (1.6,0.0)$) -- ++(0.75,0.0) node[xshift=10pt] (x) {$\begin{bmatrix}\lat0 \\ \lat1 \\ \lat2 \\ \lat3\end{bmatrix}$};
        \draw[arrow] (x.east) -- ++(0.5,0.0) node (x_out){};
        
        \node[right=0.1cm of wb_in] (wb) {$\mathbf{wb}$};
        
        \node[draw, thick, rounded corners=2pt, fill=white, right=0pt of x_out, minimum width=1.2cm, minimum height=0.8cm] (mlp) {{\textbf{MLP}}};
        \draw[arrow] ($(wb.east) + (0.1,0.0)$) -| ($(mlp.south) + (0.0,-0.05)$);

        \draw[arrow] ($(mlp.east) + (0.1,0.0)$) -- ++(0.7,0.0) node (y_out) {};

        \node[draw,thick,inner sep=0pt, anchor=west] (output_arm) at ($(y_out.east) + (0.4,-0.05)$) { \includegraphics[width=2cm]{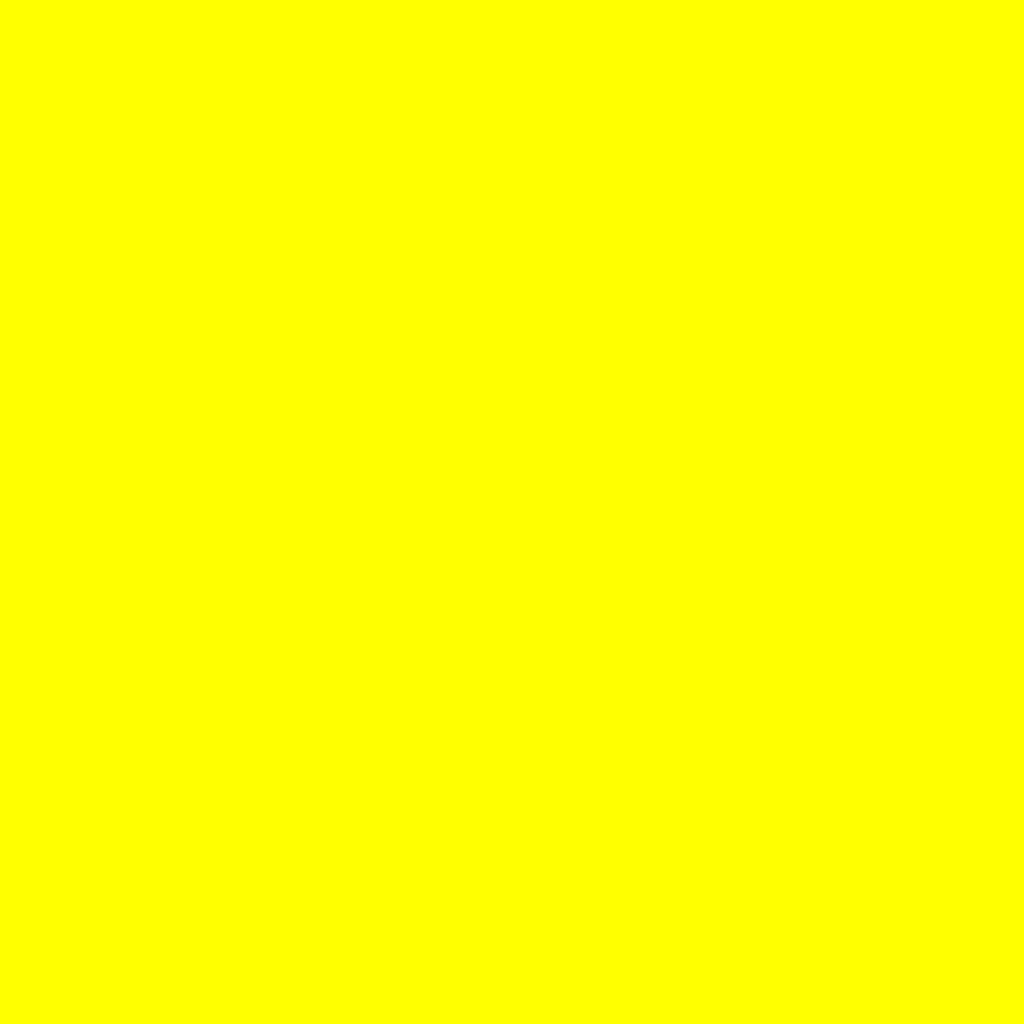} };
        \node[draw,thick,inner sep=0pt] (output_nor) at ($(output_arm) + (-0.2,-0.2)$) { \includegraphics[width=2cm]{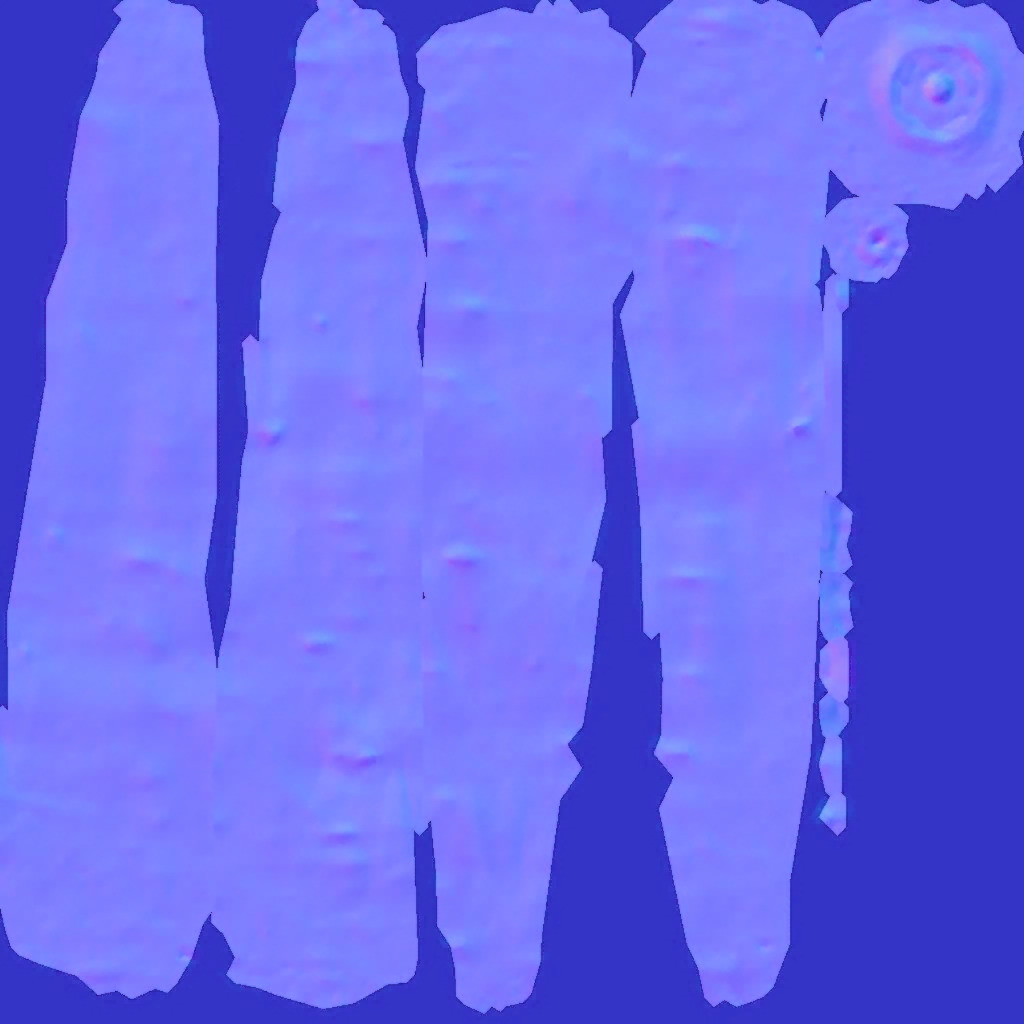} };
        \node[draw,thick,inner sep=0pt] (output_diff) at ($(output_nor) + (-0.2,-0.2)$) { \includegraphics[width=2cm]{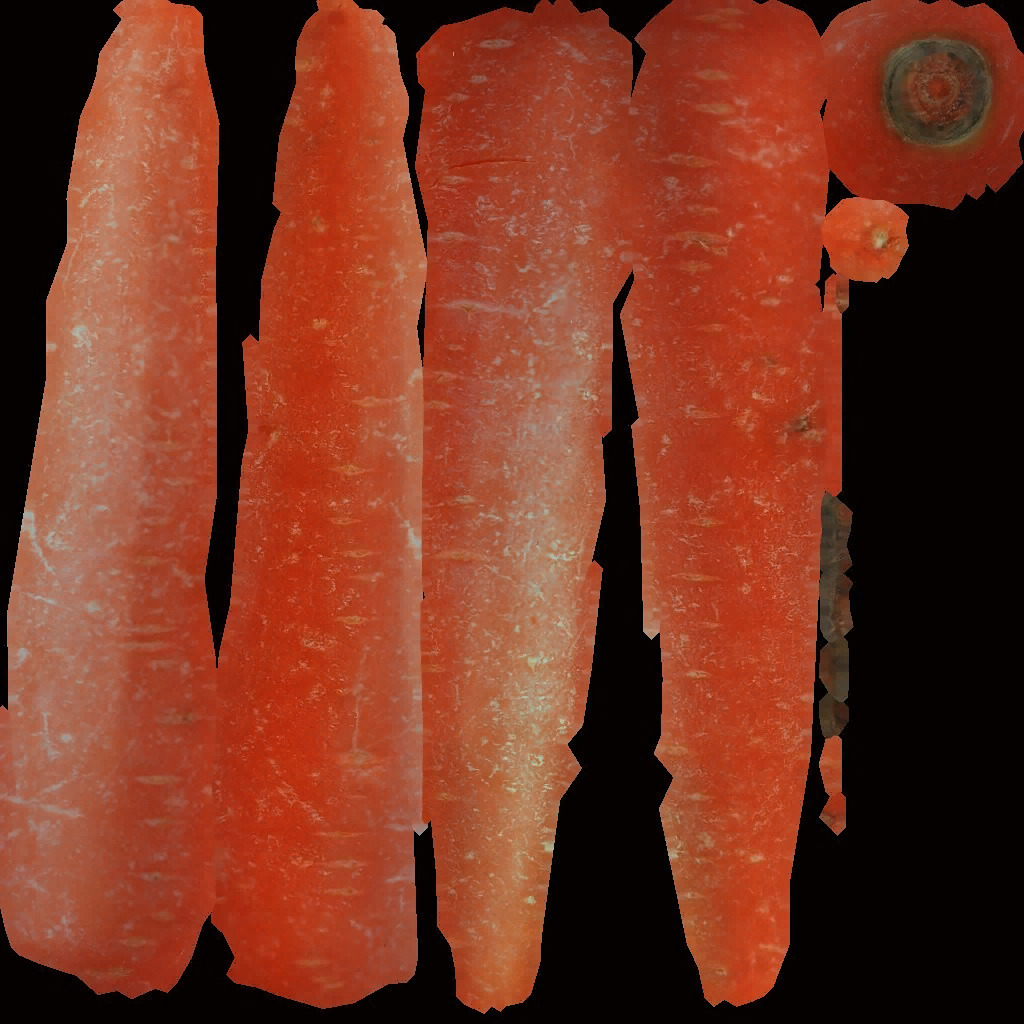} };

        \node[fit={(output_diff.south west) (output_arm.north east)}] (output) {};
        \node[below=0pt of output] {output textures};
        
    \end{tikzpicture}
    \vspace{-5pt}
    \caption{\textbf{Hypernetwork for neural texture compression.} We train a neural network (denoted Hypernetwork) to output the latents, $\lat0, .., \lat3$ and the weights and biases, $\mathbf{wb}$ of a neural model that decompresses to the target texture. Contrary to previous work that optimized $\lat0, .., \lat3, \mathbf{wb}$ with gradient descent, we only regress the Hypernetwork's weights.
    \label{fig:hypernetwork}
    \vspace{-20pt}
    }
\end{figure*}

\section{Previous Work}

\subsection{Neural texture and material compression}
Modern real-time renderers rely on hardware block-compressed textures to reduce bandwidth and storage while preserving random access and mipmapping~\cite{delp1979image,franti1994compression,nystad2012adaptive}. Using small neural networks, it is possible to further compress these textures while maintaining hardware block-compressed evaluation~\cite{fujieda2024ntbc}. Neural texture compression extends this idea by learning compact, per-material representations together with lightweight decoders that can be evaluated on demand.
Vaidyanathan et al.~\cite{vaidyanathan2023ntc} introduced a random-access neural representation that jointly compresses multiple material texture maps, while supporting on-demand decoding similar to block texture formats.

In parallel, Weinreich et al.~\cite{weinreich2024nbc} demonstrated that learned neural materials can be integrated into production pipelines by storing the latent features directly as \emph{block-compressed} texture mip hierarchies and executing a small decoder MLP in a shader, preserving filtering behavior.
Belcour and Benyoub~\cite{belcour2025tsnc} extended this idea by leveraging hardware matrix-multiplication support to accelerate the MLP evaluation, enabling fast rendering of compressed texture sets. However, this method introduces an additional pixel classification step to prevent having multiple MLPs in the same workgroup, which can impede performance. Other methods, such as using 2D Gaussian splats~\cite{zhang2025image}, have been proposed, but they do not match the adaptation to graphics hardware and locality of evaluation that both NTC and NBC provide.

\subsection{Hypernetworks}
Our work is related to the field of hypernetworks, where a model is trained to produce the weights of another network~\cite{ha2016hypernetworks}.
Convolutional hypernetworks have been used to learn the parameters of small MLP decoders to implicitly reproduce textures~\cite{klocek2019hypernetwork}. In computer graphics, hypernetworks have been used to perform style transfer~\cite{chiang2022stylizing, ruta2022hypernst}, to produce Neural Radiance Fields~\cite{sen2023hyp}, and even to fit neural BRDF models~\cite{gokbudak2024hypernetworks}.

\subsection{Neural Enhancement of PBR Materials}

\paragraph*{Super-Resolution}
We show how to retarget our work to increase the resolution of SVBRDFs while maintaining their appearance under relighting.
MatUp repurposes pretrained RGB upsamplers to guide material upsampling in the radiance domain, optimizing a compact per-material MLP filter that enhances fine detail without requiring paired low/high-resolution SVBRDF datasets \cite{gauthier2024matup}.


\section{A Hypernetwork for Material Compression}
\subsection{Problem Statement}

\paragraph*{Texture Sets}
A texture set is defined as a collection of $m$ correlated textures (e.g., diffuse, normal, roughness, metallic):
\begin{align}
\mathcal{T}_{i,j} = \{T_{i,j}^0, T_{i,j}^1, \ldots, T_{i,j}^m\}
\end{align}
where $i,j \geq 0$ are the pixel coordinates. In the following, we will assume normalized texture values $T_{i,j}^0 \in [0,1]$.

\paragraph*{Neural Compression}
The goal of our method is to produce a small neural network $\mlp(\mathbf{x};\;\wb)$ and a set of latent textures $\lat{i}, i\in \left[0, ..., N\right]$ such that:
\begin{align}
\mlp(\left[ \lat{0}^{(i,j)}, \ldots, \lat{N}^{(i,j)} \right];\;\wb) \simeq \mathcal{T}_{i,j}
\end{align}
where $\wb$ are the weights and biases of $\mlp$. The goal is that the cumulative size of $\mlp$ and $\lat{i}, i\in \left[0, ..., N\right]$ is much smaller than the original texture set $\mathcal{T}$.

For that, we train a hypernetwork $\net$ that takes as input a texture set $\mathcal{T}$ and outputs both the latent features $\lat{i}$ and the weights and biases $\wb$ of the MLP:
\begin{align}
\left(\left[\lat{0}, ...\lat{N}\right], \wb\right) = \net(\mathcal{T})
\end{align}

\subsection{Neural Block Compression}
In this work, we use the neural block compression (NBC) of Weinreich et al.~\cite{weinreich2024nbc} as our neural compression method. Specifically, we use the BCF1 variant of Belcour and Benyoub~\cite{belcour2025tsnc}. This defines the latent features $\lat{i}$ as a set of block-compressed textures (using BC1) and the decoder is a Multi-Layer Perceptron (MLP) with 1 hidden layer of $N$ neurons (typically $16$, $32$, or $64$).

\begin{figure*}[h]
    \centering
    \resizebox{\linewidth}{!}{

\begin{tikzpicture}[
    font=\small,
    arrow/.style={-Latex, thick},
    dashedArrow/.style={-Latex, thick, dashed},
    blk/.style={draw, very thick, minimum height=10mm, minimum width=22mm, align=center, fill=black!2},
    sblk/.style={draw, thick, minimum height=8mm, minimum width=18mm, align=center, fill=black!1},
    tinylabel/.style={font=\scriptsize, align=center},
]





\begin{scope}[xshift=-8cm]
    \node (zdit_label) at (0.4, 1.6) { \large \textbf{Hypernetwork} };
    \draw[thick, rounded corners=2pt, fill=white!90!black] (0,1.7) ++(-1.8,-0.63) rectangle ++(4.3,-7.92);

    \draw [arrow] ( 1.1,1.30) node[fill=white] {$\mathcal{T}$}
     -- ++(0,-0.6) node (A0){}
     -- ++(0,-1.4) node (A1){}; 
    \draw [arrow] (A0) -| ++(-1.6,-0.4) node (A2){}; 

    \node[below=0.1cm of A2, draw, thick, rounded corners=2pt, fill=white!90!orange, minimum width=18mm, minimum height=8mm, yshift=5pt] (tenc) {\textbf{t-Encoder}};
    \draw [arrow] (tenc.south) -- ++(0.0,-0.6) node[fill=white!90!black] (A3) {$ \mathbf{t} $} -- ++(0,-0.8) node (dit_t_in) {};

    \node[below=0.1cm of A1, draw, thick, rounded corners=2pt, fill=white, minimum width=18mm, minimum height=8mm, yshift=5pt,align=center] (tokenizer) {\textbf{Tokenizer} \\ {\tiny $4\times 4$ patches}};
    \draw [arrow] (tokenizer.south) -- ++(0.0,-0.4) node (dit0_x_in){}; 
    \node at ($(dit0_x_in)+(0.25,0.15)$) { $\tokenx$ };

    \node[below=0.1cm of dit0_x_in, draw, thick, rounded corners=2pt, fill=white!90!olive, minimum width=28mm, minimum height=8mm, yshift=5pt, xshift=-0.8cm] (dit0) {};
    \begin{scope}
        \clip (dit0.north west) rectangle (dit0.east);
        \node[below=0.1cm of dit0_x_in, draw, thick, rounded corners=2pt, fill=white!90!blue, minimum width=28mm, minimum height=8mm, yshift=5pt, xshift=-0.8cm] (dit0) {};
    \end{scope}
    \node[below=0.1cm of dit0_x_in, draw, thick, rounded corners=2pt, minimum width=28mm, minimum height=8mm, yshift=5pt, xshift=-0.8cm] (dit0) {\textbf{DiT Block}};
    \draw [dashed, thick] ($(dit0.south) + ( 0.8,0)$) -- ++(0,-0.3) node (dit1_x_mid){};
    \draw [dashed, thick] ($(dit0.south) + (-0.8,0)$) -- ++(0,-0.3) node (dit1_t_mid){};
    \draw [arrow] ($(dit1_x_mid.south) + (0,0.0)$) -- ++(0,-0.4) node (dit1_x_in){};
    \draw [arrow] ($(dit1_t_mid.south) + (0,0.0)$) -- ++(0,-0.4) node (dit1_t_in){};

     \node[below=0.1cm of dit1_x_in, draw, thick, rounded corners=2pt, fill=white!90!olive, minimum width=28mm, minimum height=8mm, yshift=5pt, xshift=-0.8cm] (dit1) {};
    \begin{scope}
        \clip (dit1.north west) rectangle (dit1.east);
        \node[below=0.1cm of dit1_x_in, draw, thick, rounded corners=2pt, fill=white!90!blue, minimum width=28mm, minimum height=8mm, yshift=5pt, xshift=-0.8cm] (dit1) {};
    \end{scope}
    \node[below=0.1cm of dit1_x_in, draw, thick, rounded corners=2pt, minimum width=28mm, minimum height=8mm, yshift=5pt, xshift=-0.8cm] (dit1) {\textbf{DiT Block}};
    \node at ($(dit1.north) + (0,0.3)$) {$\times N_B$};

    \draw [arrow] ($(dit1.south) + ( 0.8,0)$) -- ++(0,-0.8) node (dit2_x_mid){};
    \draw [arrow] ($(dit1.south) + (-0.8,0)$) -- ++(0,-0.8) node (dit2_t_mid){};

    \node[draw, thick, rounded corners=2pt, fill=white, below=-0.1cm of dit2_t_mid] (proj_t_in) { Linear };
    \draw [arrow] (proj_t_in.south) -- ++(0,-1.4) node[xshift=10pt,yshift=4pt] (wb){ $\mathbf{wb}$};

    \node[draw, thick, rounded corners=2pt, fill=white!90!purple, below=-0.1cm of dit2_x_mid, minimum width=18mm, minimum height=8mm] (bc1_0) { \textbf{BC1 Conv.} };
    \draw [arrow] ($(bc1_0.south) + (-0.7,0)$) -- ++(0,-1.0) node[xshift=7pt,yshift=3pt] (bc1){ $\lat0$};
    \draw [arrow] ($(bc1_0.south) + (-0.2,0)$) -- ++(0,-1.0) node[xshift=7pt,yshift=3pt] (bc1){ $\lat1$};
    \draw [arrow] ($(bc1_0.south) + (+0.3,0)$) -- ++(0,-1.0) node[xshift=7pt,yshift=3pt] (bc1){ $\lat2$};
    \draw [arrow] ($(bc1_0.south) + (+0.8,0)$) -- ++(0,-1.0) node[xshift=7pt,yshift=3pt] (bc1){ $\lat3$};
\end{scope}

\begin{scope}[xshift=-3cm]
    \node (zdit_label) at (0.4, 1.6) { \large \textbf{t-Encoder} };

    \draw[thick, rounded corners=2pt, fill=white!90!orange] (0,1.7) ++(-1.8,-0.63) rectangle ++(4.3,-7.92);
    \draw [arrow] ( 1.0,1.30) node[fill=white] {$\mathcal{T}$} -- ++(0,-0.6) node (A){};
    \draw [arrow] (-0.4,0.80) node[fill=white!90!orange] {$\tokent_0$} -- ++(0,-1.05) node (B) {};

    \node[draw,trapezium, shape border rotate=180, rounded corners=2pt, fill=white, yshift=-3pt] (C) at (A.south) {\textbf{Conv2D}};
    \draw [arrow] (C.south) -- ++(0,-0.1) -| ++(-0.35,-0.4) node (A0){};
    \draw [arrow] (C.south) -- ++(0,-0.1) -| ++(+0.35,-0.4) node (A1){};
    \draw [dashed,thick] (C.south) -- ++(0,-1.4)  node (A2){};
    \draw [arrow] (A2.south) -- ++(0,-0.4)  node (A3){};
    
    \draw (B.south) node[draw, thick, rounded corners=2pt, fill=white, yshift=-8pt, xshift=0.8cm, minimum width=3.0cm, minimum height=0.7cm] (zblock_1) {Cross Attention};
    
    \node[violet, xshift=-0.5cm, yshift=0.55cm] at (zblock_1) {Q};
    \node[violet, xshift=0.0cm, yshift=0.55cm] at (zblock_1) {K};
    \node[violet, xshift=1.2cm, yshift=0.55cm] at (zblock_1) {V};

    \draw[arrow] ($(zblock_1.south) + (-0.8,0.0)$) -- ++(0,-1.58) node (B2) {};
    \node[draw,trapezium, shape border rotate=180, rounded corners=2pt, fill=white, yshift=-3pt] (C1) at (A3.south) {\textbf{Conv2D}};

    \draw [arrow] (C1.south) -- ++(0,-0.1) -| ++(-0.35,-0.4) node (A20){};
    \draw [arrow] (C1.south) -- ++(0,-0.1) -| ++(+0.35,-0.4) node (A21){};
    \draw [dashed,thick] (C1.south) -- ++(0,-1.4)  node (A22){};
    \draw [arrow] (A22.south) -- ++(0,-0.4)  node (A23){};

    \draw (B2.south) node[draw, thick, rounded corners=2pt, fill=white, yshift=-8pt, xshift=0.8cm, minimum width=3.0cm, minimum height=0.7cm] (zblock_2) {Cross Attention};
    
    \node[violet, xshift=-0.5cm, yshift=0.55cm] at (zblock_2) {Q};
    \node[violet, xshift=0.0cm, yshift=0.55cm] at (zblock_2) {K};
    \node[violet, xshift=1.2cm, yshift=0.55cm] at (zblock_2) {V};

    \draw[arrow] ($(zblock_2.south) + (-0.8,0.0)$) -- ++(0,-1.58) node (B2) {};
    \node[draw,trapezium, shape border rotate=180, rounded corners=2pt, fill=white, yshift=-3pt] (C1) at (A23.south) {\textbf{Conv2D}};

    \draw [arrow] (C1.south) -- ++(0,-0.1) -| ++(-0.35,-0.4) node (A20){};
    \draw [arrow] (C1.south) -- ++(0,-0.1) -| ++(+0.35,-0.4) node (A21){};

    \draw (B2.south) node[draw, thick, rounded corners=2pt, fill=white, yshift=-8pt, xshift=0.8cm, minimum width=3.0cm, minimum height=0.7cm] (zblock_2) {Cross Attention};
    
    \node[violet, xshift=-0.5cm, yshift=0.55cm] at (zblock_2) {Q};
    \node[violet, xshift=0.0cm, yshift=0.55cm] at (zblock_2) {K};
    \node[violet, xshift=1.2cm, yshift=0.55cm] at (zblock_2) {V};

    \draw[arrow] ($(zblock_2.south) + (-0.8,0.0)$) -- ++(0,-1.50) node[xshift=7pt,yshift=4pt] (B2) {$\tokent$};
\end{scope}

\begin{scope}[xshift=2cm]
    \node (zdit_label) at (0.4, 1.6) { \large \textbf{DiT Block} };

    \draw[thick, rounded corners=2pt, fill=white!90!blue] (0,1.7) ++(-1.8,-0.63) rectangle ++(4.3,-4.05);
    \draw[thick, rounded corners=2pt, fill=white!90!olive] (0,1.7) ++(-1.8,-4.75) rectangle ++(4.3,-3.80);

    \draw [arrow] (1.2,1.3) node[fill=white] {$\tokent$} -- ++(0,-1.35) node (t_split){} -- ++(0,-0.6);
    \draw [arrow] (t_split) -- ++(0.5,0) -- ++(0,-0.6);
    \draw [dashed,thick] (t_split) ++(0,-1.0) -- ++(0,-1.5) node[fill=white!90!blue] (C) {$\tokent$};
    \draw [arrow] (C) -- ++(0,-4.70);
    \draw [arrow] (0,1.3) node[fill=white] {$\tokenx$} -- ++(0,-0.3) node (A){} -- ++(0,-0.4) 
    node[draw, thick, rounded corners=2pt, fill=white, yshift=-8pt] (zblock_0) {Layer Norm};
    \draw[arrow] (zblock_0.south) -- ++(0, -0.35) node[draw, thick, rounded corners=2pt, fill=white, yshift=-12pt, xshift=0.7cm, minimum width=3.0cm, minimum height=0.7cm] (zblock_1) {Cross Attention};
    \node[violet, xshift=-0.5cm, yshift=0.55cm] at (zblock_1) {Q};
    \node[violet, xshift=0.3cm, yshift=0.55cm] at (zblock_1) {K};
    \node[violet, xshift=1.2cm, yshift=0.55cm] at (zblock_1) {V};

    \draw[arrow] ($(zblock_1.south) + (-0.7,0)$) -- ++(0,-0.35) node[draw, thick, rounded corners=2pt, fill=white, yshift=-8pt] (zblock_2) {Layer Norm};
    
    \draw[arrow] (zblock_2.south) -- ++(0,-0.35) node[draw, circle, thick, fill=white, yshift=-9pt] (zblock_3) {$+$};

    \draw [arrow] (A) -- ++(-1.5,0) -- ++(0,-2) |- (zblock_3.west);

    \draw [arrow] (zblock_3.south) -- ++(0,-0.25) node (D) {} -- ++(0,-0.4) 
    node[draw, thick, rounded corners=2pt, fill=white, yshift=-8pt] (ff0) {Layer Norm};

    \draw [arrow] (ff0.south) -- ++(0,-0.35) 
    node[draw, thick, rounded corners=2pt, fill=white, yshift=-8pt] (ff1) {Feed Forward};

    \draw [arrow] (ff1.south) -- ++(0,-0.35) 
    node[draw, thick, rounded corners=2pt, fill=white, yshift=-8pt] (ff2) {Layer Norm};

    \draw[arrow] (ff2.south) -- ++(0,-0.35) node[draw, circle, thick, fill=white, yshift=-9pt] (p2) {$+$};

    \draw[arrow] (D) -- ++(-1.5,0.0) -- ++(0.0,-1.0) |- (p2.west);
    \draw[arrow] (p2.south) -- ++(0.0,-0.45) node (A) {};

    \node[xshift=7pt,yshift=8pt] at (A.south) (B2) {$\tokenx$};
    \node[xshift=1.4cm,yshift=8pt] at (A.south) (B2) {$\tokent$};
\end{scope}

\begin{scope}[xshift=7cm]
    \node (zdit_label) at (0.4, 1.6) { \large \textbf{BC1 Conversion} };

    \draw[thick, rounded corners=2pt, fill=white!90!purple] (0,1.7) ++(-1.8,-0.63) rectangle ++(4.3,-7.92);

    \draw [arrow] (0.4,1.3) node[fill=white] {$\tokenx$} -- ++(0,-0.8) node (A){};

    \node[draw, thick, rounded corners=2pt, fill=white, minimum width=18mm, minimum height=8mm] (reshape) at (A.south) { \tiny (B,N,C) $\to$ (B,C,H/4,W/4) };
    
    \draw[arrow] (reshape.south) -- ++(0,-0.4) node (B){};
    \node[draw, thick, rounded corners=2pt, fill=white, minimum width=18mm, minimum height=8mm, yshift=-2.8mm, align=center] (conv2d) at (B.south) { \textbf{Conv2D}};

    \draw[thick, arrow] (conv2d.south) |- ++(-1.0,-0.2) node (C){} -| ++(0,-0.4) node (D){};
    \node at (C) [xshift=-0.5cm, yshift=-0.2cm] {\tiny $d=22$};
    \node[draw, thick, rounded corners=2pt, fill=white, minimum width=18mm, minimum height=8mm, yshift=-5.0mm, align=center] (offset_sigmoid) at (D.south) {\tiny  \textbf{sigmoid} \\ \tiny  \textbf{offsets} \\ \tiny  \textbf{quantization}};
    \draw[arrow] ($(offset_sigmoid.south) + (-0.4,0)$) -- ++(0,-4.2) node (E){};
    \node[xshift=0.3cm,yshift=8pt] at (E.south) (B2) {$\lat0$};

    \draw[thick, arrow] ($(conv2d.south) + (0.4,0)$) -- ++(0,-2.0) node (D){};
    \node[draw, thick, rounded corners=2pt, fill=white, minimum width=1mm, minimum height=6mm, yshift=-2.8mm, align=center, trapezium, shape border rotate=180] (conv2d) at (D.south) { \textbf{Conv2D}};

    \draw[thick, arrow] (conv2d.south) |- ++(-0.5,-0.2) node (C){} -| ++(0,-0.4) node (D){};
    \node at (C) [xshift=-0.5cm, yshift=-0.2cm] {\tiny $d=22$};
    \node[draw, thick, rounded corners=2pt, fill=white, minimum width=18mm, minimum height=8mm, yshift=-5.0mm, align=center] (offset_sigmoid) at (D.south) {\tiny  \textbf{offsets} \\  \tiny \textbf{sigmoid} \\ \tiny  \textbf{quantization}};
    \draw[arrow] ($(offset_sigmoid.south) + (-0.4,0)$) -- ++(0,-1.5) node (E){};
    \node[xshift=0.3cm,yshift=8pt] at (E.south) (B2) {$\lat1$};

    \draw[dashed, thick] ($(conv2d.south) + (0.6,0)$) -- ++(0,-1.0) node[xshift=0.3cm] {$\times 2$} -- ++(0,-1.0) node (D){};
    \draw[arrow] (D) -- ++(0,-0.5);

    \draw[arrow] ($(D) + (-0.6,-0.7)$) -- ++(0,-0.6) node (E){};
    \node[xshift=0.3cm,yshift=8pt] at (E.south) (B2) {$\lat2$};

    \draw[arrow] ($(D) + (+0.2,-0.7)$) -- ++(0,-0.6) node (E){};
    \node[xshift=0.3cm,yshift=8pt] at (E.south) (B2) {$\lat3$};
\end{scope}
\end{tikzpicture}
}
    \vspace{-10pt}
    \caption{\textbf{The Hypernetwork architecture.} A DiT-style~\cite{dosovitskiy2020vit,peebles2022dit} architecture processes a normalized input tensor $\mathcal{T}$ of shape $(B, 9, H, W)$ to produce $4$ latent textures and the weights and biases of an MLP. We decompose it into 3 main blocks: first the input image is fed to the \textbf{t-Encoder} to extract a small set of global tokens $\tokent$; second the \textbf{DiT Block} processes the tokenized input $\tokenx$ using many cross-attention blocks; last, we project the output of the DiT block into (i) BC1 latents $\{\lat{\ell}\}_{\ell=0}^3$ and (ii) the weights $\mathbf{wb}$ of the per-material MLP decoder. During that phase, we replicate the differentiable BC1~\cite{belcour2025tsnc} by extracting tensors with $22$ dimensions from $\textbf{x}$, applying to them a sigmoid function along with an offset, and quantizing the result.
    \label{fig:architecture_zdit}
    \vspace{-10pt}
    }
\end{figure*}
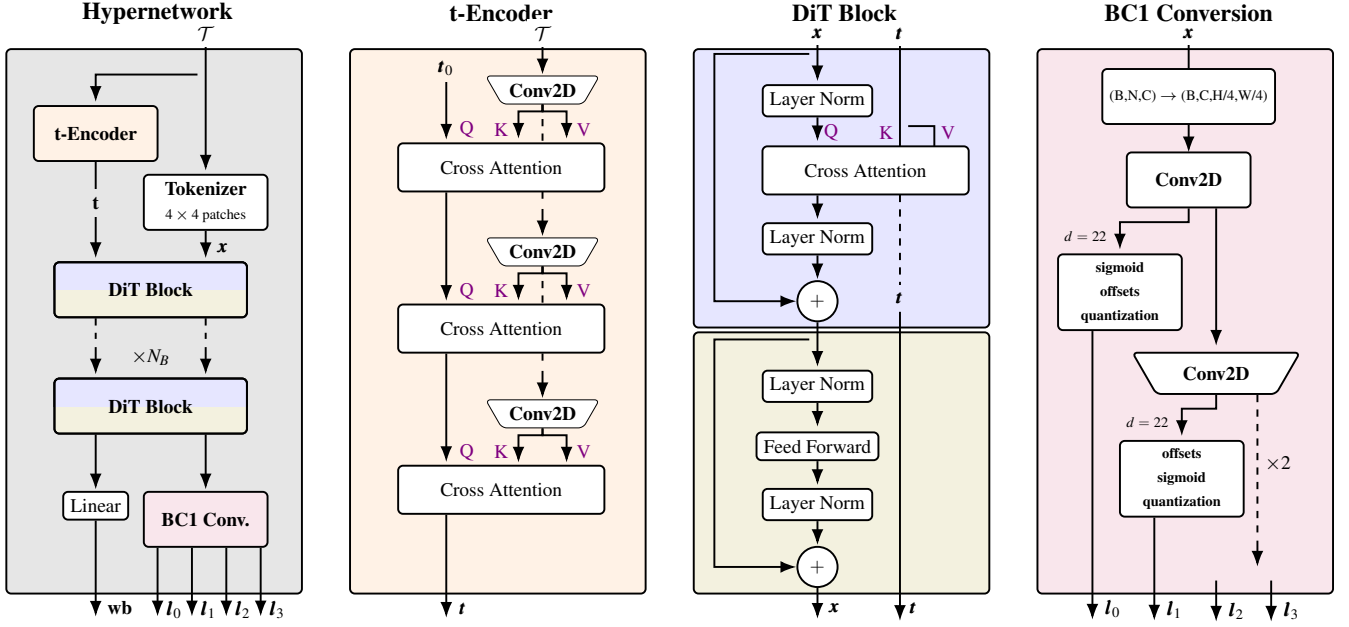

Instead of relying on gradient descent to optimize the latent features and the MLP weights for each texture set, we train a monolithic hypernetwork $\net$ that takes as input a texture set $\mathcal{T}$ and outputs both the latent features $\lat{i}$ and the weights $\wb$ of the MLP (see Fig.~\ref{fig:hypernetwork} for an illustration of this pipeline).

In the following, we assume that the input texture set is normalized:
$
\Mean[\mathcal{T}^k] = 0 \, \mbox{and}\, \Var[\mathcal{T}^k] = 1
$. Prior to forwarding $\mathcal{T}$ to the Hypernetwork, we normalize it and later scale and shift the last layer of the generated MLP to include this bias and scale.

\subsection{The Hypernetwork Architecture}

We tested several architectures for the hypernetwork $\net$ (U-Net, ConvNext, ...). The architecture working best in this case is a Diffusion Transformer architecture~\cite{peebles2022dit} with cross-attention (see Figure~\ref{fig:architecture_zdit}-left). In the following, $D \in \mathbb{N}$ is the network's inner depth. The overall architecture decomposes into three major sections:

\begin{itemize}
    \item First, the input texture set is encoded to a set of tokens $\tokent \in \mathbb{R}^{N \times D}$ using the \textbf{t-Encoder}. We use a number of tokens large enough to represent the MLP (i.e. $|\tokent| \geq |\wb|$).
    \item Then, we convert the input texture set into a set of tokens $\tokenx \in \mathbb{R}^{M \times D}$ by grouping together blocks of $4 \times 4$ texels and applying several transformer blocks (\textbf{DiT blocks}) with cross-attention between the image tokens $\tokenx$ and the latent tokens $\tokent$. As such, $M = {W \times H \over 4 \times 4}$ is fixed.
    \item Finally, we decode $\tokent$ into $\wb$ and $\lat{}$ into $\lat{i}$ using a \textbf{BC1 conversion} module.
\end{itemize}

\paragraph*{t-Encoder.} We convert the texture set $\ts$ to a latent representation $\tokent$ using a mixed convolutional and attention architecture (Figure~\ref{fig:architecture_zdit}-orange). We downscale the input texture set using several convolutional layers and apply a cross-attention block with a learned default token $\tokent_0$ as the queries (and outputs) after each convolution. The activations from the texture set are used for the keys and values. At the end of the t-Encoder, we have a set of tokens $\tokent$ that contains global information about the input texture set.

\paragraph*{DiT Blocks.} The main part of the Hypernetwork (Figure~\ref{fig:architecture_zdit}-blue and green) follows a standard cross-attention architecture. We apply $N_B$ blocks of cross-attention between the image tokens $\tokenx$ and the latent tokens $\tokent$. We use a sandwiched normalization~\cite{kim2025peri} for both the attention and the feed-forward parts to avoid dead signals.

\paragraph*{BC1 Conversion}
The last step of the Hypernetwork is to convert the latent tokens $\tokent$ into the weights and biases $\wb$ of the MLP and the image tokens $\tokenx$ into the latent textures $\lat{i}$. We use a simple linear layer to convert $\tokent$ into $\wb$. For the latent textures, we use the BC1 conversion module (Figure~\ref{fig:architecture_zdit}-pink). This module reshapes $\tokenx$ from a set of tokens $(N,C)$ to a 3D tensor $(W/4, H/4, C)$. For each latent texture, a convolution layer changes the $C$ channels into $22$ channels that produce a BC1 block: 2 RGB endpoints $e_0$, $e_1$ and a grid of $4 \times 4$ blending parameters $\alpha_{u,v}$.

Depending on the BCF1 variant used (the \textsc{varA} or \textsc{varB} defined by the BCF1 method~\cite{belcour2025tsnc}), the convolutions can have different stride and kernel size. For example, \textsc{varA} outputs latents of $[(W,H), (W,H), (W/2,H/2), (W/2,H/2)]$ resolution resulting in strides of $[1,1,2,1]$. \textsc{varB} outputs latents of $[(W,H), (W/2,H/2), (W/4,H/4), (W/8,H/8)]$ resolution resulting in strides of $[1,2,2,2]$.

To avoid collapse of the endpoints of BC blocks, we offset $e_0$ and $e_1$ by $-3$ and $3$ respectively before passing them through a sigmoid function and quantization (last block in Fig.~\ref{fig:architecture_zdit}-pink). Each $\alpha_{u,v}$ is also passed through a sigmoid function and quantized to $2$ bits.

\begin{figure*}[t!]
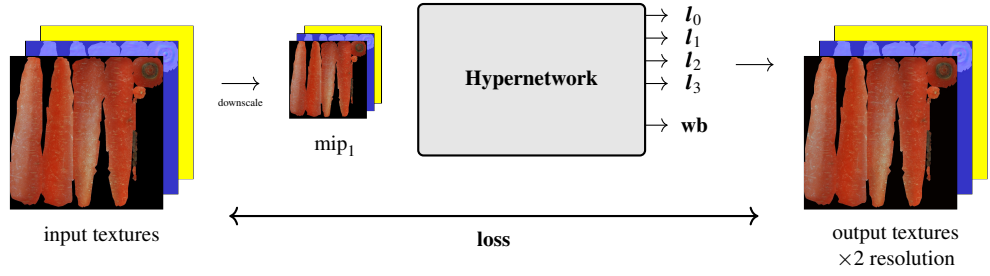

    \centering
    \begin{tikzpicture}[
    font=\small,
    arrow/.style={-Latex, thick},
    dashedArrow/.style={-Latex, very thick, dashed},
    blk/.style={draw, very thick, minimum height=10mm, minimum width=22mm, align=center, fill=black!2},
    sblk/.style={draw, thick, minimum height=8mm, minimum width=18mm, align=center, fill=black!1},
    tinylabel/.style={font=\scriptsize, align=center},
    ]

    \begin{scope}[yshift=-0.5cm]
        \node[draw,thick,inner sep=0pt, anchor=west] (input_arm) { \includegraphics[width=2cm]{fig/architecture/img/neural_0_ref_arm.jpg} };
        \node[draw,thick,inner sep=0pt] (input_nor) at ($(input_arm) + (-0.2,-0.2)$) { \includegraphics[width=2cm]{fig/architecture/img/neural_0_ref_norm.jpg} };
        \node[draw,thick,inner sep=0pt] (input_diff) at ($(input_nor) + (-0.2,-0.2)$) { \includegraphics[width=2cm]{fig/architecture/img/neural_0_ref_diff.jpg} };
        \node[fit={(input_diff.south west) (input_arm.north east)}] (input) {};
        \node[below=0pt of input] {input textures};
    \end{scope}

    \begin{scope}[xshift=3.5cm, scale=0.5, transform shape]
        \node[draw,thick,inner sep=0pt, anchor=west] (input_arm) { \includegraphics[width=2cm]{fig/architecture/img/neural_0_ref_arm.jpg} };
        \node[draw,thick,inner sep=0pt] (input_nor) at ($(input_arm) + (-0.2,-0.2)$) { \includegraphics[width=2cm]{fig/architecture/img/neural_0_ref_norm.jpg} };
        \node[draw,thick,inner sep=0pt] (input_diff) at ($(input_nor) + (-0.2,-0.2)$) { \includegraphics[width=2cm]{fig/architecture/img/neural_0_ref_diff.jpg} };
        \node[fit={(input_diff.south west) (input_arm.north east)}] (input) {};
        \draw[->] (-2.2,-0.5) -- +(1,0);
        \node at (-1.7,-1.0) {downscale};
    \end{scope}
    \node[below=0pt of input] {$\mbox{mip}_1$};

    \begin{scope}[xshift=2.4cm]
        \draw[thick, rounded corners=2pt, fill=white!90!black] (2.6,0.8) rectangle ++(3.0,-2.0);
        \node at (4.1,-0.2) {\textbf{Hypernetwork}};
        \draw[->] (5.6,+0.65) -- +(0.3,0.0) node[xshift=10pt] (A) {$\lat{0}$};
        \draw[->] (5.6,+0.35) -- +(0.3,0.0) node[xshift=10pt] (A) {$\lat{1}$};
        \draw[->] (5.6,+0.05) -- +(0.3,0.0) node[xshift=10pt] (A) {$\lat{2}$};
        \draw[->] (5.6,-0.25) -- +(0.3,0.0) node[xshift=10pt] (A) {$\lat{3}$};
        \draw[->] (5.6,-0.8) -- +(0.3,0.0) node[xshift=10pt] (A) {$\mathbf{wb}$};

        \draw[->] (6.8,+0.0) -- +(0.5,0.0);
    \end{scope}

    \begin{scope}[xshift=10.5cm, yshift=-0.5cm]
        \node[draw,thick,inner sep=0pt, anchor=west] (output_arm) { \includegraphics[width=2cm]{fig/architecture/img/neural_0_ours_arm.jpg} };
        \node[draw,thick,inner sep=0pt] (output_nor) at ($(output_arm) + (-0.2,-0.2)$) { \includegraphics[width=2cm]{fig/architecture/img/neural_0_ours_norm.jpg} };
        \node[draw,thick,inner sep=0pt] (output_diff) at ($(output_nor) + (-0.2,-0.2)$) { \includegraphics[width=2cm]{fig/architecture/img/neural_0_ours_diff.jpg} };

        \node[fit={(output_diff.south west) (output_arm.north east)}] (output) {};
        \node[below=0pt of output, align=center] {output textures \\ $\times 2$ resolution};
    \end{scope}
    \draw[<->,thick] (2.5,-2.0) -- (9.5,-2.0) ;
    \node at (6.0,-2.3) { $\mathbf{loss}$ } ;
    \end{tikzpicture}
    \vspace{-5pt}
    \caption{
        \textbf{Hypernetwork for super-resolution.}
        We learn to encode neural texture sets with super-resolution capabilities by repurposing the hypernetwork. In practice, it suffices to input a downscaled texture and apply the loss on the full-resolution one.
        \label{fig:superresolution}
        \vspace{-20pt}
    }
\end{figure*}
\subsection{Training The Hypernetwork}
We train the hypernetwork using a dataset of texture sets. For each texture set $\mathcal{T}$, we first evaluate the $\net$
\begin{align}
(\lat{0}, \lat{1}, \lat{2}, \lat{3}, \wb) = \net(\mathcal{T};\,\mathbf{\phi}).
\end{align}
Using the weights and latents, we can compute the decompressed texture set $\tilde{\mathcal{T}}$:
\begin{align}
\tilde{\mathcal{T}} = \mlp(\left[\lat{0}, ..., \lat{3}\right];\;\wb).
\end{align}
We can backpropagate the error between $\tilde{\mathcal{T}}$ and $\mathcal{T}$ to optimize the parameters $\mathbf{\phi}$ of the hypernetwork:
\begin{align}
\underset{\phi}{\mbox{argmin}} \; \mbox{loss}\left[\tilde{\mathcal{T}}, \mathcal{T} \right]
\label{eq:optimization_ref}
\end{align}
We use a blend of the $l_1$ norm and DISTS~\cite{ding2020dists} using random permutation of channels~\cite{chambon2021passing}:
\begin{align}
\mbox{loss}\left[\tilde{\mathcal{T}}, \mathcal{T} \right] = \sum_{i,j} \|\tilde{\mathcal{T}}_{i,j} - \mathcal{T}_{i,j}\|_1 + \mbox{dists}(\tilde{\mathcal{T}}^{a,b,c}, \mathcal{T}^{a,b,c}),
\end{align}
where $a,b,c$ are random numbers in $\left[0, \dots, 9 \right]$. We also apply some random jitter to evaluate the decompressed image $\tilde{\mathcal{T}}$ and use bilinear interpolation with the same jitter on the target $\mathcal{T}$.

We optimize the Hypernetwork parameters $\mathbf{\phi}$ using stochastic gradient descent and the \textsc{Adam}~\cite{kingma2014adam} optimizer with weight decay~\cite{loshchilov2017decoupled}. We set the learning rate to $lr = 5 \times 10^{-5}$ and kept the other parameters unchanged.

We train our network on a dataset composed of crops from textures from diverse sources. We used the Polyhaven Textures (814 unique texture sets), Polyhaven Assets (431 unique texture sets) and random PBR assets from Sketchfab (902 unique texture sets). Most textures in our dataset are $4096 \times 4096$. During training we randomly crop a batch of texture sets at a random resolution ranging from $32 \times 32$ to $512 \times 512$. We also apply random color transformations to the diffuse texture. We trained our model on a single Nvidia RTX A6000 Ada GPU for $100$K iterations with a batch size of $16$, taking approximately 10 hours to complete.

\begin{figure*}[h!]
    \center
    \vspace{-0.1cm}
    \resizebox{\linewidth}{!}{
    \begin{tikzpicture}[font=\footnotesize]
        \node[rotate=90] at (-3.6, 1) {\textbf{Reference}};
        \node[rotate=90] at (-3.6,-1) {\textbf{Outputs}};

        \begin{scope}
            \node { \includegraphics[width=6.8cm]{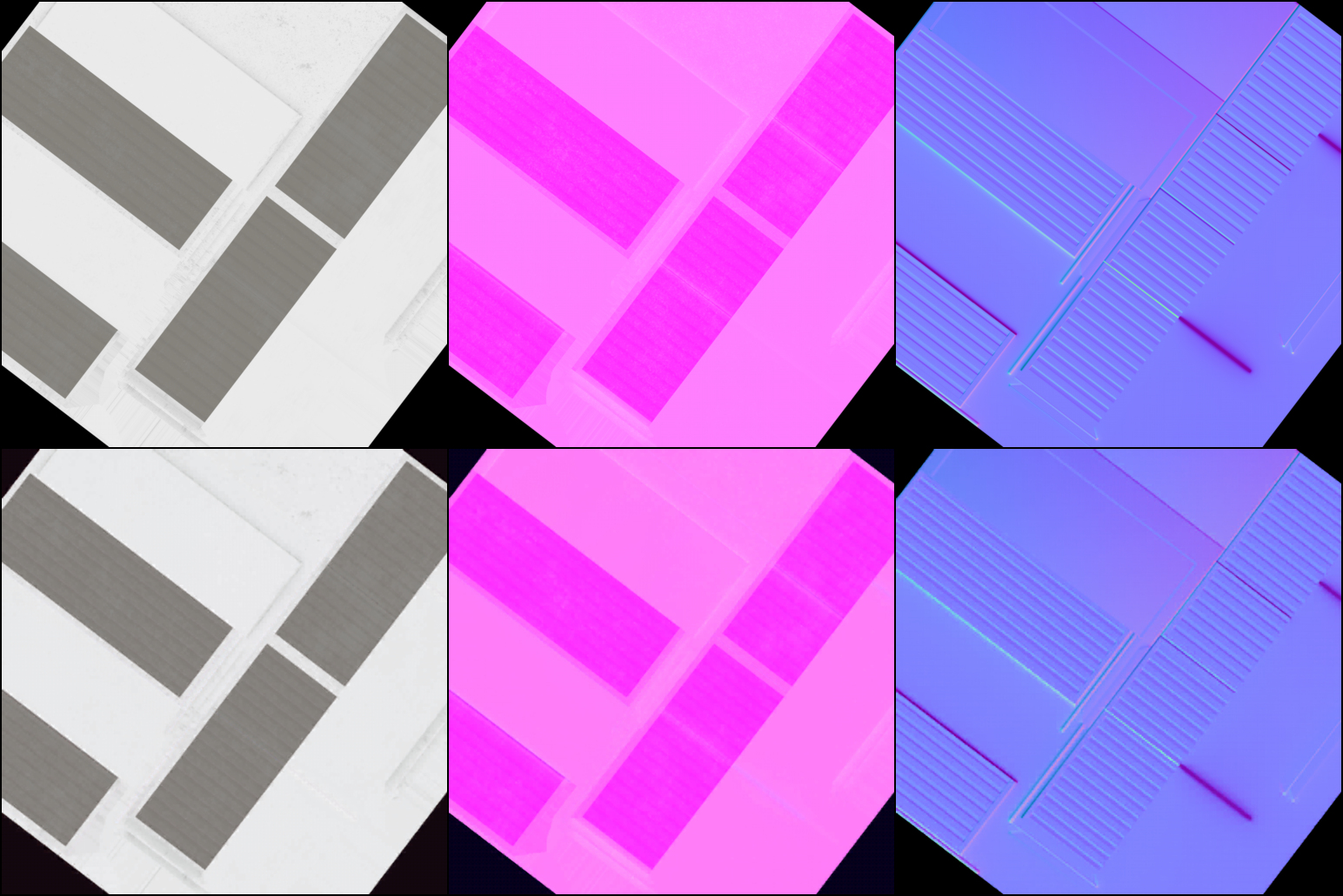} };
            \node[] at (-2.2, 2.5) {\textbf{Albedo}};
            \node[] at ( 0.0, 2.5) {\textbf{ARM}};
            \node[] at (2.2, 2.5) {\textbf{Normal}};
            \node[] at ( 0.0, -2.5) {\textit{Texture set in dataset}};
        \end{scope}

        \begin{scope}[xshift=7cm]
            \node { \includegraphics[width=6.8cm]{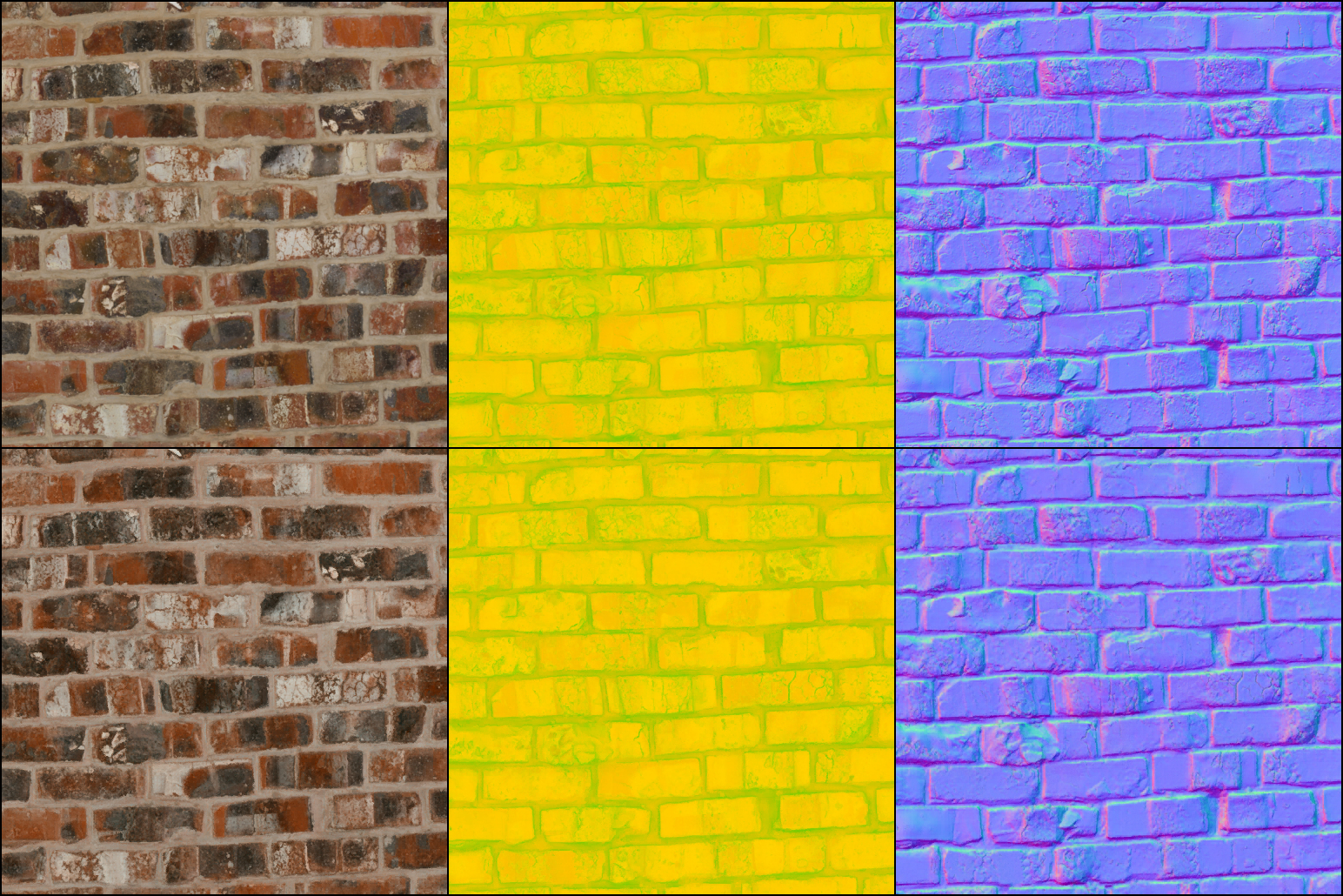} };
            \node[] at (-2.2, 2.5) {\textbf{Albedo}};
            \node[] at ( 0.0, 2.5) {\textbf{ARM}};
            \node[] at (2.2, 2.5) {\textbf{Normal}};
            \node[] at ( 0.0, -2.5) {\textit{Texture set outside of training set}};
        \end{scope}

        \begin{scope}[xshift=14cm]
            \node { \includegraphics[width=6.8cm]{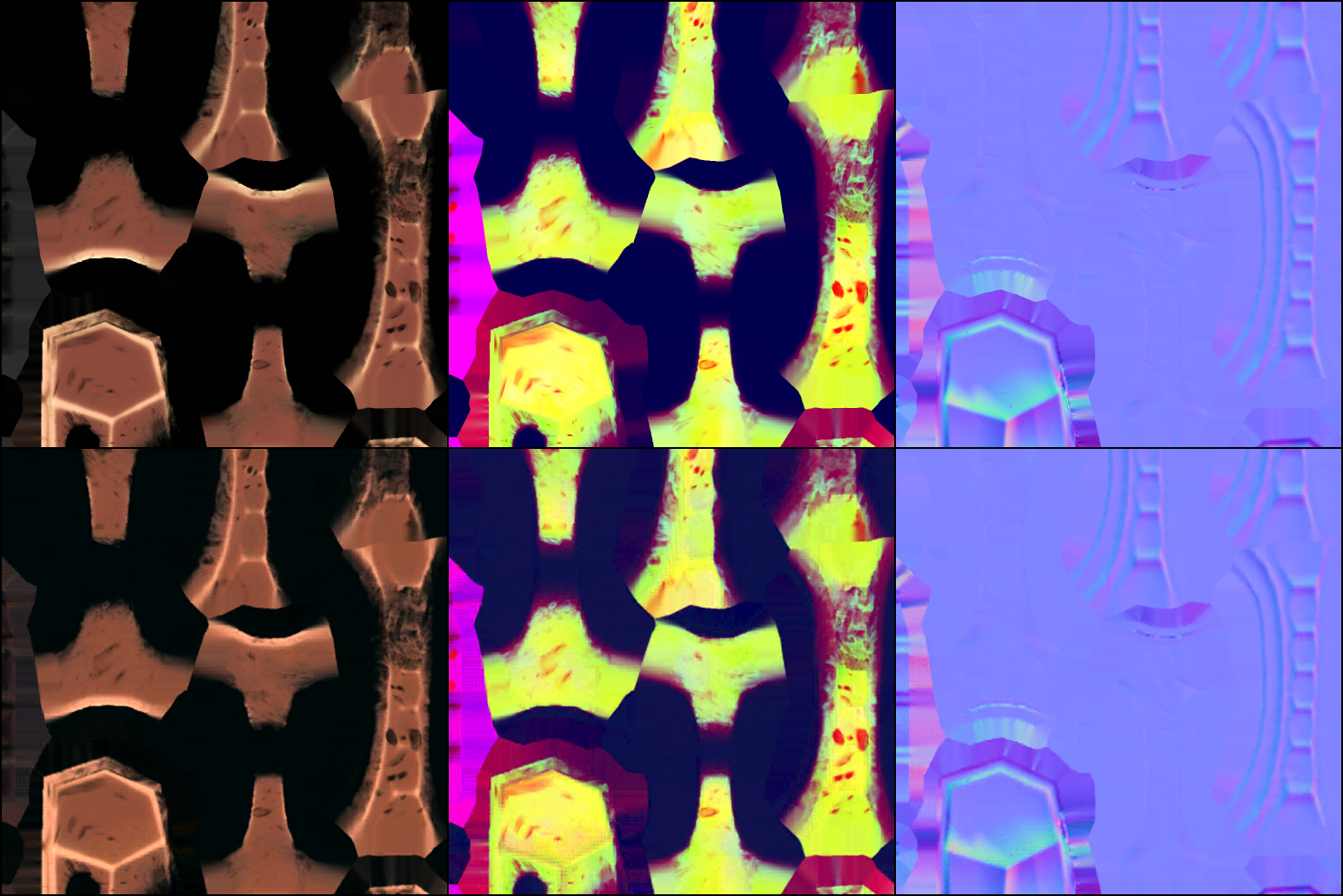} };
            \node[] at (-2.2, 2.5) {\textbf{Albedo}};
            \node[] at ( 0.0, 2.5) {\textbf{ARM}};
            \node[] at (2.2, 2.5) {\textbf{Normal}};
            \node[] at ( 0.0, -2.5) {\textit{Texture set outside of dataset and training set}};
        \end{scope}
    \end{tikzpicture}
    }
    \vspace{-20pt}
    \caption{
        \textbf{Outputs of Hypernetwork after decompression.}
        We show outputs of our Hypernetwork (bottom row), compared to its input (top row). We show that we obtain visually close images whether the input is in the training set, in the test set or outside of the dataset. This demonstrates that our model generalizes to other data than what it has been trained with.
        \label{fig:results_polyhaven_images}
        \vspace{-10pt}
    }
\end{figure*}

\section{Repurposing the Hypernetwork}
In the following, we show how the hypernetwork can be repurposed to accomplish more tasks. For example, we can leverage it to encode a single MLP for all the compressed texture sets, or output multiple MLPs for a single texture set. We can even extend it for tasks other than compression, such as super-resolution.

\subsection{From Tokens to MLPs}
\label{sec:tokens_to_mlps}
In BCF1~\cite{belcour2025tsnc}, each texture set is compressed along its own MLP. This means that the runtime decompression must account for divergent MLPs in the same workgroup. Also, during optimization, the MLP's hidden depth is fixed. To produce different quality/performance tradeoffs, it is necessary to optimize multiple MLPs, which either requires running the optimization multiple times or complicating the optimization routine, negatively impacting training time.

Using our hypernetwork, it is possible to train it to output a single MLP for all texture sets. This means that the weights and biases of the MLP are shared across all texture sets, and only the latent textures $\lat{i}$ are different. To demonstrate this, we took our hypernetwork and fine-tuned it to bypass the \textbf{t-Encoder} step and output jointly $\mathbf{t}_0$ in place of $\mathbf{t}$ (see Fig.~\ref{fig:architecture_zdit}). We found that it is better for this unique MLP to work on the normalized values of the texture set and to correct the mean and variance at runtime after decompression. Sharing the MLP weights and biases avoids the necessity of performing tile-based classification at runtime in order to homogenize the MLPs per workgroup~\cite{belcour2025tsnc}.

Similarly, we can also fine-tune multiple heads that convert token $\mathbf{t}$ to different MLP configurations. For example, we can have a single network that outputs both $16$, $32$ and $64$ hidden-depth MLPs for the same latent. While it is possible to repurpose BCF1's gradient descent optimizer, it would have a negative impact on optimization performance. Thanks to the monolithic structure of the hypernetwork, multiple outputs marginally impact its performance.

\subsection{Material Super-Resolution}
To obtain a neural texture set that can evaluate at higher resolution, we train the hypernetwork using a downscaled version of $\mathcal{L}$ as input to $\mbox{net}(\dot)$ but keep the original texture set to compute the loss (see Figure~\ref{fig:superresolution}). The optimization of Equation~\ref{eq:optimization_ref} can be written as:
\begin{align}
\underset{\phi}{\mbox{argmin}} \; \mbox{loss}\left[\mlp\left( \net(\mathcal{T}^h;\,\mathbf{\phi}) \right), \mathcal{T}^l \right]
\label{eq:optimization_down}
\end{align}
where $\mathcal{T}^l$ is the downscaled version of $\mathcal{T}^h$. In our tests, we use $\mbox{mip}_1$ of the input texture set for $\mathcal{T}^l$ and $\mbox{mip}_0$ for $\mathcal{T}^h$.

As a result, the hypernetwork learns to produce super-resolution texture sets even for texture sets where no higher resolution is known.
\section{Results}
In this section, we present both qualitative and quantitative results. Unless mentioned, we produced our results on texture sets of $4096 \times 4096$ resolution. For our results, we set up the Hypernetwork as follows: we use a dimension of $512$ for the tokens; the \textbf{t-Encoder} produces $16$ tokens for $\mathbf{t}$;  $\mathbf{x}$ passes through $12$ \textbf{DiT Blocks} with $32$ heads to compute the attention. Unless noted, we trained the Hypernetwork to output an MLP with $32$ hidden channels and \textsc{SiLU} activation functions~\cite{hendrycks2016gaussian}. This corresponds to the \textsc{VarA} variant of Belcour and Benyoub~\cite{belcour2025tsnc}. Using those settings, our model's unquantized weights were $294$MB on disk.

\paragraph*{Output of the Hypernetwork.} First, we provide qualitative results by showing the decompressed output of our network. With Figures~\ref{fig:results_polyhaven_images} we provide visual examples of outputs (bottom row) of our Hypernetwork after decompression for $512 \times 512$ texture sets compared to the input of the network (top row). To validate that our model does not overfit the training set, we display texture sets from the dataset, the test set, and a texture set that is neither in the dataset nor the training set. We can see in the images that our model generalizes correctly. We provide more visual results in the supplementary material.

\begin{figure}[b!]
    \vspace{-15pt}
    \resizebox{\linewidth}{!}{
    \begin{tikzpicture}[ font=\tiny ]
        \node[draw,thick,inner sep=0] (ref) at (0,0) {\includegraphics[width=0.5\linewidth]{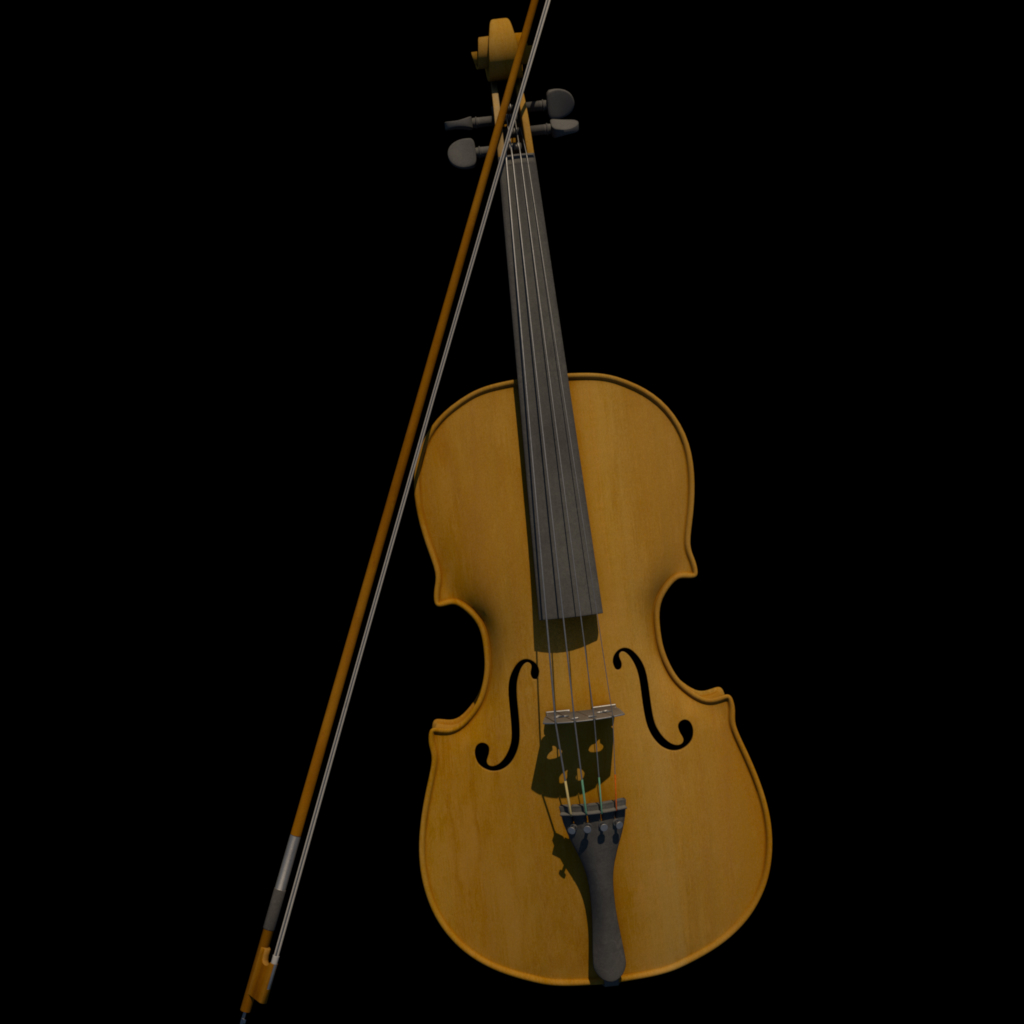}};
        \node[draw,thick,anchor=north west,inner sep=0,xshift=0.2pt] (ref_zoom) at (ref.north east) {\includegraphics[width=0.248\linewidth]{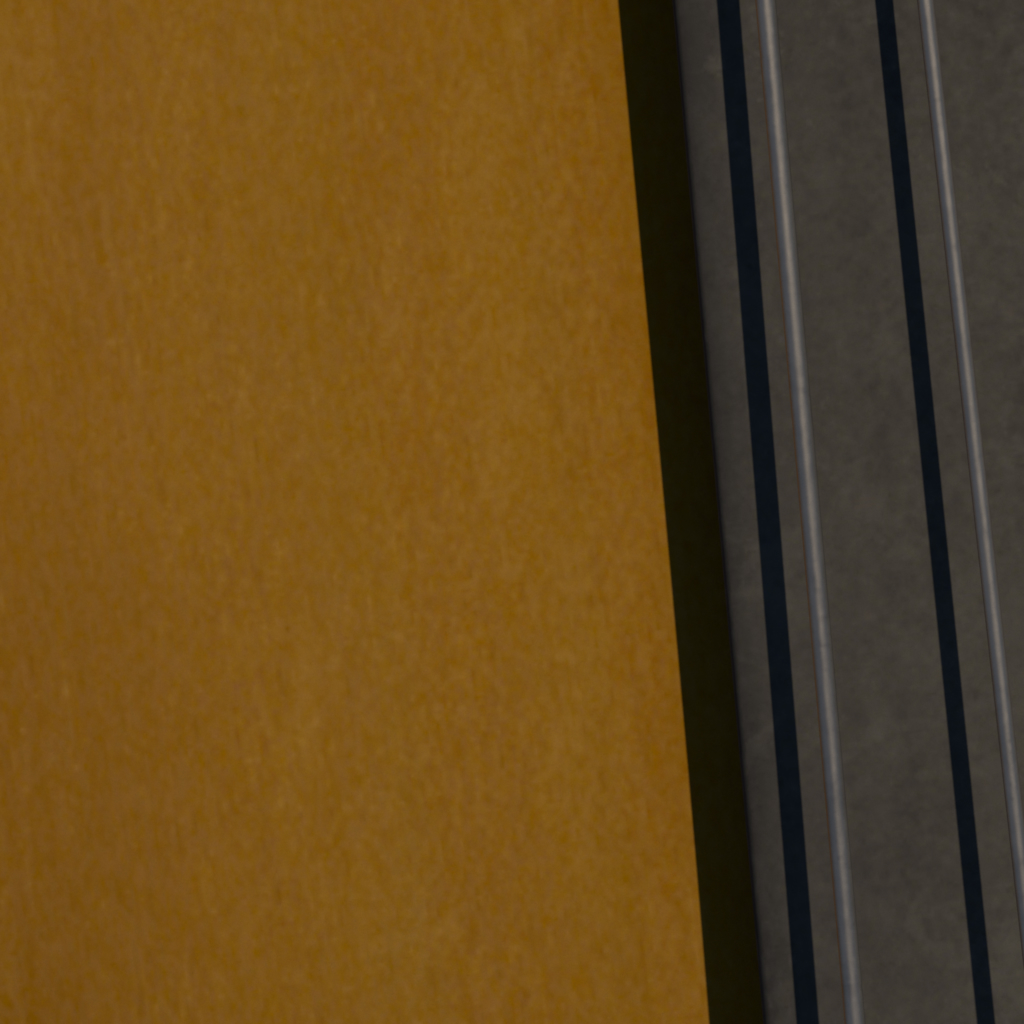}};
        \node[below=-10pt of ref_zoom, white] {\textbf{Reference}};
        \node[draw,thick,anchor=south west,inner sep=0,xshift=0.2pt] (ours_zoom) at (ref.south east) {\includegraphics[width=0.248\linewidth]{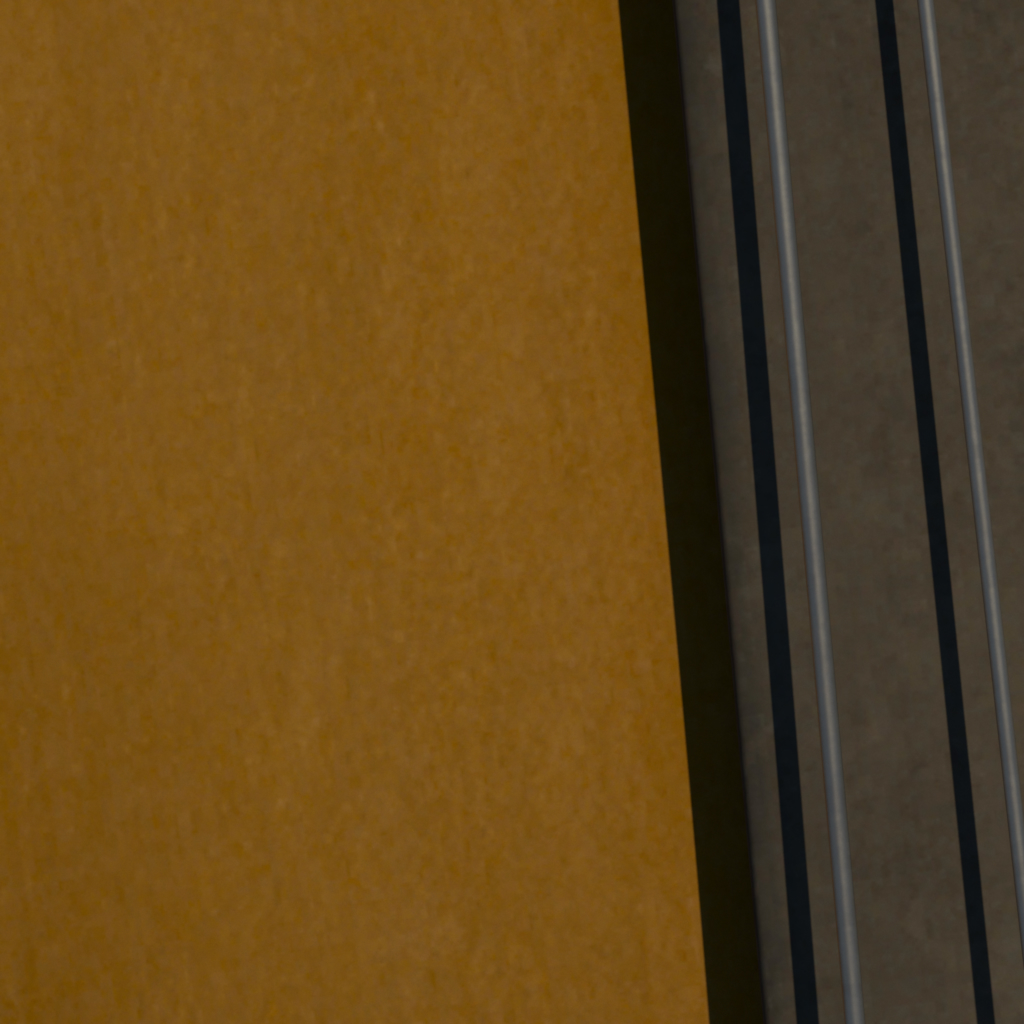}};
        \node[below=-10pt of ours_zoom, white] {\textbf{Ours}};
        \node[draw,thick,anchor=south west,inner sep=0,xshift=0.2pt] (ours_zoom) at (ours_zoom.south east) {\includegraphics[width=0.248\linewidth]{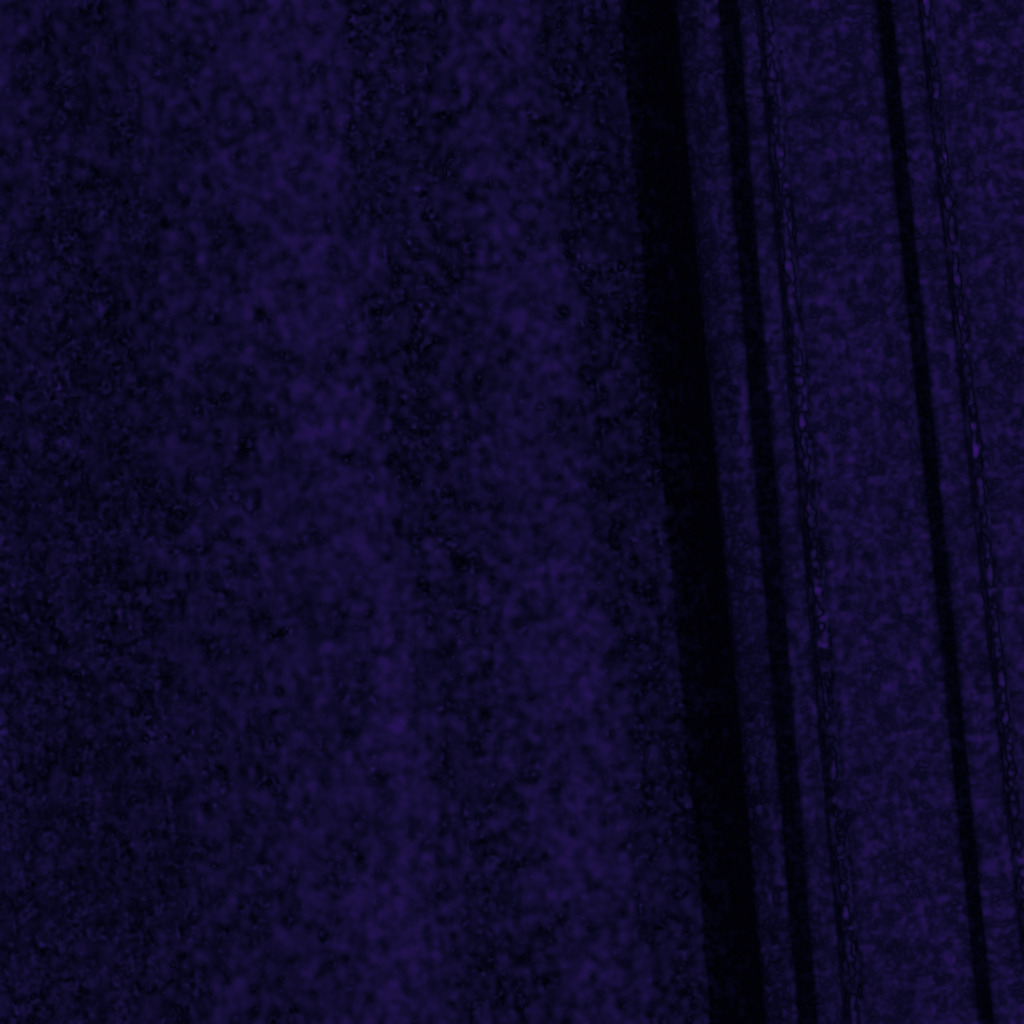}};
        \node[below=-10pt of ours_zoom, white] {\textbf{\reflectbox{F}LIP}};
    \end{tikzpicture}
    }
    \vspace{-0.5cm}
    \centering
    \caption{\textbf{Neural texture compression on assets}. We show that once applied to assets, BCF1 generated by our method is visually close to the reference. For example, on this violin asset the grain of the wood is still visible in our method. As shown by the \reflectbox{F}LIP difference image, the difference between renderings is small.
        \label{fig:mitsuba_violin}
        \vspace{-8pt}
    }
\end{figure}

With Figure~\ref{fig:teaser} and~\ref{fig:mitsuba_violin}, we show the decompressed texture set after rendering. For that, we rendered assets found in Sketchfab using Mitsuba~\cite{jakob2022mitsuba3}. We can see with the insets that our compressor preserves the high-frequency visual elements of the original textures such as grain and scratches.

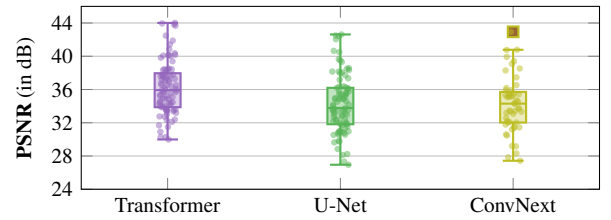
\begin{figure}[h!]

\definecolor{tab_blue}{rgb}{0.12156862745098039,0.4666666666666667,0.7058823529411765}
\definecolor{tab_orange}{rgb}{1.0,0.4980392156862745,0.054901960784313725}
\definecolor{tab_green}{rgb}{0.17254901960784313,0.6274509803921569,0.17254901960784313}
\definecolor{tab_red}{rgb}{0.8392156862745098,0.15294117647058825,0.1568627450980392}
\definecolor{tab_purple}{rgb}{0.5803921568627451,0.403921568627451,0.7411764705882353}
\definecolor{tab_brown}{rgb}{0.5490196078431373,0.33725490196078434,0.29411764705882354}
\definecolor{tab_pink}{rgb}{0.8901960784313725,0.4666666666666667,0.7607843137254902}
\definecolor{tab_gray}{rgb}{0.4980392156862745,0.4980392156862745,0.4980392156862745}
\definecolor{tab_olive}{rgb}{0.7372549019607844,0.7411764705882353,0.13333333333333333}
\definecolor{tab_cyan}{rgb}{0.09019607843137255,0.7450980392156863,0.8117647058823529}          

\begin{tikzpicture}[font=\small]
    \pgfplotsset{height=4cm, width=\linewidth}
    \begin{axis}[
        boxplot/draw direction=y, 
        ytick distance=4.0, 
        xmin=1.5, xmax=4.5,
        xticklabels={
            $N =  4$,
            Transformer,
            U-Net,
            ConvNext,
        },
        ymin=24.0, ymax=46.0,
        ymajorgrids, 
        xtick={1,2,3,4}, 
        ylabel={\textbf{PSNR} (in dB)},
    ]
\addplot+ [boxplot, thick, solid, tab_blue, fill=tab_blue!30!white,boxplot/box extend=0.15,] table [row sep=\\,y index=0] {
24.92 \\22.35 \\19.29 \\20.08 \\23.09 \\17.23 \\24.06 \\20.25 \\23.64 \\19.46 \\18.07 \\17.66 \\18.56 \\16.49 \\28.24 \\21.06 \\19.57 \\19.37 \\30.33 \\21.56 \\17.70 \\23.03 \\21.29 \\29.03 \\18.54 \\15.17 \\18.01 \\17.74 \\16.53 \\16.03 \\20.49 \\27.00 \\21.72 \\18.78 \\21.60 \\25.30 \\30.94 \\16.32 \\22.85 \\31.84 \\17.41 \\17.52 \\19.84 \\16.07 \\20.78 \\18.67 \\20.85 \\18.93 \\16.03 \\18.05 \\26.67 \\26.55 \\21.94 \\20.48 \\22.59 \\19.15 \\20.57 \\19.00 \\19.75 \\22.88 \\20.19 \\19.76 \\17.33 \\17.54 \\23.39 \\22.31 \\20.54 \\21.02 \\22.89 \\18.50 \\19.86 \\22.52 \\19.66 \\33.29 \\19.88 \\25.83 \\18.88 \\19.52 \\19.02 \\20.25 \\26.23 \\16.44 \\21.81 \\27.02 \\17.82 \\21.13 \\23.44 \\18.53 \\34.04 \\};
\addplot [tab_blue,only marks, mark=*, mark size=1pt, opacity=0.5] coordinates {
	(1+ 0.05*rand, 24.92)
	(1+ 0.05*rand, 22.35)
	(1+ 0.05*rand, 19.29)
	(1+ 0.05*rand, 20.08)
	(1+ 0.05*rand, 23.09)
	(1+ 0.05*rand, 17.23)
	(1+ 0.05*rand, 24.06)
	(1+ 0.05*rand, 20.25)
	(1+ 0.05*rand, 23.64)
	(1+ 0.05*rand, 19.46)
	(1+ 0.05*rand, 18.07)
	(1+ 0.05*rand, 17.66)
	(1+ 0.05*rand, 18.56)
	(1+ 0.05*rand, 16.49)
	(1+ 0.05*rand, 28.24)
	(1+ 0.05*rand, 21.06)
	(1+ 0.05*rand, 19.57)
	(1+ 0.05*rand, 19.37)
	(1+ 0.05*rand, 30.33)
	(1+ 0.05*rand, 21.56)
	(1+ 0.05*rand, 17.70)
	(1+ 0.05*rand, 23.03)
	(1+ 0.05*rand, 21.29)
	(1+ 0.05*rand, 29.03)
	(1+ 0.05*rand, 18.54)
	(1+ 0.05*rand, 15.17)
	(1+ 0.05*rand, 18.01)
	(1+ 0.05*rand, 17.74)
	(1+ 0.05*rand, 16.53)
	(1+ 0.05*rand, 16.03)
	(1+ 0.05*rand, 20.49)
	(1+ 0.05*rand, 27.00)
	(1+ 0.05*rand, 21.72)
	(1+ 0.05*rand, 18.78)
	(1+ 0.05*rand, 21.60)
	(1+ 0.05*rand, 25.30)
	(1+ 0.05*rand, 30.94)
	(1+ 0.05*rand, 16.32)
	(1+ 0.05*rand, 22.85)
	(1+ 0.05*rand, 31.84)
	(1+ 0.05*rand, 17.41)
	(1+ 0.05*rand, 17.52)
	(1+ 0.05*rand, 19.84)
	(1+ 0.05*rand, 16.07)
	(1+ 0.05*rand, 20.78)
	(1+ 0.05*rand, 18.67)
	(1+ 0.05*rand, 20.85)
	(1+ 0.05*rand, 18.93)
	(1+ 0.05*rand, 16.03)
	(1+ 0.05*rand, 18.05)
	(1+ 0.05*rand, 26.67)
	(1+ 0.05*rand, 26.55)
	(1+ 0.05*rand, 21.94)
	(1+ 0.05*rand, 20.48)
	(1+ 0.05*rand, 22.59)
	(1+ 0.05*rand, 19.15)
	(1+ 0.05*rand, 20.57)
	(1+ 0.05*rand, 19.00)
	(1+ 0.05*rand, 19.75)
	(1+ 0.05*rand, 22.88)
	(1+ 0.05*rand, 20.19)
	(1+ 0.05*rand, 19.76)
	(1+ 0.05*rand, 17.33)
	(1+ 0.05*rand, 17.54)
	(1+ 0.05*rand, 23.39)
	(1+ 0.05*rand, 22.31)
	(1+ 0.05*rand, 20.54)
	(1+ 0.05*rand, 21.02)
	(1+ 0.05*rand, 22.89)
	(1+ 0.05*rand, 18.50)
	(1+ 0.05*rand, 19.86)
	(1+ 0.05*rand, 22.52)
	(1+ 0.05*rand, 19.66)
	(1+ 0.05*rand, 33.29)
	(1+ 0.05*rand, 19.88)
	(1+ 0.05*rand, 25.83)
	(1+ 0.05*rand, 18.88)
	(1+ 0.05*rand, 19.52)
	(1+ 0.05*rand, 19.02)
	(1+ 0.05*rand, 20.25)
	(1+ 0.05*rand, 26.23)
	(1+ 0.05*rand, 16.44)
	(1+ 0.05*rand, 21.81)
	(1+ 0.05*rand, 27.02)
	(1+ 0.05*rand, 17.82)
	(1+ 0.05*rand, 21.13)
	(1+ 0.05*rand, 23.44)
	(1+ 0.05*rand, 18.53)
	(1+ 0.05*rand, 34.04)
};
\addplot+ [boxplot, thick, solid, tab_purple, fill=tab_purple!30!white,boxplot/box extend=0.15,] table [row sep=\\,y index=0] {
39.40 \\33.86 \\33.02 \\37.15 \\32.76 \\35.28 \\40.15 \\34.09 \\38.32 \\33.70 \\38.42 \\32.11 \\33.98 \\37.28 \\40.02 \\36.73 \\38.45 \\37.31 \\43.95 \\33.59 \\33.39 \\35.93 \\34.59 \\41.88 \\33.56 \\37.96 \\37.53 \\34.48 \\30.27 \\33.47 \\36.53 \\40.08 \\33.87 \\34.16 \\38.51 \\40.36 \\41.42 \\30.92 \\35.93 \\42.48 \\33.39 \\35.68 \\35.71 \\34.38 \\35.34 \\31.69 \\37.61 \\33.56 \\30.29 \\29.99 \\37.78 \\38.33 \\37.07 \\36.55 \\35.07 \\33.19 \\34.92 \\32.34 \\34.92 \\36.19 \\36.37 \\39.83 \\34.47 \\36.07 \\35.69 \\36.91 \\35.55 \\37.51 \\37.19 \\32.39 \\34.38 \\37.73 \\37.64 \\44.01 \\34.25 \\43.83 \\40.17 \\35.87 \\35.17 \\32.74 \\38.45 \\33.58 \\40.14 \\40.84 \\34.85 \\36.56 \\37.75 \\31.50 \\43.68 \\};
\addplot [tab_purple,only marks, mark=*, mark size=1pt, opacity=0.5] coordinates {
	(2+ 0.05*rand, 39.40)
	(2+ 0.05*rand, 33.86)
	(2+ 0.05*rand, 33.02)
	(2+ 0.05*rand, 37.15)
	(2+ 0.05*rand, 32.76)
	(2+ 0.05*rand, 35.28)
	(2+ 0.05*rand, 40.15)
	(2+ 0.05*rand, 34.09)
	(2+ 0.05*rand, 38.32)
	(2+ 0.05*rand, 33.70)
	(2+ 0.05*rand, 38.42)
	(2+ 0.05*rand, 32.11)
	(2+ 0.05*rand, 33.98)
	(2+ 0.05*rand, 37.28)
	(2+ 0.05*rand, 40.02)
	(2+ 0.05*rand, 36.73)
	(2+ 0.05*rand, 38.45)
	(2+ 0.05*rand, 37.31)
	(2+ 0.05*rand, 43.95)
	(2+ 0.05*rand, 33.59)
	(2+ 0.05*rand, 33.39)
	(2+ 0.05*rand, 35.93)
	(2+ 0.05*rand, 34.59)
	(2+ 0.05*rand, 41.88)
	(2+ 0.05*rand, 33.56)
	(2+ 0.05*rand, 37.96)
	(2+ 0.05*rand, 37.53)
	(2+ 0.05*rand, 34.48)
	(2+ 0.05*rand, 30.27)
	(2+ 0.05*rand, 33.47)
	(2+ 0.05*rand, 36.53)
	(2+ 0.05*rand, 40.08)
	(2+ 0.05*rand, 33.87)
	(2+ 0.05*rand, 34.16)
	(2+ 0.05*rand, 38.51)
	(2+ 0.05*rand, 40.36)
	(2+ 0.05*rand, 41.42)
	(2+ 0.05*rand, 30.92)
	(2+ 0.05*rand, 35.93)
	(2+ 0.05*rand, 42.48)
	(2+ 0.05*rand, 33.39)
	(2+ 0.05*rand, 35.68)
	(2+ 0.05*rand, 35.71)
	(2+ 0.05*rand, 34.38)
	(2+ 0.05*rand, 35.34)
	(2+ 0.05*rand, 31.69)
	(2+ 0.05*rand, 37.61)
	(2+ 0.05*rand, 33.56)
	(2+ 0.05*rand, 30.29)
	(2+ 0.05*rand, 29.99)
	(2+ 0.05*rand, 37.78)
	(2+ 0.05*rand, 38.33)
	(2+ 0.05*rand, 37.07)
	(2+ 0.05*rand, 36.55)
	(2+ 0.05*rand, 35.07)
	(2+ 0.05*rand, 33.19)
	(2+ 0.05*rand, 34.92)
	(2+ 0.05*rand, 32.34)
	(2+ 0.05*rand, 34.92)
	(2+ 0.05*rand, 36.19)
	(2+ 0.05*rand, 36.37)
	(2+ 0.05*rand, 39.83)
	(2+ 0.05*rand, 34.47)
	(2+ 0.05*rand, 36.07)
	(2+ 0.05*rand, 35.69)
	(2+ 0.05*rand, 36.91)
	(2+ 0.05*rand, 35.55)
	(2+ 0.05*rand, 37.51)
	(2+ 0.05*rand, 37.19)
	(2+ 0.05*rand, 32.39)
	(2+ 0.05*rand, 34.38)
	(2+ 0.05*rand, 37.73)
	(2+ 0.05*rand, 37.64)
	(2+ 0.05*rand, 44.01)
	(2+ 0.05*rand, 34.25)
	(2+ 0.05*rand, 43.83)
	(2+ 0.05*rand, 40.17)
	(2+ 0.05*rand, 35.87)
	(2+ 0.05*rand, 35.17)
	(2+ 0.05*rand, 32.74)
	(2+ 0.05*rand, 38.45)
	(2+ 0.05*rand, 33.58)
	(2+ 0.05*rand, 40.14)
	(2+ 0.05*rand, 40.84)
	(2+ 0.05*rand, 34.85)
	(2+ 0.05*rand, 36.56)
	(2+ 0.05*rand, 37.75)
	(2+ 0.05*rand, 31.50)
	(2+ 0.05*rand, 43.68)
};
\addplot+ [boxplot, thick, solid, tab_green!80!white, fill=tab_green!30!white,boxplot/box extend=0.15,] table [row sep=\\,y index=0] {
38.16 \\31.82 \\32.00 \\34.84 \\30.93 \\34.83 \\38.57 \\31.51 \\36.38 \\31.31 \\37.18 \\29.83 \\32.15 \\33.44 \\38.07 \\35.08 \\37.34 \\36.15 \\42.50 \\30.74 \\30.96 \\33.62 \\32.78 \\40.65 \\32.03 \\36.20 \\35.37 \\31.73 \\27.32 \\30.76 \\34.87 \\39.32 \\31.86 \\32.34 \\36.35 \\37.98 \\39.50 \\28.24 \\34.02 \\40.50 \\31.02 \\34.18 \\33.78 \\32.36 \\32.96 \\29.74 \\35.60 \\30.90 \\28.12 \\26.95 \\34.97 \\38.49 \\34.57 \\34.18 \\33.44 \\30.57 \\32.91 \\30.92 \\32.70 \\33.69 \\36.05 \\37.72 \\33.44 \\33.59 \\32.97 \\35.83 \\33.57 \\35.42 \\34.54 \\29.64 \\31.79 \\35.79 \\35.52 \\42.42 \\32.40 \\42.05 \\38.35 \\33.14 \\33.22 \\30.54 \\36.27 \\30.57 \\38.16 \\39.65 \\34.88 \\35.23 \\35.45 \\28.84 \\42.62 \\};
\addplot [tab_green!80!white,only marks, mark=*, mark size=1pt, opacity=0.5] coordinates {
	(3+ 0.05*rand, 38.16)
	(3+ 0.05*rand, 31.82)
	(3+ 0.05*rand, 32.00)
	(3+ 0.05*rand, 34.84)
	(3+ 0.05*rand, 30.93)
	(3+ 0.05*rand, 34.83)
	(3+ 0.05*rand, 38.57)
	(3+ 0.05*rand, 31.51)
	(3+ 0.05*rand, 36.38)
	(3+ 0.05*rand, 31.31)
	(3+ 0.05*rand, 37.18)
	(3+ 0.05*rand, 29.83)
	(3+ 0.05*rand, 32.15)
	(3+ 0.05*rand, 33.44)
	(3+ 0.05*rand, 38.07)
	(3+ 0.05*rand, 35.08)
	(3+ 0.05*rand, 37.34)
	(3+ 0.05*rand, 36.15)
	(3+ 0.05*rand, 42.50)
	(3+ 0.05*rand, 30.74)
	(3+ 0.05*rand, 30.96)
	(3+ 0.05*rand, 33.62)
	(3+ 0.05*rand, 32.78)
	(3+ 0.05*rand, 40.65)
	(3+ 0.05*rand, 32.03)
	(3+ 0.05*rand, 36.20)
	(3+ 0.05*rand, 35.37)
	(3+ 0.05*rand, 31.73)
	(3+ 0.05*rand, 27.32)
	(3+ 0.05*rand, 30.76)
	(3+ 0.05*rand, 34.87)
	(3+ 0.05*rand, 39.32)
	(3+ 0.05*rand, 31.86)
	(3+ 0.05*rand, 32.34)
	(3+ 0.05*rand, 36.35)
	(3+ 0.05*rand, 37.98)
	(3+ 0.05*rand, 39.50)
	(3+ 0.05*rand, 28.24)
	(3+ 0.05*rand, 34.02)
	(3+ 0.05*rand, 40.50)
	(3+ 0.05*rand, 31.02)
	(3+ 0.05*rand, 34.18)
	(3+ 0.05*rand, 33.78)
	(3+ 0.05*rand, 32.36)
	(3+ 0.05*rand, 32.96)
	(3+ 0.05*rand, 29.74)
	(3+ 0.05*rand, 35.60)
	(3+ 0.05*rand, 30.90)
	(3+ 0.05*rand, 28.12)
	(3+ 0.05*rand, 26.95)
	(3+ 0.05*rand, 34.97)
	(3+ 0.05*rand, 38.49)
	(3+ 0.05*rand, 34.57)
	(3+ 0.05*rand, 34.18)
	(3+ 0.05*rand, 33.44)
	(3+ 0.05*rand, 30.57)
	(3+ 0.05*rand, 32.91)
	(3+ 0.05*rand, 30.92)
	(3+ 0.05*rand, 32.70)
	(3+ 0.05*rand, 33.69)
	(3+ 0.05*rand, 36.05)
	(3+ 0.05*rand, 37.72)
	(3+ 0.05*rand, 33.44)
	(3+ 0.05*rand, 33.59)
	(3+ 0.05*rand, 32.97)
	(3+ 0.05*rand, 35.83)
	(3+ 0.05*rand, 33.57)
	(3+ 0.05*rand, 35.42)
	(3+ 0.05*rand, 34.54)
	(3+ 0.05*rand, 29.64)
	(3+ 0.05*rand, 31.79)
	(3+ 0.05*rand, 35.79)
	(3+ 0.05*rand, 35.52)
	(3+ 0.05*rand, 42.42)
	(3+ 0.05*rand, 32.40)
	(3+ 0.05*rand, 42.05)
	(3+ 0.05*rand, 38.35)
	(3+ 0.05*rand, 33.14)
	(3+ 0.05*rand, 33.22)
	(3+ 0.05*rand, 30.54)
	(3+ 0.05*rand, 36.27)
	(3+ 0.05*rand, 30.57)
	(3+ 0.05*rand, 38.16)
	(3+ 0.05*rand, 39.65)
	(3+ 0.05*rand, 34.88)
	(3+ 0.05*rand, 35.23)
	(3+ 0.05*rand, 35.45)
	(3+ 0.05*rand, 28.84)
	(3+ 0.05*rand, 42.62)
};
\addplot+ [boxplot, thick, solid, tab_olive, fill=tab_olive!30!white,boxplot/box extend=0.15,] table [row sep=\\,y index=0] {
32.20 \\32.03 \\35.23 \\31.44 \\35.13 \\39.09 \\32.03 \\30.48 \\32.59 \\35.51 \\35.44 \\37.46 \\42.94 \\31.50 \\34.30 \\33.36 \\40.78 \\35.91 \\32.08 \\27.85 \\31.78 \\35.30 \\39.41 \\33.07 \\38.29 \\29.16 \\40.76 \\34.56 \\28.31 \\27.43 \\35.06 \\39.93 \\35.23 \\31.19 \\33.40 \\33.47 \\34.34 \\38.53 \\33.79 \\33.67 \\35.92 \\34.70 \\35.51 \\30.59 \\36.52 \\33.19 \\33.33 \\31.39 \\35.19 \\36.19 \\29.20 \\};
\addplot [tab_olive,only marks, mark=*, mark size=1pt, opacity=0.5] coordinates {
	(4+ 0.05*rand, 32.20)
	(4+ 0.05*rand, 32.03)
	(4+ 0.05*rand, 35.23)
	(4+ 0.05*rand, 31.44)
	(4+ 0.05*rand, 35.13)
	(4+ 0.05*rand, 39.09)
	(4+ 0.05*rand, 32.03)
	(4+ 0.05*rand, 30.48)
	(4+ 0.05*rand, 32.59)
	(4+ 0.05*rand, 35.51)
	(4+ 0.05*rand, 35.44)
	(4+ 0.05*rand, 37.46)
	(4+ 0.05*rand, 42.94)
	(4+ 0.05*rand, 31.50)
	(4+ 0.05*rand, 34.30)
	(4+ 0.05*rand, 33.36)
	(4+ 0.05*rand, 40.78)
	(4+ 0.05*rand, 35.91)
	(4+ 0.05*rand, 32.08)
	(4+ 0.05*rand, 27.85)
	(4+ 0.05*rand, 31.78)
	(4+ 0.05*rand, 35.30)
	(4+ 0.05*rand, 39.41)
	(4+ 0.05*rand, 33.07)
	(4+ 0.05*rand, 38.29)
	(4+ 0.05*rand, 29.16)
	(4+ 0.05*rand, 40.76)
	(4+ 0.05*rand, 34.56)
	(4+ 0.05*rand, 28.31)
	(4+ 0.05*rand, 27.43)
	(4+ 0.05*rand, 35.06)
	(4+ 0.05*rand, 39.93)
	(4+ 0.05*rand, 35.23)
	(4+ 0.05*rand, 31.19)
	(4+ 0.05*rand, 33.40)
	(4+ 0.05*rand, 33.47)
	(4+ 0.05*rand, 34.34)
	(4+ 0.05*rand, 38.53)
	(4+ 0.05*rand, 33.79)
	(4+ 0.05*rand, 33.67)
	(4+ 0.05*rand, 35.92)
	(4+ 0.05*rand, 34.70)
	(4+ 0.05*rand, 35.51)
	(4+ 0.05*rand, 30.59)
	(4+ 0.05*rand, 36.52)
	(4+ 0.05*rand, 33.19)
	(4+ 0.05*rand, 33.33)
	(4+ 0.05*rand, 31.39)
	(4+ 0.05*rand, 35.19)
	(4+ 0.05*rand, 36.19)
	(4+ 0.05*rand, 29.20)
};
    \end{axis}
\end{tikzpicture}
\vspace{-0.25cm}
\caption{
    \label{fig:ablation_architectures}
    \textbf{Different Architectures.}
	We compared three architectures: a Transformer (left), a U-Net (middle), and a ConvNext (right). Using the same training hyper-parameters, we experienced that for roughly the same number of parameters the Transformer architecture provides the best quality.
	\vspace{-0.4cm}
}
\end{figure}

\paragraph*{Ablation Study.} We validated that our transformer-based hypernetwork performs better than a ConvNext-based or a U-Net-based hypernetwork. We report the PSNR and SSIM metrics for the different architectures in Figure~\ref{fig:ablation_architectures}. We matched all architectures to have an equal number of parameters and trained them on the same dataset with the same hyperparameters.


\begin{figure}[t!]

\definecolor{tab_blue}{rgb}{0.12156862745098039,0.4666666666666667,0.7058823529411765}
\definecolor{tab_orange}{rgb}{1.0,0.4980392156862745,0.054901960784313725}
\definecolor{tab_green}{rgb}{0.17254901960784313,0.6274509803921569,0.17254901960784313}
\definecolor{tab_red}{rgb}{0.8392156862745098,0.15294117647058825,0.1568627450980392}
\definecolor{tab_purple}{rgb}{0.5803921568627451,0.403921568627451,0.7411764705882353}
\definecolor{tab_brown}{rgb}{0.5490196078431373,0.33725490196078434,0.29411764705882354}
\definecolor{tab_pink}{rgb}{0.8901960784313725,0.4666666666666667,0.7607843137254902}
\definecolor{tab_gray}{rgb}{0.4980392156862745,0.4980392156862745,0.4980392156862745}
\definecolor{tab_olive}{rgb}{0.7372549019607844,0.7411764705882353,0.13333333333333333}
\definecolor{tab_cyan}{rgb}{0.09019607843137255,0.7450980392156863,0.8117647058823529}          

\begin{tikzpicture}[font=\small]
    \pgfplotsset{height=4cm, width=\linewidth}
    \begin{axis}[
        boxplot/draw direction=y, 
        ytick distance=2.0, 
        xmin=0.5, xmax=4.5,
        xticklabels={
            $N =  4$,
            $N = 16$,
            $N = 32$,
            $N = 64$,
        },
        ymin=28.0, ymax=44.0,
        ymajorgrids, 
        xtick={1,2,3,4}, 
        ylabel={\textbf{PSNR} (in dB)},
    ]

\addplot+ [boxplot, thick, solid, tab_blue, fill=tab_blue!30!white,boxplot/box extend=0.25,] table [row sep=\\,y index=0] {
34.39 \\37.24 \\36.59 \\33.45 \\36.63 \\39.92 \\33.93 \\32.30 \\33.64 \\34.17 \\37.07 \\37.25 \\43.95 \\33.94 \\36.45 \\34.30 \\41.95 \\37.30 \\33.75 \\30.25 \\32.21 \\35.79 \\40.35 \\34.26 \\40.19 \\30.73 \\43.21 \\35.54 \\30.27 \\30.15 \\38.52 \\38.91 \\36.92 \\33.19 \\33.02 \\34.55 \\36.23 \\39.61 \\31.54 \\35.33 \\37.06 \\35.60 \\37.24 \\32.17 \\37.39 \\33.98 \\35.76 \\32.81 \\35.14 \\38.02 \\31.61 \\};
\addplot [tab_blue,only marks, mark=*, mark size=1pt, opacity=0.5] coordinates {
	(1+ 0.1*rand, 34.39)
	(1+ 0.1*rand, 37.24)
	(1+ 0.1*rand, 36.59)
	(1+ 0.1*rand, 33.45)
	(1+ 0.1*rand, 36.63)
	(1+ 0.1*rand, 39.92)
	(1+ 0.1*rand, 33.93)
	(1+ 0.1*rand, 32.30)
	(1+ 0.1*rand, 33.64)
	(1+ 0.1*rand, 34.17)
	(1+ 0.1*rand, 37.07)
	(1+ 0.1*rand, 37.25)
	(1+ 0.1*rand, 43.95)
	(1+ 0.1*rand, 33.94)
	(1+ 0.1*rand, 36.45)
	(1+ 0.1*rand, 34.30)
	(1+ 0.1*rand, 41.95)
	(1+ 0.1*rand, 37.30)
	(1+ 0.1*rand, 33.75)
	(1+ 0.1*rand, 30.25)
	(1+ 0.1*rand, 32.21)
	(1+ 0.1*rand, 35.79)
	(1+ 0.1*rand, 40.35)
	(1+ 0.1*rand, 34.26)
	(1+ 0.1*rand, 40.19)
	(1+ 0.1*rand, 30.73)
	(1+ 0.1*rand, 43.21)
	(1+ 0.1*rand, 35.54)
	(1+ 0.1*rand, 30.27)
	(1+ 0.1*rand, 30.15)
	(1+ 0.1*rand, 38.52)
	(1+ 0.1*rand, 38.91)
	(1+ 0.1*rand, 36.92)
	(1+ 0.1*rand, 33.19)
	(1+ 0.1*rand, 33.02)
	(1+ 0.1*rand, 34.55)
	(1+ 0.1*rand, 36.23)
	(1+ 0.1*rand, 39.61)
	(1+ 0.1*rand, 31.54)
	(1+ 0.1*rand, 35.33)
	(1+ 0.1*rand, 37.06)
	(1+ 0.1*rand, 35.60)
	(1+ 0.1*rand, 37.24)
	(1+ 0.1*rand, 32.17)
	(1+ 0.1*rand, 37.39)
	(1+ 0.1*rand, 33.98)
	(1+ 0.1*rand, 35.76)
	(1+ 0.1*rand, 32.81)
	(1+ 0.1*rand, 35.14)
	(1+ 0.1*rand, 38.02)
	(1+ 0.1*rand, 31.61)
};

\addplot+ [boxplot, thick, solid, tab_blue, fill=tab_blue!30!white,boxplot/box extend=0.25,] table [row sep=\\,y index=0] {
34.25 \\35.12 \\36.63 \\32.57 \\35.65 \\40.28 \\34.05 \\31.79 \\33.45 \\34.47 \\35.97 \\37.12 \\43.76 \\34.00 \\36.15 \\34.22 \\41.90 \\36.94 \\34.12 \\29.86 \\32.90 \\35.72 \\39.87 \\33.78 \\39.86 \\29.95 \\42.61 \\34.55 \\29.98 \\30.24 \\38.17 \\37.46 \\36.94 \\33.13 \\33.63 \\34.27 \\36.30 \\39.11 \\31.72 \\35.36 \\36.61 \\34.45 \\37.69 \\31.92 \\37.16 \\33.94 \\35.80 \\32.62 \\34.56 \\37.54 \\31.47 \\};
\addplot [tab_blue,only marks, mark=*, mark size=1pt, opacity=0.5] coordinates {
	(2+ 0.1*rand, 34.25)
	(2+ 0.1*rand, 35.12)
	(2+ 0.1*rand, 36.63)
	(2+ 0.1*rand, 32.57)
	(2+ 0.1*rand, 35.65)
	(2+ 0.1*rand, 40.28)
	(2+ 0.1*rand, 34.05)
	(2+ 0.1*rand, 31.79)
	(2+ 0.1*rand, 33.45)
	(2+ 0.1*rand, 34.47)
	(2+ 0.1*rand, 35.97)
	(2+ 0.1*rand, 37.12)
	(2+ 0.1*rand, 43.76)
	(2+ 0.1*rand, 34.00)
	(2+ 0.1*rand, 36.15)
	(2+ 0.1*rand, 34.22)
	(2+ 0.1*rand, 41.90)
	(2+ 0.1*rand, 36.94)
	(2+ 0.1*rand, 34.12)
	(2+ 0.1*rand, 29.86)
	(2+ 0.1*rand, 32.90)
	(2+ 0.1*rand, 35.72)
	(2+ 0.1*rand, 39.87)
	(2+ 0.1*rand, 33.78)
	(2+ 0.1*rand, 39.86)
	(2+ 0.1*rand, 29.95)
	(2+ 0.1*rand, 42.61)
	(2+ 0.1*rand, 34.55)
	(2+ 0.1*rand, 29.98)
	(2+ 0.1*rand, 30.24)
	(2+ 0.1*rand, 38.17)
	(2+ 0.1*rand, 37.46)
	(2+ 0.1*rand, 36.94)
	(2+ 0.1*rand, 33.13)
	(2+ 0.1*rand, 33.63)
	(2+ 0.1*rand, 34.27)
	(2+ 0.1*rand, 36.30)
	(2+ 0.1*rand, 39.11)
	(2+ 0.1*rand, 31.72)
	(2+ 0.1*rand, 35.36)
	(2+ 0.1*rand, 36.61)
	(2+ 0.1*rand, 34.45)
	(2+ 0.1*rand, 37.69)
	(2+ 0.1*rand, 31.92)
	(2+ 0.1*rand, 37.16)
	(2+ 0.1*rand, 33.94)
	(2+ 0.1*rand, 35.80)
	(2+ 0.1*rand, 32.62)
	(2+ 0.1*rand, 34.56)
	(2+ 0.1*rand, 37.54)
	(2+ 0.1*rand, 31.47)
};

\addplot+ [boxplot, thick, solid, tab_blue, fill=tab_blue!30!white,boxplot/box extend=0.25,] table [row sep=\\,y index=0] {
34.47 \\32.52 \\37.05 \\33.01 \\35.94 \\38.99 \\34.44 \\31.98 \\33.73 \\34.58 \\37.22 \\36.95 \\43.48 \\34.43 \\36.39 \\33.88 \\41.69 \\37.07 \\34.21 \\30.57 \\32.40 \\36.05 \\39.74 \\33.69 \\40.46 \\30.15 \\42.38 \\35.32 \\30.50 \\30.31 \\39.11 \\38.46 \\37.10 \\33.36 \\34.53 \\34.29 \\36.67 \\39.35 \\32.70 \\35.80 \\36.48 \\33.39 \\38.23 \\31.93 \\37.23 \\33.96 \\36.02 \\32.77 \\34.12 \\37.86 \\31.72 \\};
\addplot [tab_blue,only marks, mark=*, mark size=1pt, opacity=0.5] coordinates {
	(3+ 0.1*rand, 34.47)
	(3+ 0.1*rand, 32.52)
	(3+ 0.1*rand, 37.05)
	(3+ 0.1*rand, 33.01)
	(3+ 0.1*rand, 35.94)
	(3+ 0.1*rand, 38.99)
	(3+ 0.1*rand, 34.44)
	(3+ 0.1*rand, 31.98)
	(3+ 0.1*rand, 33.73)
	(3+ 0.1*rand, 34.58)
	(3+ 0.1*rand, 37.22)
	(3+ 0.1*rand, 36.95)
	(3+ 0.1*rand, 43.48)
	(3+ 0.1*rand, 34.43)
	(3+ 0.1*rand, 36.39)
	(3+ 0.1*rand, 33.88)
	(3+ 0.1*rand, 41.69)
	(3+ 0.1*rand, 37.07)
	(3+ 0.1*rand, 34.21)
	(3+ 0.1*rand, 30.57)
	(3+ 0.1*rand, 32.40)
	(3+ 0.1*rand, 36.05)
	(3+ 0.1*rand, 39.74)
	(3+ 0.1*rand, 33.69)
	(3+ 0.1*rand, 40.46)
	(3+ 0.1*rand, 30.15)
	(3+ 0.1*rand, 42.38)
	(3+ 0.1*rand, 35.32)
	(3+ 0.1*rand, 30.50)
	(3+ 0.1*rand, 30.31)
	(3+ 0.1*rand, 39.11)
	(3+ 0.1*rand, 38.46)
	(3+ 0.1*rand, 37.10)
	(3+ 0.1*rand, 33.36)
	(3+ 0.1*rand, 34.53)
	(3+ 0.1*rand, 34.29)
	(3+ 0.1*rand, 36.67)
	(3+ 0.1*rand, 39.35)
	(3+ 0.1*rand, 32.70)
	(3+ 0.1*rand, 35.80)
	(3+ 0.1*rand, 36.48)
	(3+ 0.1*rand, 33.39)
	(3+ 0.1*rand, 38.23)
	(3+ 0.1*rand, 31.93)
	(3+ 0.1*rand, 37.23)
	(3+ 0.1*rand, 33.96)
	(3+ 0.1*rand, 36.02)
	(3+ 0.1*rand, 32.77)
	(3+ 0.1*rand, 34.12)
	(3+ 0.1*rand, 37.86)
	(3+ 0.1*rand, 31.72)
};

\addplot+ [boxplot, thick, solid, tab_blue, fill=tab_blue!30!white,boxplot/box extend=0.25,] table [row sep=\\,y index=0] {
34.09 \\31.92 \\36.42 \\32.46 \\36.01 \\38.74 \\34.04 \\31.75 \\33.22 \\34.69 \\36.43 \\36.50 \\43.34 \\34.07 \\35.87 \\33.61 \\40.94 \\36.70 \\34.14 \\30.52 \\32.54 \\34.99 \\38.62 \\33.33 \\38.80 \\30.44 \\42.01 \\35.06 \\30.04 \\29.92 \\38.73 \\37.51 \\36.74 \\32.92 \\32.70 \\33.88 \\36.26 \\39.02 \\32.87 \\35.72 \\36.47 \\34.11 \\37.46 \\31.54 \\36.88 \\33.60 \\35.53 \\32.30 \\34.88 \\37.08 \\31.56 \\};
\addplot [tab_blue,only marks, mark=*, mark size=1pt, opacity=0.5] coordinates {
	(4+ 0.1*rand, 34.09)
	(4+ 0.1*rand, 31.92)
	(4+ 0.1*rand, 36.42)
	(4+ 0.1*rand, 32.46)
	(4+ 0.1*rand, 36.01)
	(4+ 0.1*rand, 38.74)
	(4+ 0.1*rand, 34.04)
	(4+ 0.1*rand, 31.75)
	(4+ 0.1*rand, 33.22)
	(4+ 0.1*rand, 34.69)
	(4+ 0.1*rand, 36.43)
	(4+ 0.1*rand, 36.50)
	(4+ 0.1*rand, 43.34)
	(4+ 0.1*rand, 34.07)
	(4+ 0.1*rand, 35.87)
	(4+ 0.1*rand, 33.61)
	(4+ 0.1*rand, 40.94)
	(4+ 0.1*rand, 36.70)
	(4+ 0.1*rand, 34.14)
	(4+ 0.1*rand, 30.52)
	(4+ 0.1*rand, 32.54)
	(4+ 0.1*rand, 34.99)
	(4+ 0.1*rand, 38.62)
	(4+ 0.1*rand, 33.33)
	(4+ 0.1*rand, 38.80)
	(4+ 0.1*rand, 30.44)
	(4+ 0.1*rand, 42.01)
	(4+ 0.1*rand, 35.06)
	(4+ 0.1*rand, 30.04)
	(4+ 0.1*rand, 29.92)
	(4+ 0.1*rand, 38.73)
	(4+ 0.1*rand, 37.51)
	(4+ 0.1*rand, 36.74)
	(4+ 0.1*rand, 32.92)
	(4+ 0.1*rand, 32.70)
	(4+ 0.1*rand, 33.88)
	(4+ 0.1*rand, 36.26)
	(4+ 0.1*rand, 39.02)
	(4+ 0.1*rand, 32.87)
	(4+ 0.1*rand, 35.72)
	(4+ 0.1*rand, 36.47)
	(4+ 0.1*rand, 34.11)
	(4+ 0.1*rand, 37.46)
	(4+ 0.1*rand, 31.54)
	(4+ 0.1*rand, 36.88)
	(4+ 0.1*rand, 33.60)
	(4+ 0.1*rand, 35.53)
	(4+ 0.1*rand, 32.30)
	(4+ 0.1*rand, 34.88)
	(4+ 0.1*rand, 37.08)
	(4+ 0.1*rand, 31.56)
};
    \end{axis}
\end{tikzpicture}
\vspace{-0.25cm}
\caption{
    \label{fig:ablation_number_tokent}
    \textbf{Number of tokens.}
    Our method does not seem to be highly sensitive to the number of tokens $\tokent$.
    Between tokens ranging from $N=4$ to $N=64$, we see little change in PSNR quantiles.
	\vspace{-0.50cm}
}
\end{figure}
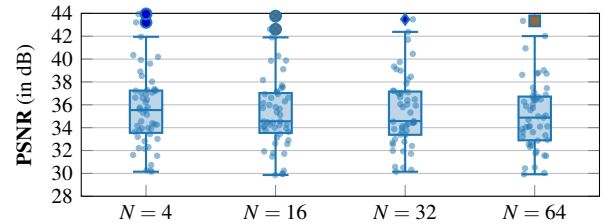

We also tested different configurations for our architecture. We benchmarked the use of different numbers of tokens $\tokent$. We trained multiple Hypernetworks with $N$ ranging from $4$ to $64$ and found that increasing the number of tokens $\tokent$ did not improve quality. We report those statistics in Figure~\ref{fig:ablation_number_tokent}. For this test, we use the same training parameters and kept the other hyperparameters constant.

\begin{figure*}[t]
    \center
    \begin{tikzpicture}
        \node[draw,thick,anchor=south west,inner sep=0] (image) at (0,0) {\includegraphics[height=5cm]{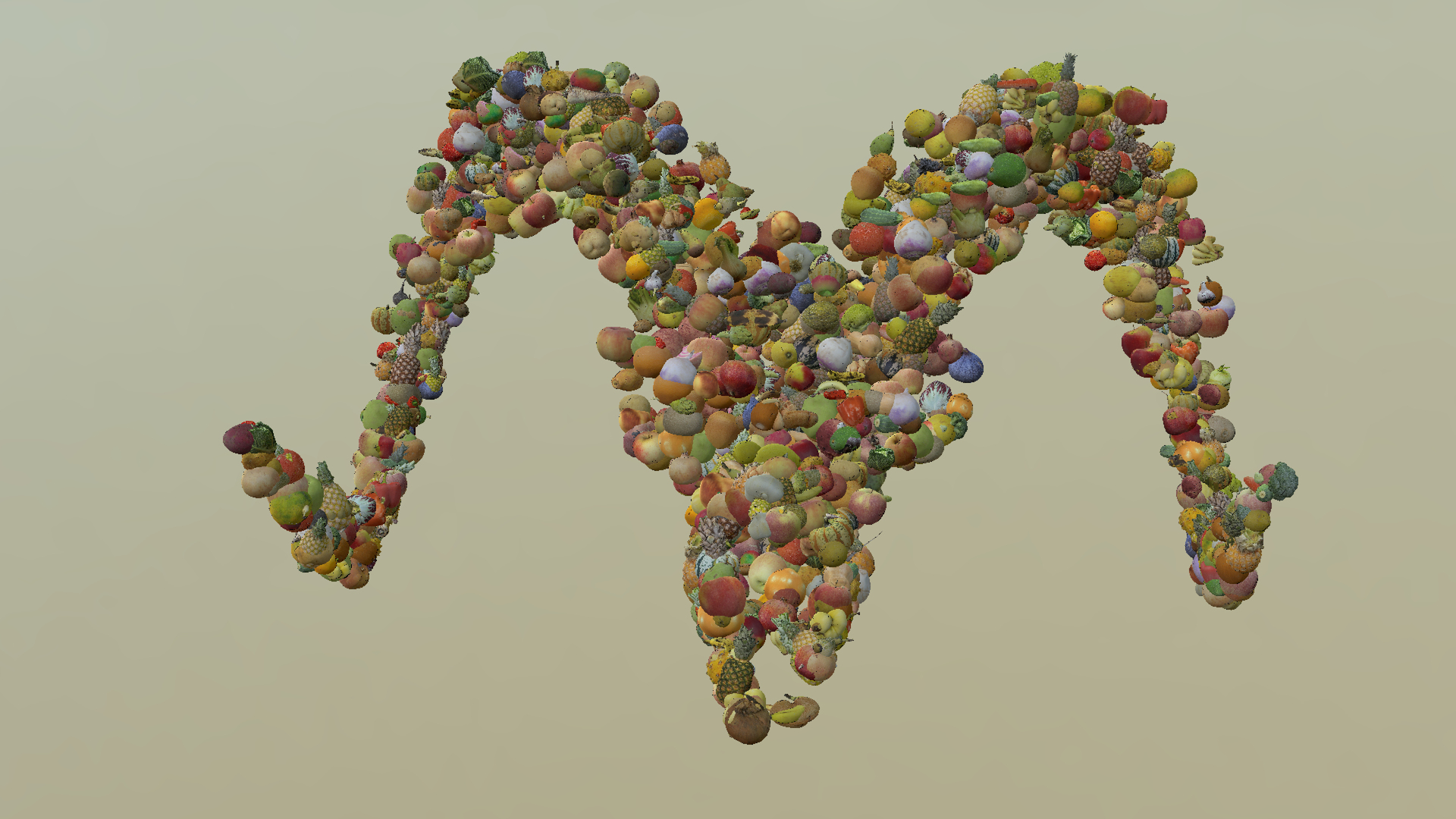}};

        \node[black,draw,thick,inner sep=0pt, right=0.02cm of image.north east, anchor=north west] (closeup_0) { \includegraphics[width=2.48cm]{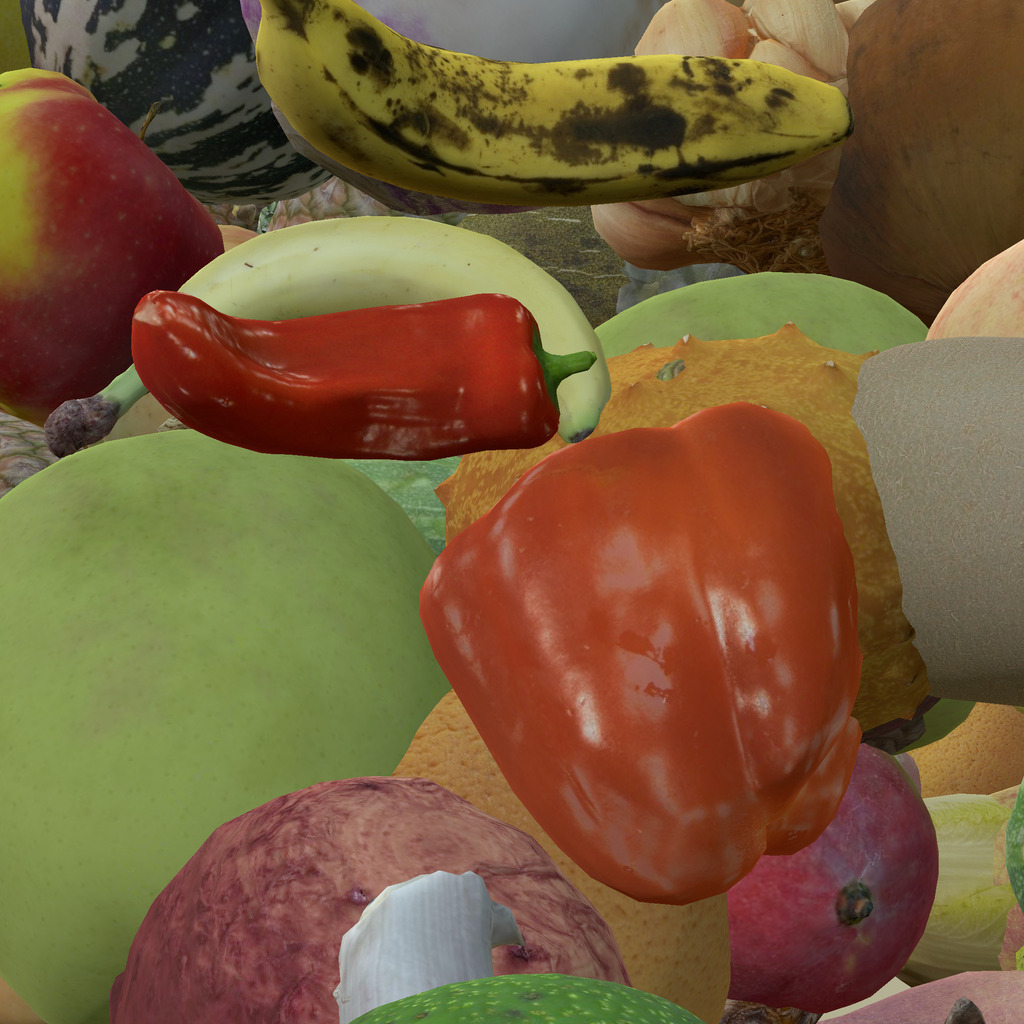} };
        \node[black,draw,thick,inner sep=0pt, right=0.02cm of image.south east, anchor=south west] (closeup_0_b) { \includegraphics[width=2.48cm]{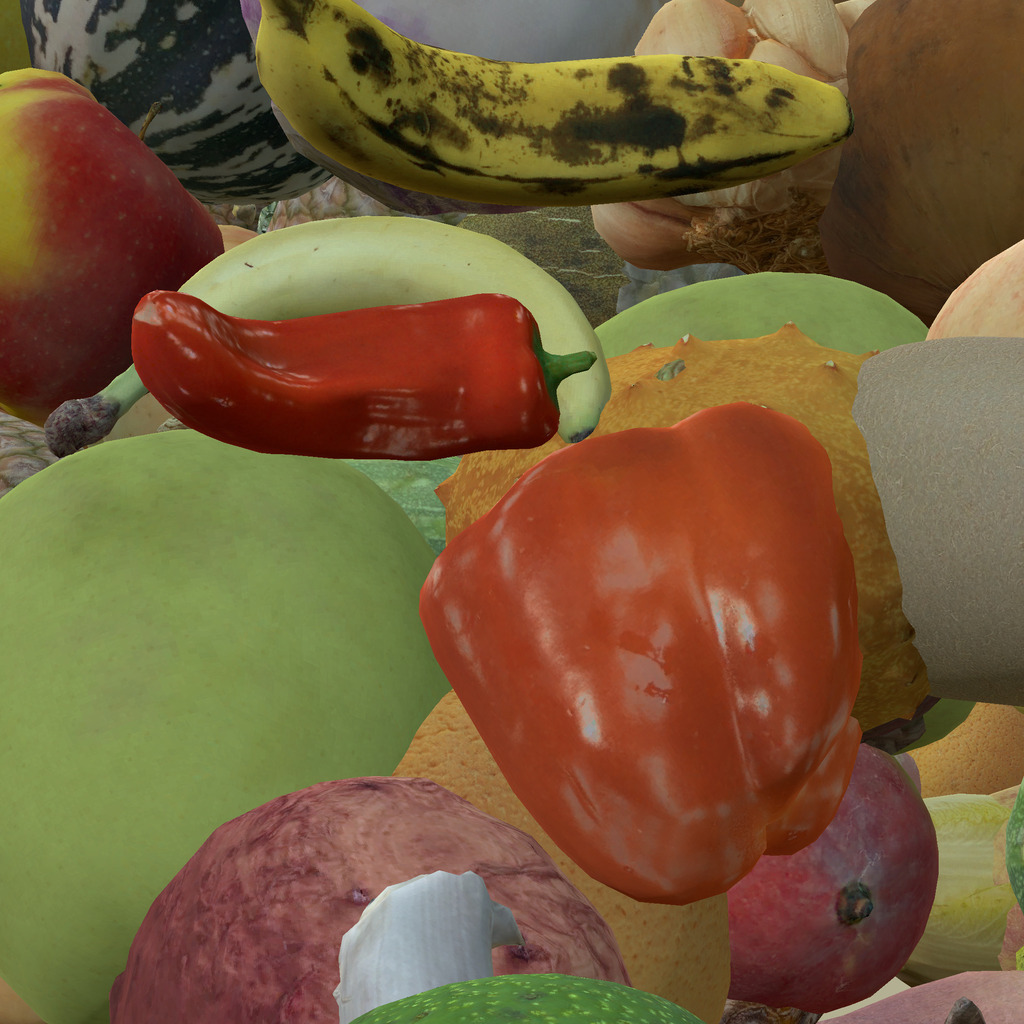} };   
        
        \node[black,draw,thick,inner sep=0pt, left=0.02cm of image.north west, anchor=north east] (closeup_1) { \includegraphics[width=2.48cm]{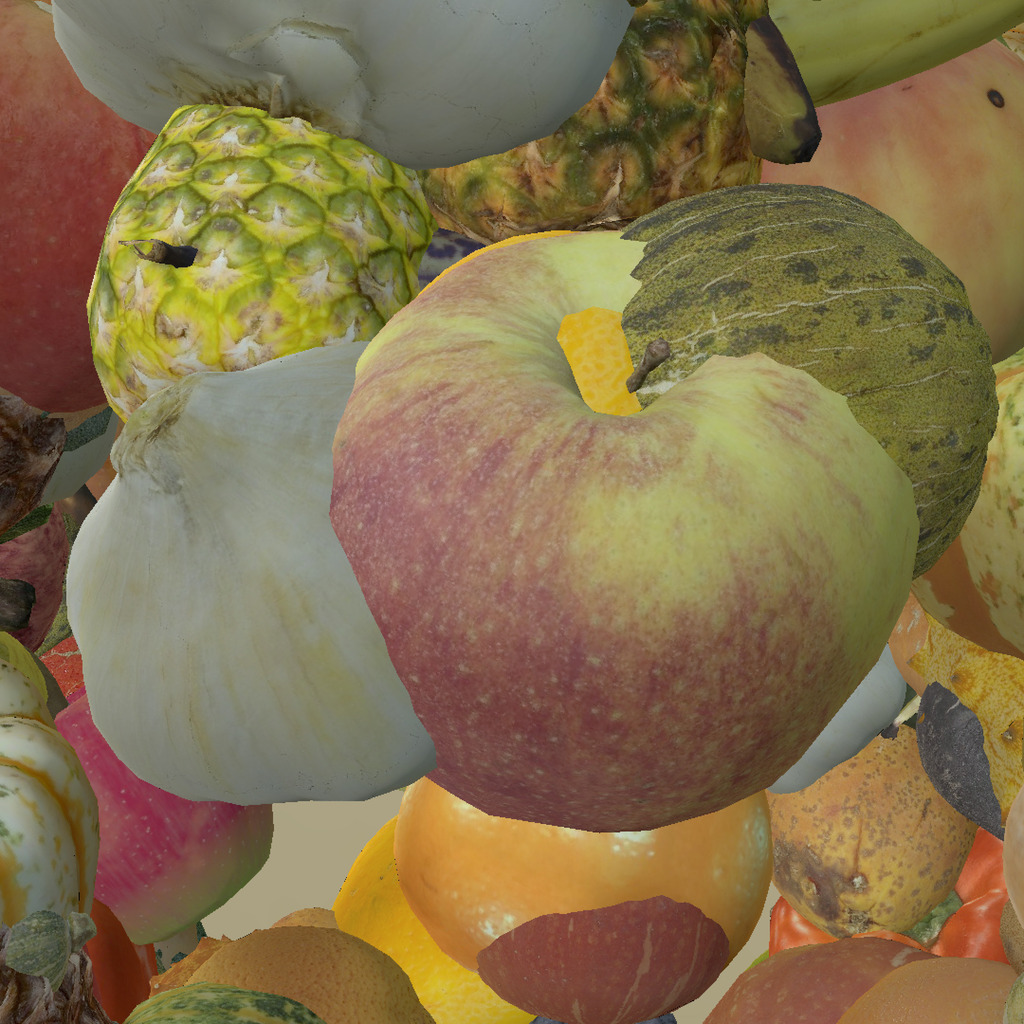} };
        \node[black,draw,thick,inner sep=0pt, left=0.02cm of image.south west, anchor=south east] (closeup_1_b) { \includegraphics[width=2.48cm]{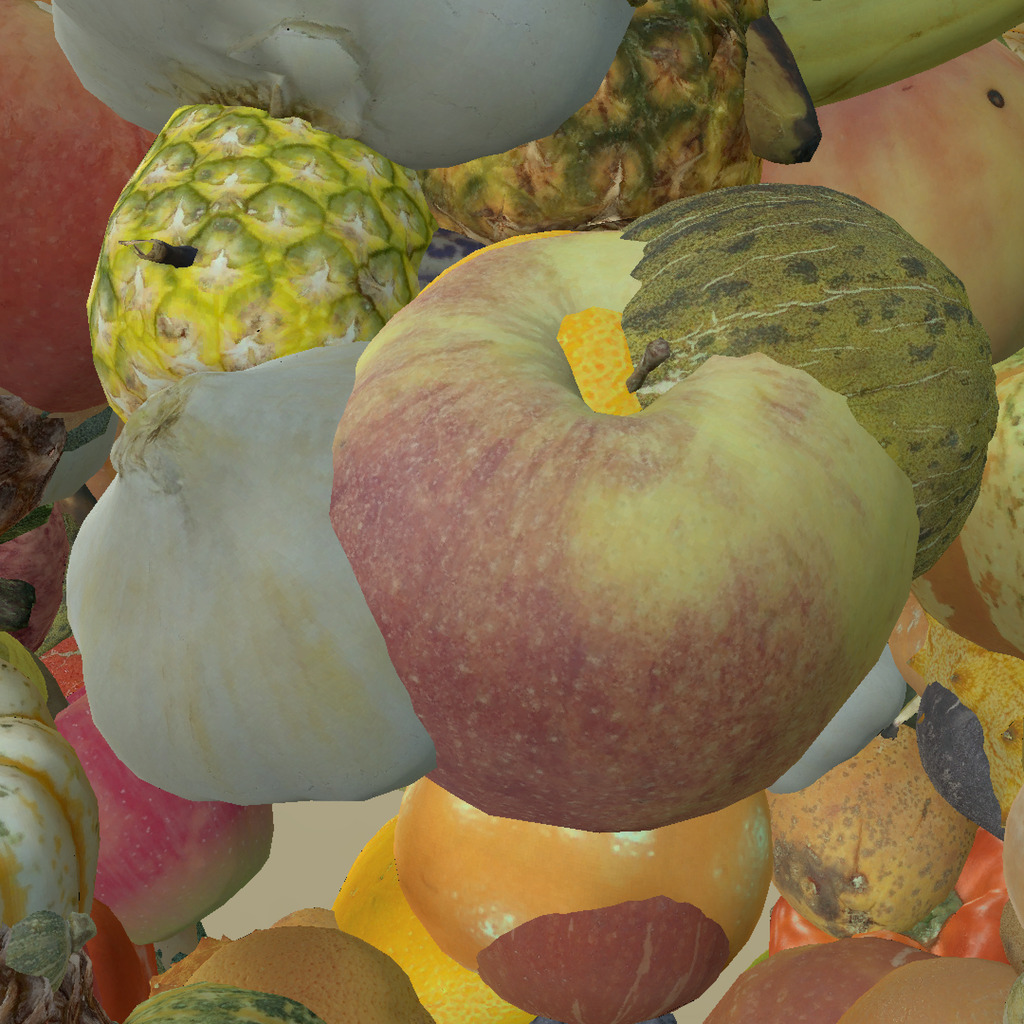} };

        \node[black,draw,thick,inner sep=0pt, left=0.02cm of closeup_1.north west, anchor=north east] (closeup_1) { \includegraphics[width=2.48cm]{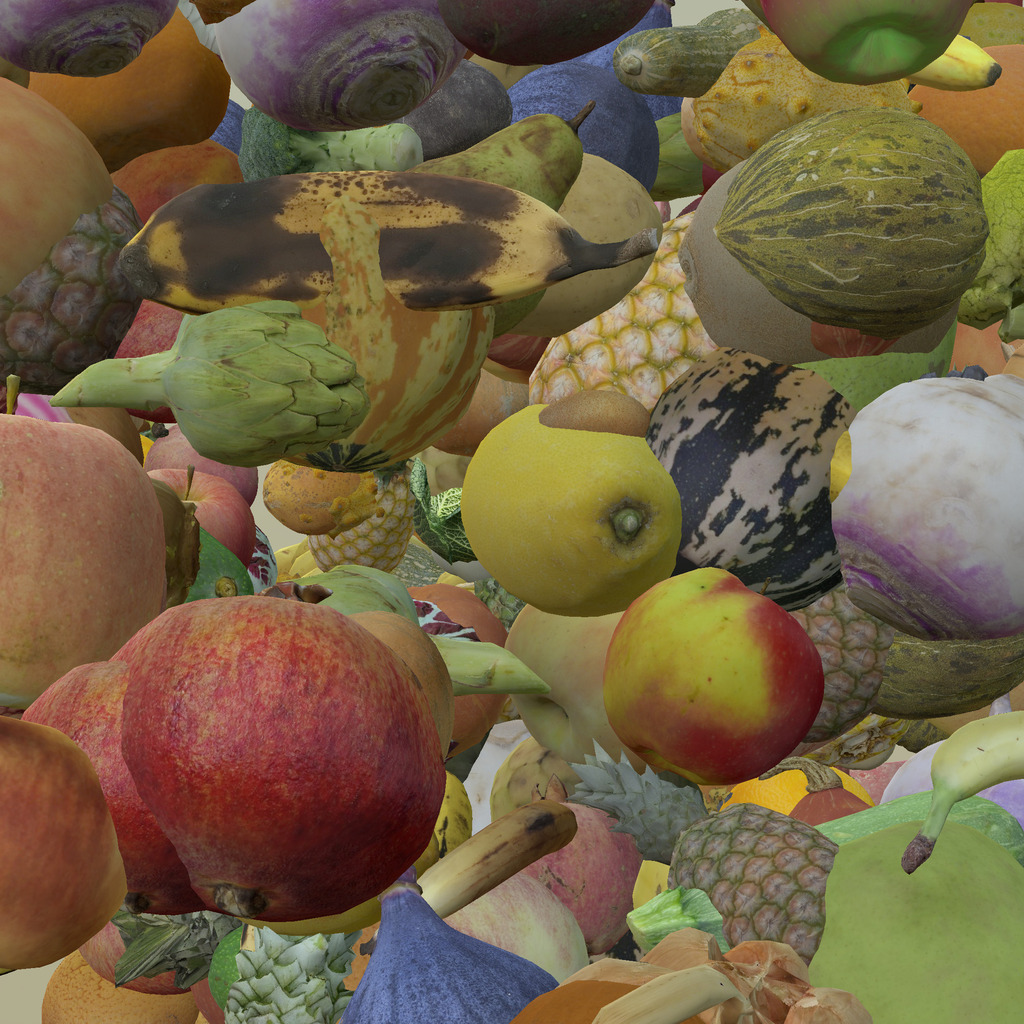} };
        \node[left=0pt of closeup_1,rotate=90,anchor=south] {\tiny\textbf{Reference}};
        \node[black,draw,thick,inner sep=0pt, left=0.02cm of closeup_1_b.south west, anchor=south east] (closeup_1_b) { \includegraphics[width=2.48cm]{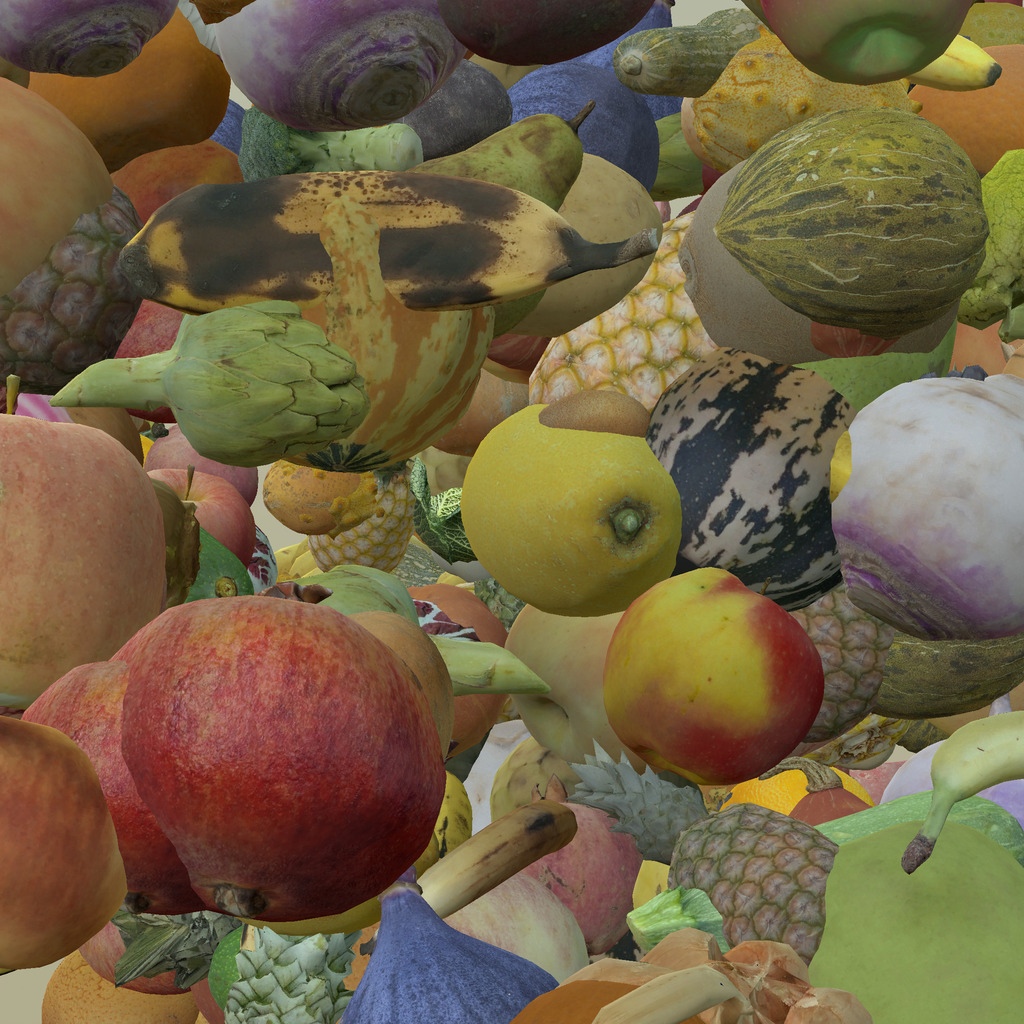} };
        \node[left=0pt of closeup_1_b,rotate=90,anchor=south] {\tiny\textbf{Ours}};
    \end{tikzpicture}
    \vspace{-5pt}
    \caption{\textbf{Batching of Neural Compression.}
    We use the Hypernetwork to quickly convert a scene composed of 100 different PBR materials.
    Including time to load and write files, it took us $8\mbox{min}\,32\mbox{sec}$ to completely convert this asset to neural texture sets.
    \label{fig:results_demo}
    \vspace{-0.50cm}
    }
\end{figure*}

\paragraph*{Comparison with BCF1.} We compared our Hypernetwork approach with a gradient descent optimizer for BCF1~\cite{belcour2025tsnc}. We applied both methods to a random subset of texture sets from the Polyhaven dataset (including textures and models) and we selected \textsc{Variant A} with an MLP of $32$ hidden dimensions. We ran the BCF1 method for $40$K iterations, which already takes $14$ minutes to run per texture set. In comparison, our hypernetwork takes $2.4$s per texture set on the same hardware (Nvidia A6000 Ada). We notice that we report an average PSNR for the BCF1 method that is lower than the one reported in the original paper. This is because we use a different dataset. Indeed, we experienced that texture sets from Models (and not stochastic or repetitive textures) are harder to compress.

A benefit of using the hypernetwork, is that we can efficiently many compress texture sets. In Figure~\ref{fig:results_demo}, we show the rendering of a scene composed of hundreds of individual texture sets of $1024 \times 1024$ pixels. Our hypernetwork converted this scene in less than $10$ minutes. It would take more than half a day to convert the same scene using the original BCF1 optimizer. Our approach permits to avoid long baking times.

\begin{figure}[b]
    \vspace{-0.2cm}
        \center
    \resizebox{\linewidth}{!}{
    \begin{tikzpicture}[font=\tiny]

        \node[draw,thick,inner sep=0pt] (DR) at (0,0) { \includegraphics[width=2.0cm]{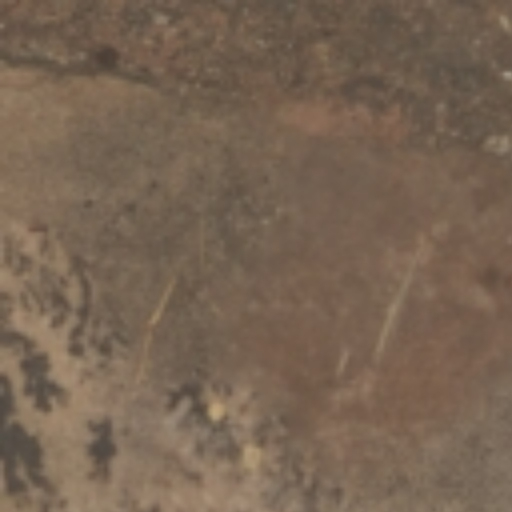} };
        \node[draw,thick,inner sep=0pt,below=1pt of DR] (DO) { \includegraphics[width=2.0cm]{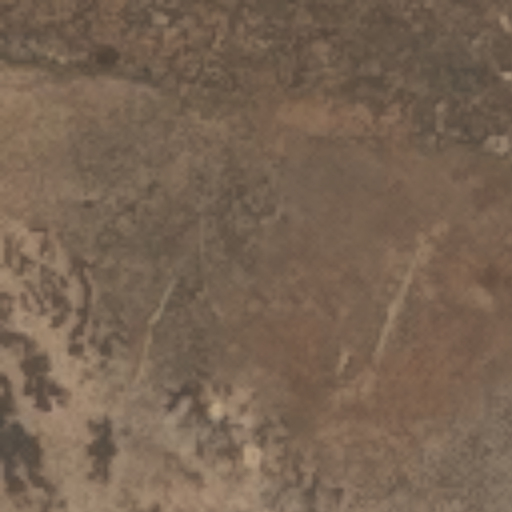} };

        \node[draw,thick,inner sep=0pt,right=1pt of DR,anchor=west] (AR) { \includegraphics[width=2.0cm]{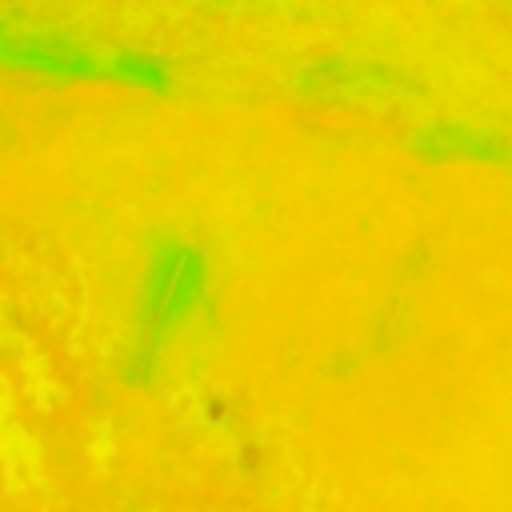} };
        \node[draw,thick,inner sep=0pt,below=1pt of AR] (AO) { \includegraphics[width=2.0cm]{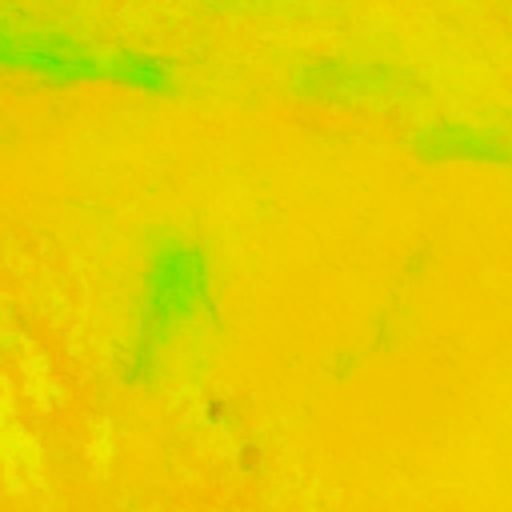} };

        \node[draw,thick,inner sep=0pt,right=1pt of AR,anchor=west] (NR) { \includegraphics[width=2.0cm]{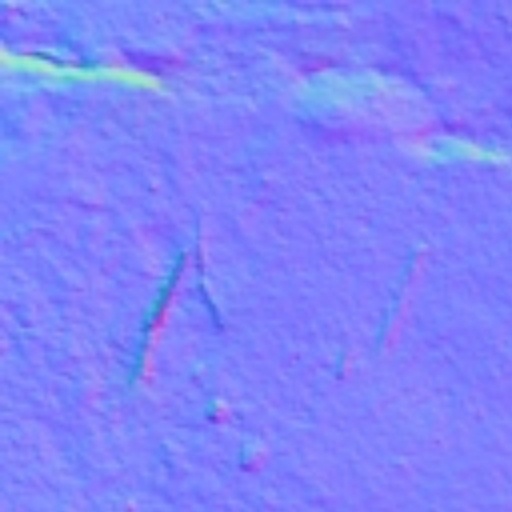} };
        \node[draw,thick,inner sep=0pt,below=1pt of NR] (NO) { \includegraphics[width=2.0cm]{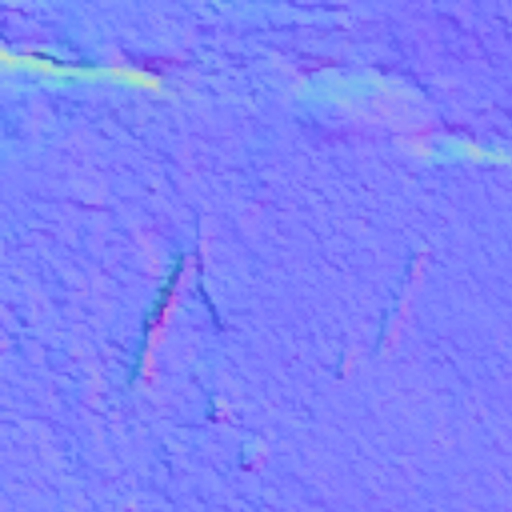} };
        \node[] at ($(DR.north)+(0,0.1)$) {\textbf{Albedo}};
        \node[] at ($(AR.north)+(0,0.1)$) {\textbf{ARM}};
        \node[] at ($(NR.north)+(0,0.1)$) {\textbf{Normal}};

        \node[rotate=90] at ($(DR.west)+(-0.15,0.0)$) {\textbf{Reference}};
        \node[rotate=90] at ($(DO.west)+(-0.15,0.0)$) {\textbf{Super Resolution}};
        
    \end{tikzpicture}
    }
    \caption{\textbf{Material Super-Resolution.} When the Hypernetwork is targeted for super-resolution, the decoder enhances texture details while preserving the material's overall appearance.
        \label{fig:results_super_resolution}
    }
\end{figure}  

In Figure~\ref{fig:results_metrics_polyhaven}, we compare the quality statistics (for both the PSNR and the MS-SSIM metrics) of our method compared to the original BCF1 optimizer~\cite{belcour2025tsnc}. Our method produces a lower average PSNR over all texture sets, but has a higher average MS-SSIM. This is expected since the BCF1 optimizer uses an $l_2$ loss but we incorporate a perceptual loss in our training. In the same figure, we also report the statistics of using a single global token for all texture sets (see Sec.~\ref{sec:tokens_to_mlps}). As expected, we see a degradation in PSNR compared to using one token per texture set. However, this drop in quality is smaller than $1$dB in PSNR and $0.02$ in MS-SSIM.

\paragraph*{Material Super-Resolution} In Figure~\ref{fig:results_super_resolution} we show the results of using our hypernetwork for material super-resolution. For that example, we trained our model targeting a MLP with an inner dimension of $64$. We can see that our method is able to enhance the details of the textures while preserving the overall appearance of the material. In particular, we can see that our method is able to enhance the grain and scratches of the original textures, which are important for the visual quality of the material.

\begin{figure}[b]
    \vspace{-0.0cm}
        \center
    \begin{tikzpicture}[font=\tiny]

        \node[draw,thick,inner sep=0pt] (D16) at (0,0) { \includegraphics[width=1.8cm]{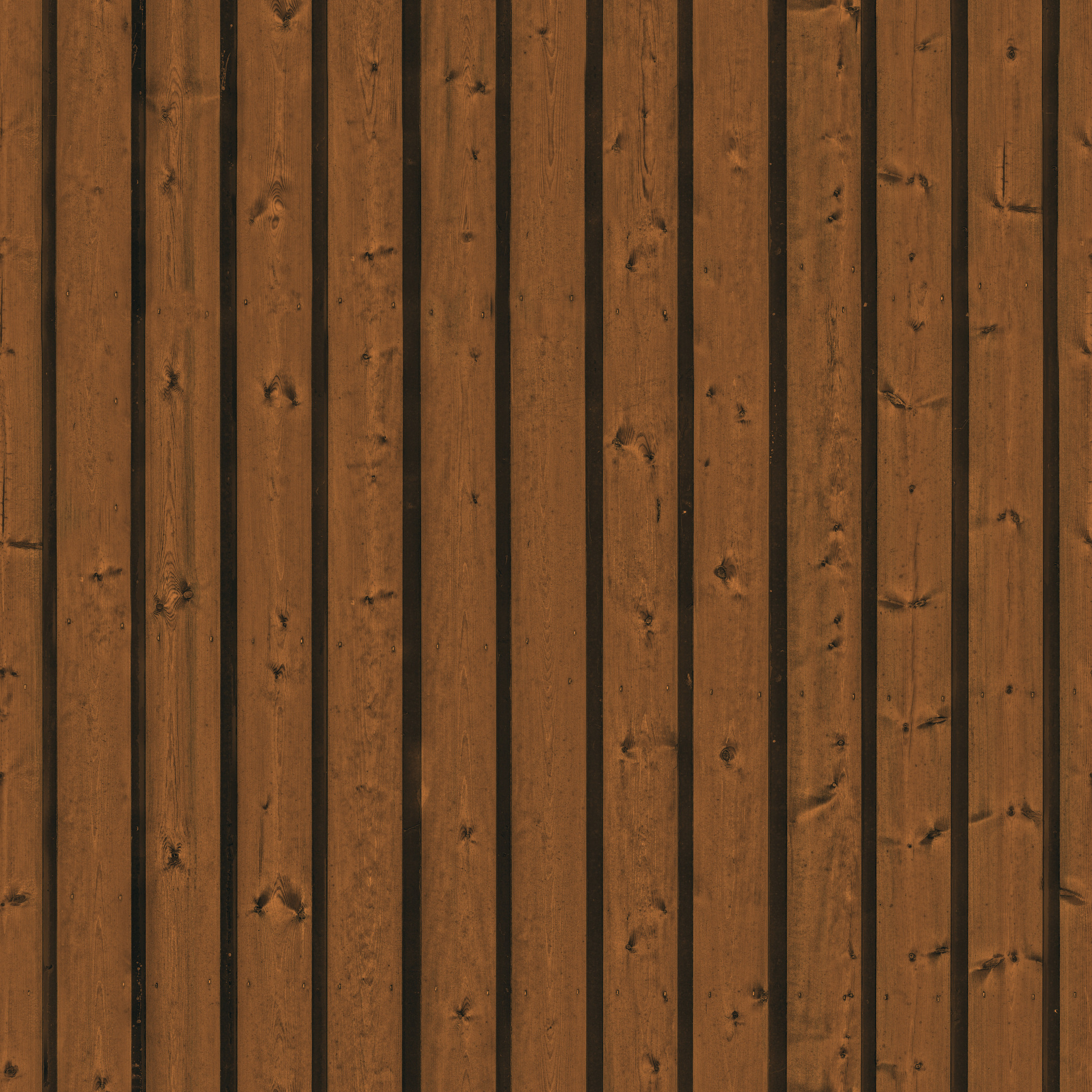} };
        \node[draw,thick,inner sep=0pt,right=1pt of D16] (D32) { \includegraphics[width=1.8cm]{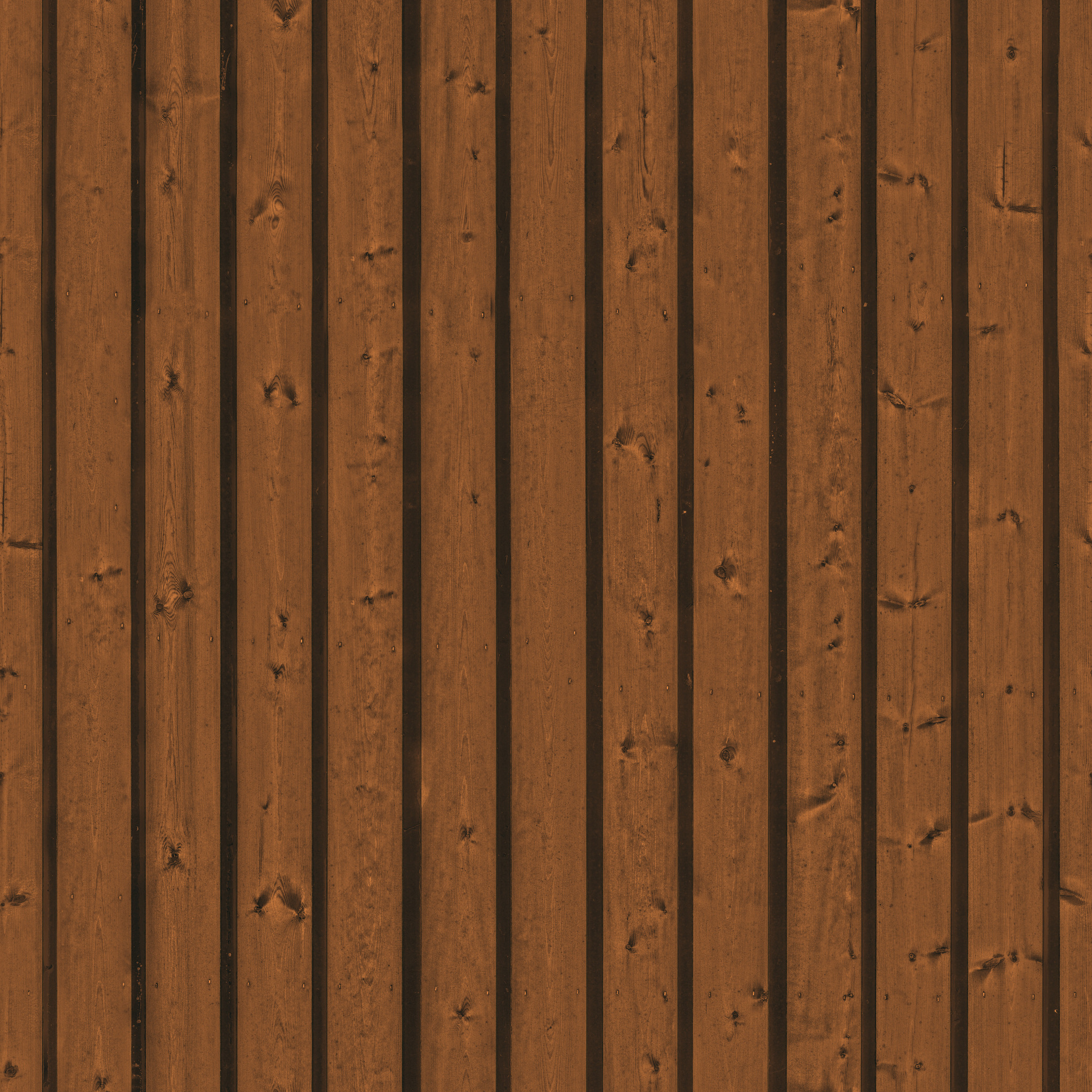} };
        \node[draw,thick,inner sep=0pt,right=1pt of D32] (D64) { \includegraphics[width=1.8cm]{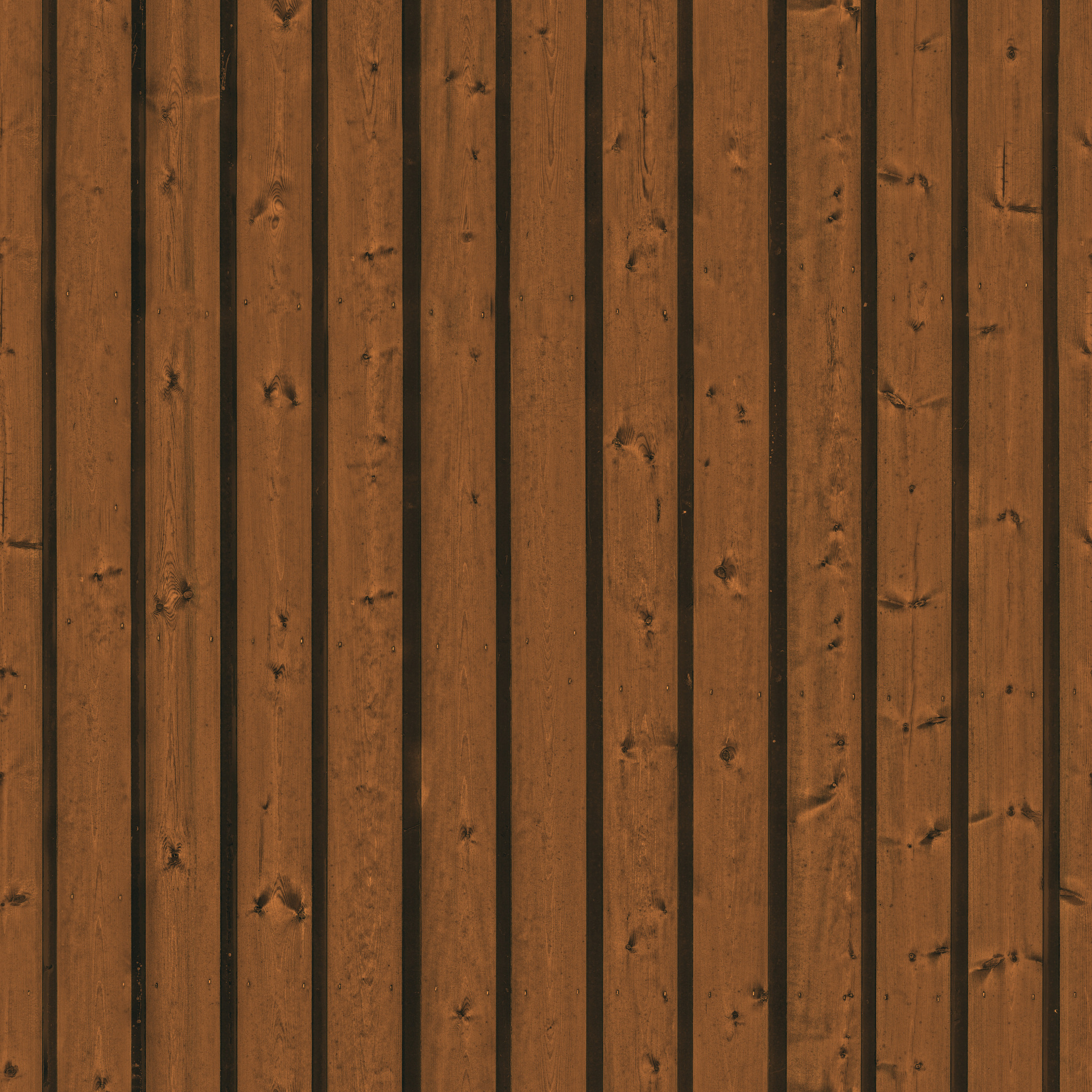} };
        \node[draw,thick,inner sep=0pt,right=1pt of D64] (ref) { \includegraphics[width=1.8cm]{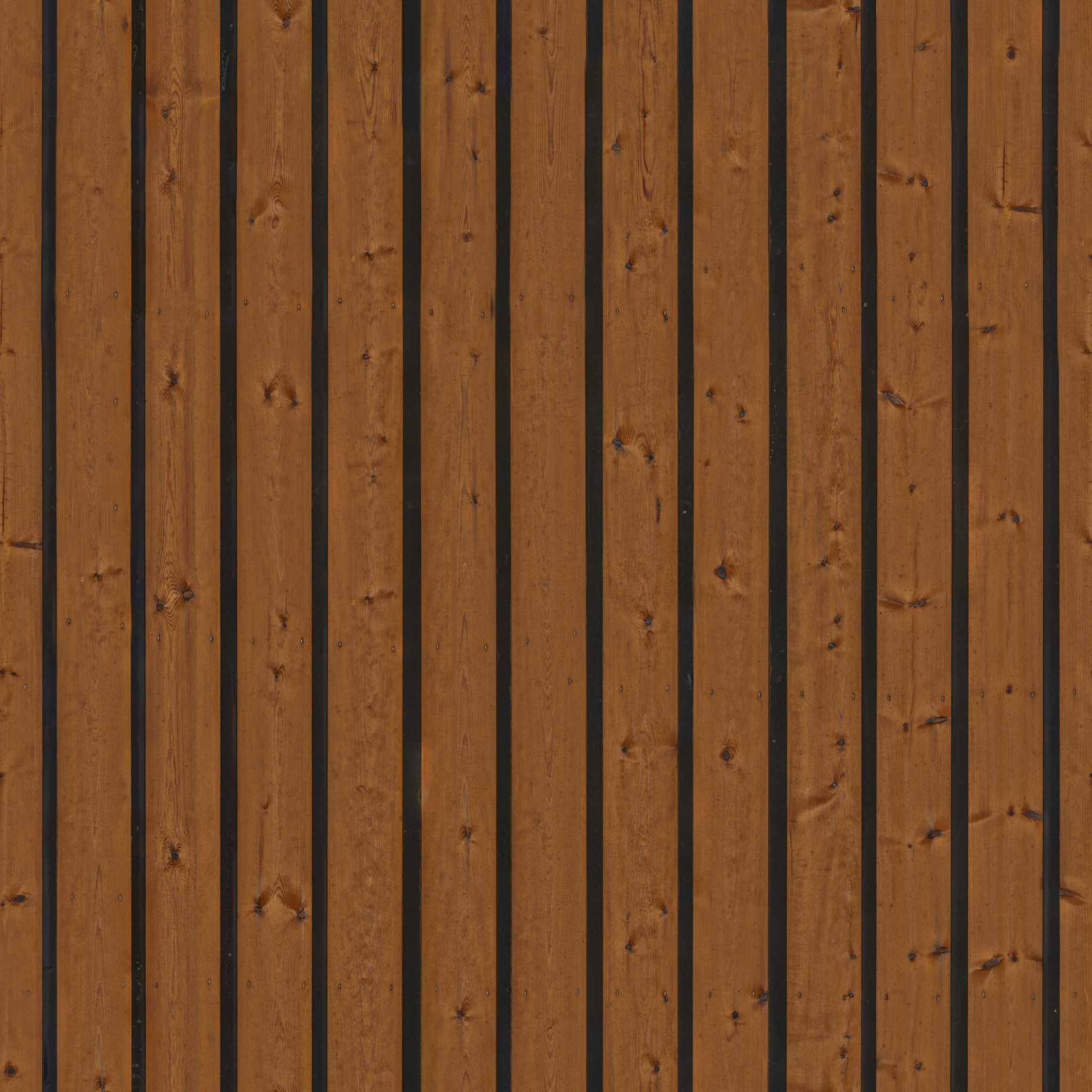} };

        \node[rotate=90] at ($(D16.west)+(-0.14,0.0)$) {\textbf{Albedo}};
        \node[] at ($(D16.north)+(0.0,0.13)$) {$D_h = 16$};
        \node[] at ($(D32.north)+(0.0,0.13)$) {$D_h = 32$};
        \node[] at ($(D64.north)+(0.0,0.13)$) {$D_h = 64$};
        \node[] at ($(ref.north)+(0.0,0.13)$) {\textbf{Reference}};

        \node[draw,thick,inner sep=0pt,below=1pt of D16] (D16) { \includegraphics[width=1.8cm]{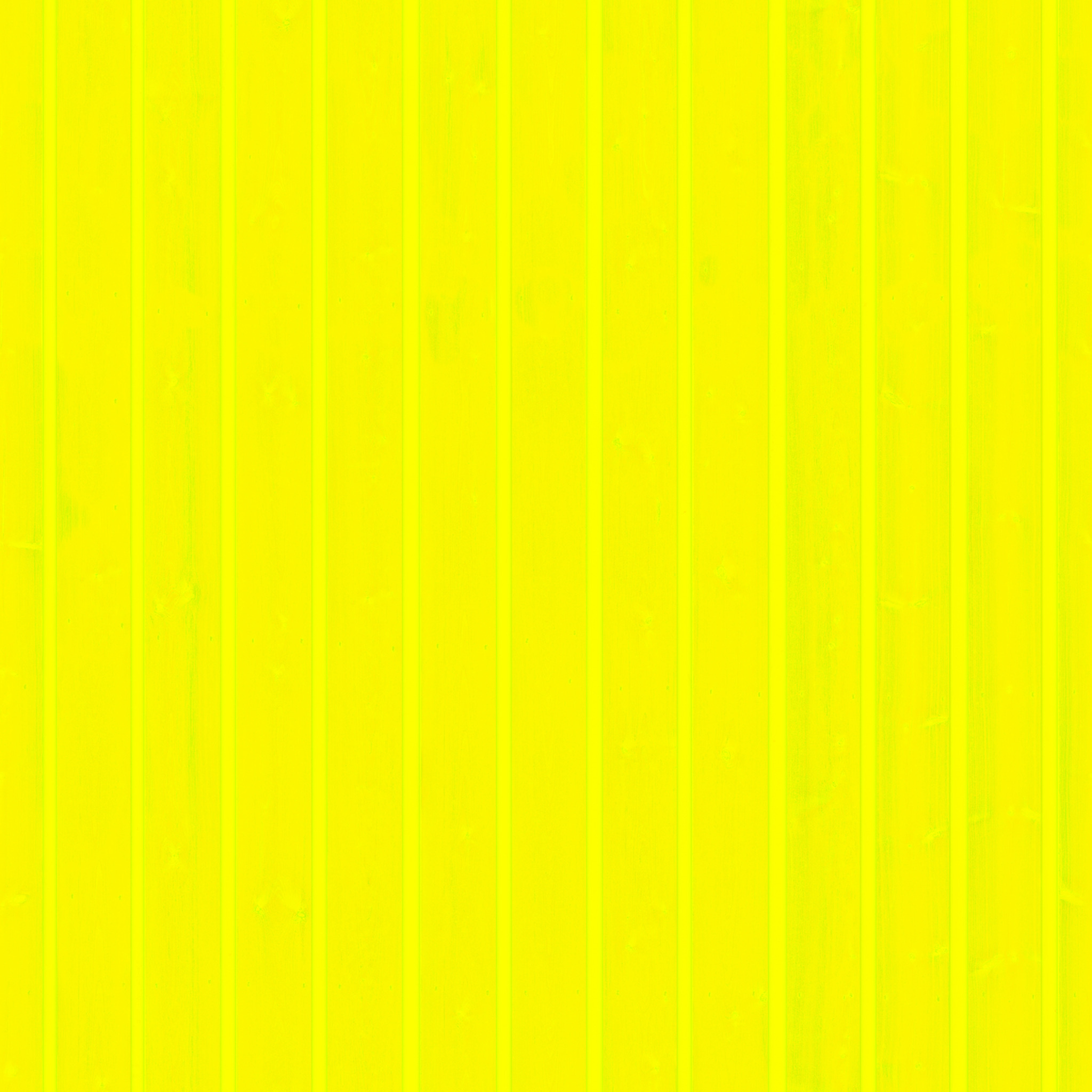} };
        \node[draw,thick,inner sep=0pt,right=1pt of D16] (D32) { \includegraphics[width=1.8cm]{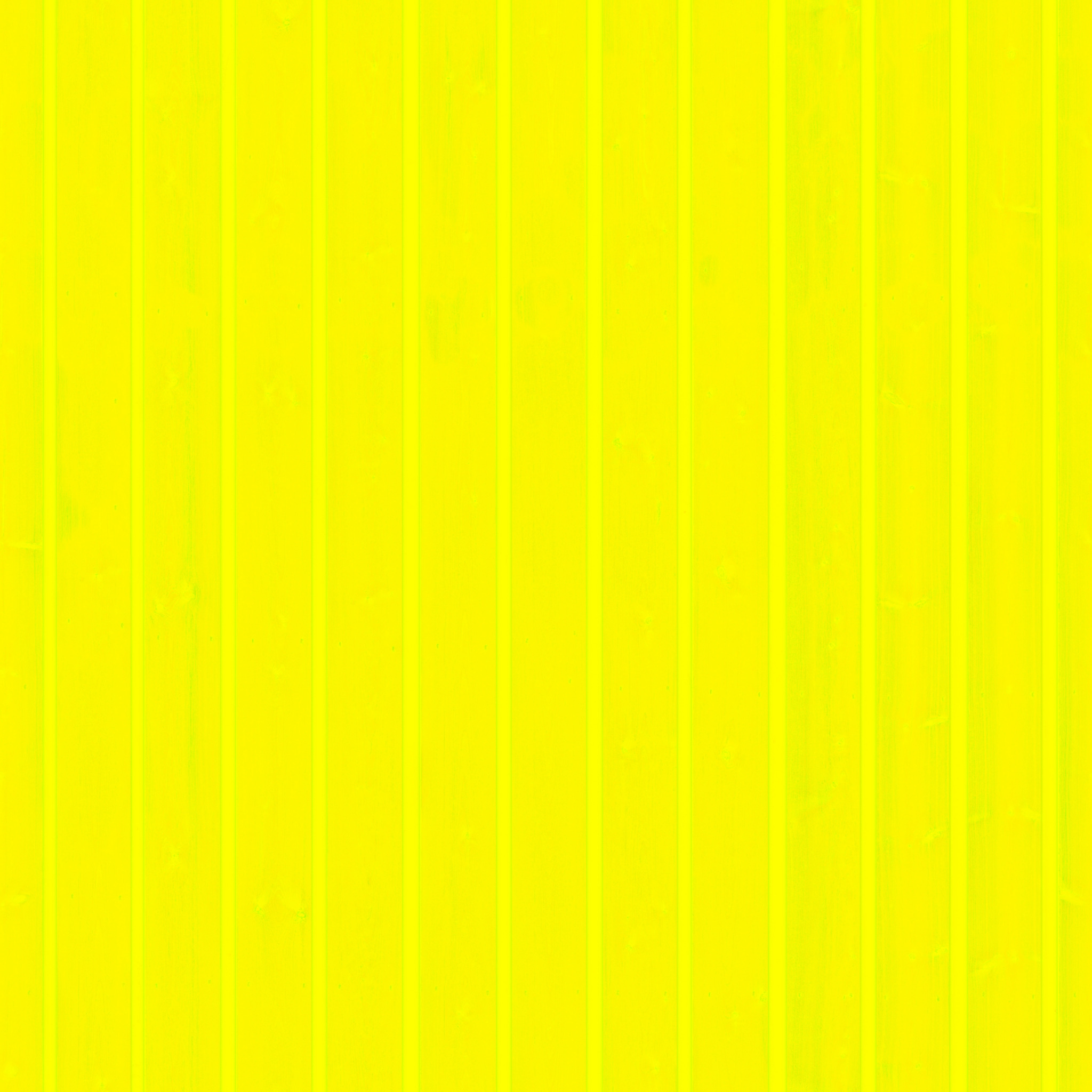} };
        \node[draw,thick,inner sep=0pt,right=1pt of D32] (D64) { \includegraphics[width=1.8cm]{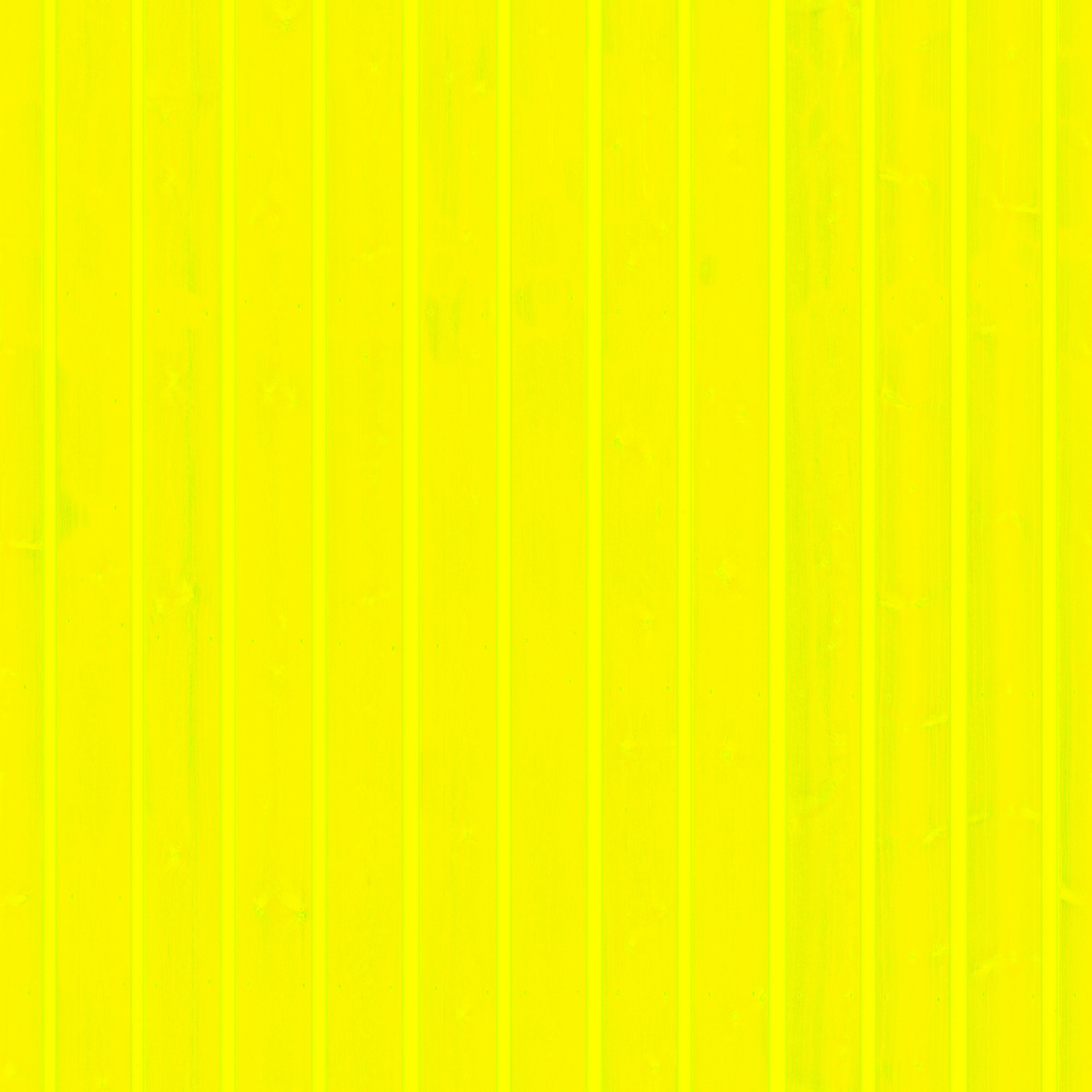} };
        \node[draw,thick,inner sep=0pt,right=1pt of D64] (ref) { \includegraphics[width=1.8cm]{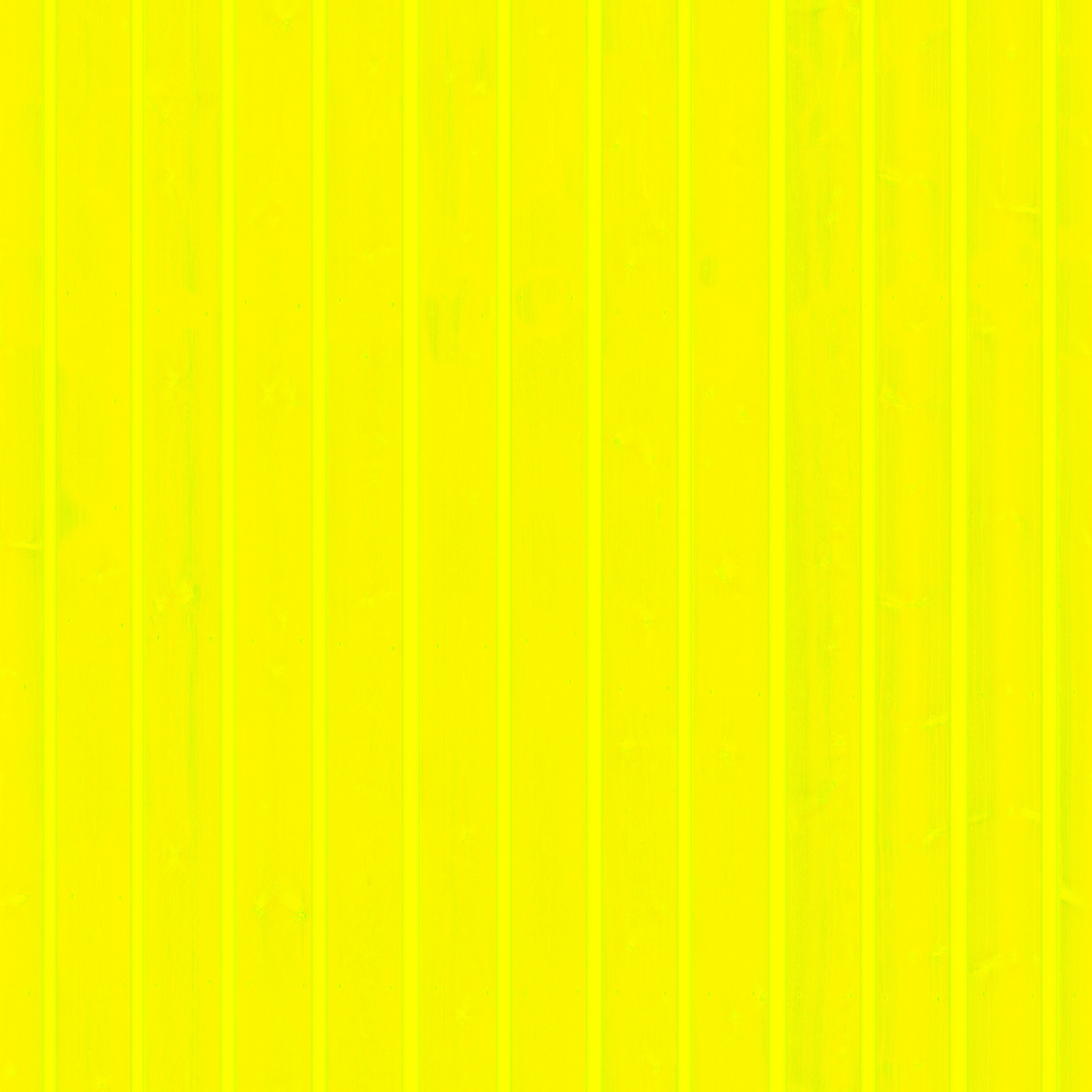} };
        \node[rotate=90] at ($(D16.west)+(-0.14,0.0)$) {\textbf{ARM}};

        \node[draw,thick,inner sep=0pt,below=1pt of D16] (D16) { \includegraphics[width=1.8cm]{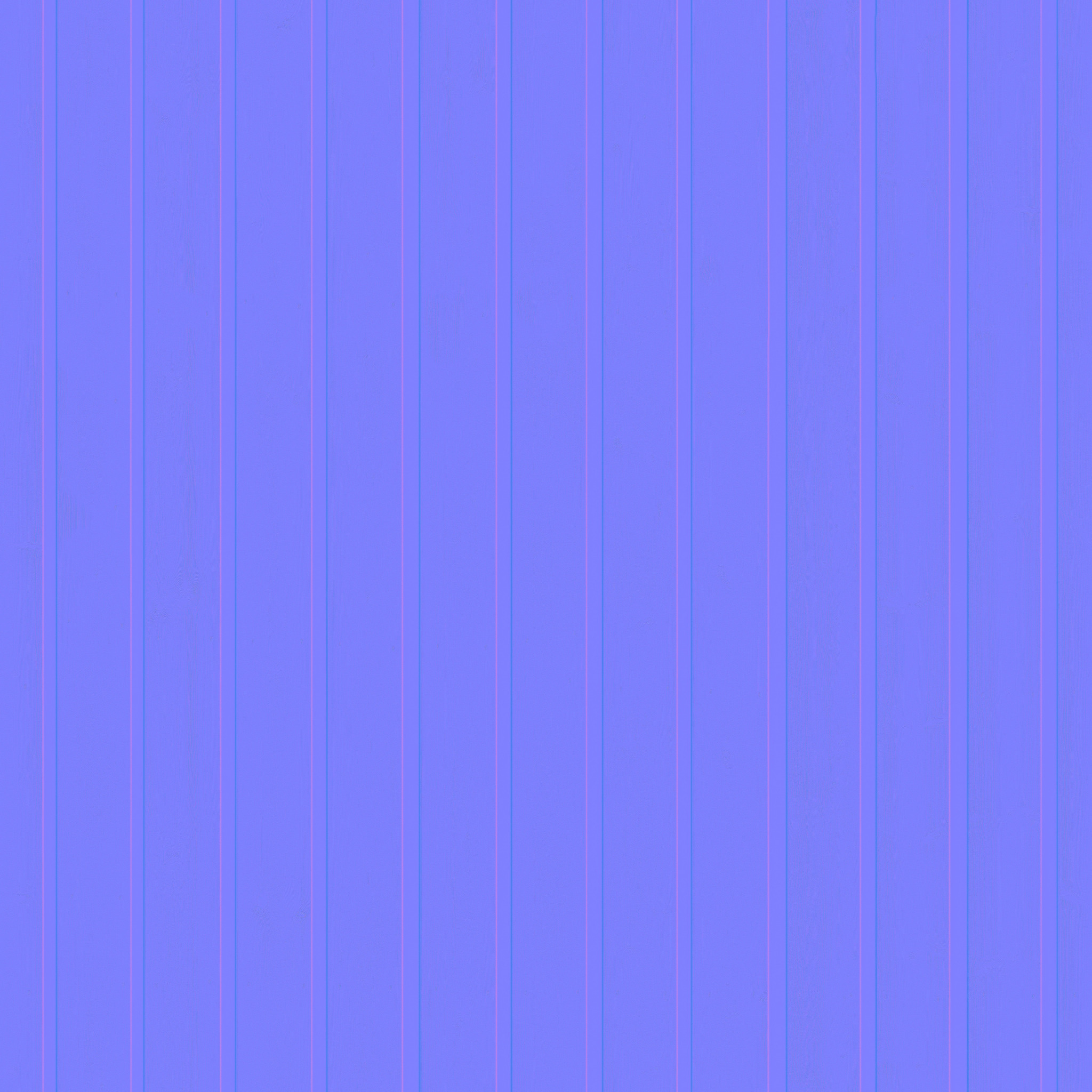} };
        \node[draw,thick,inner sep=0pt,right=1pt of D16] (D32) { \includegraphics[width=1.8cm]{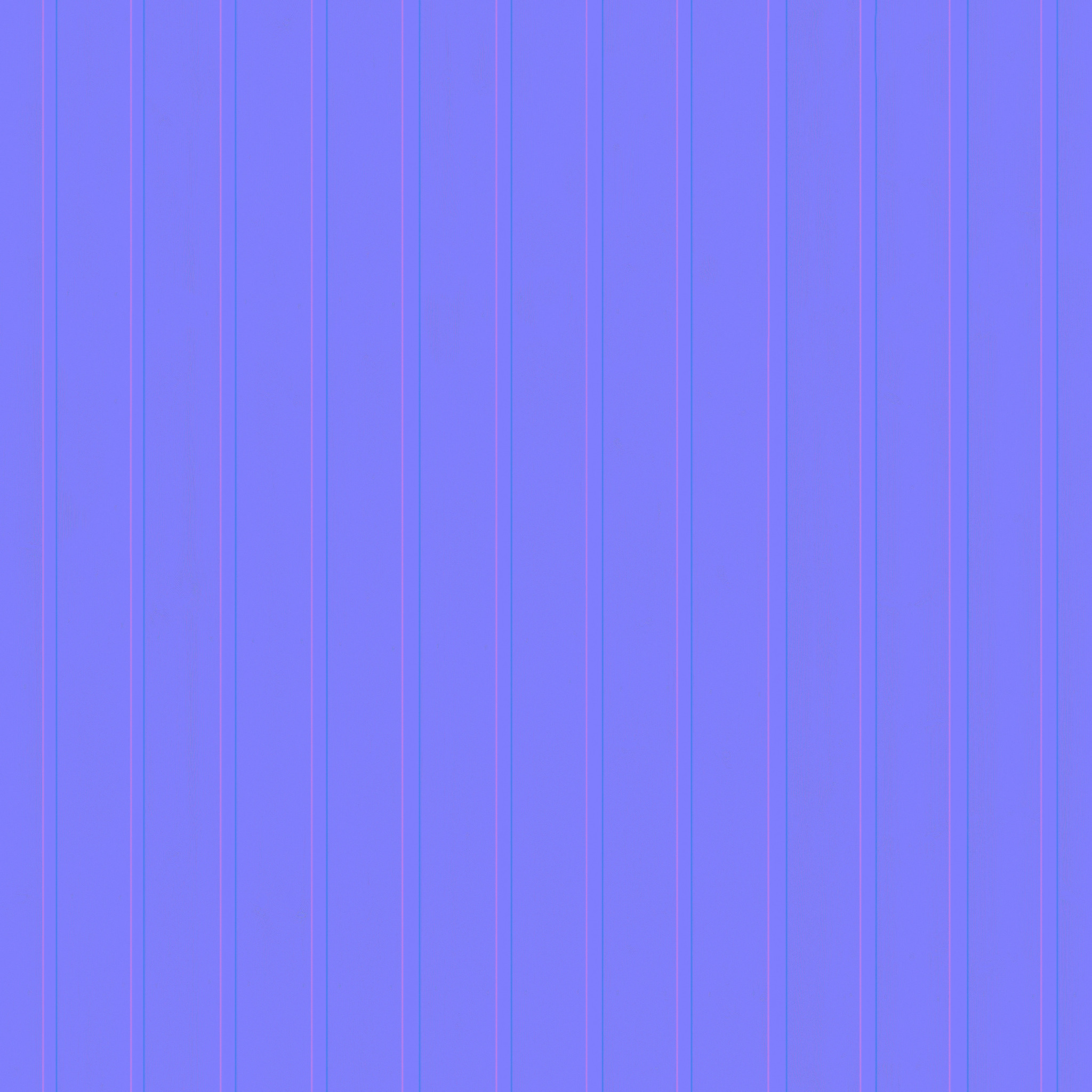} };
        \node[draw,thick,inner sep=0pt,right=1pt of D32] (D64) { \includegraphics[width=1.8cm]{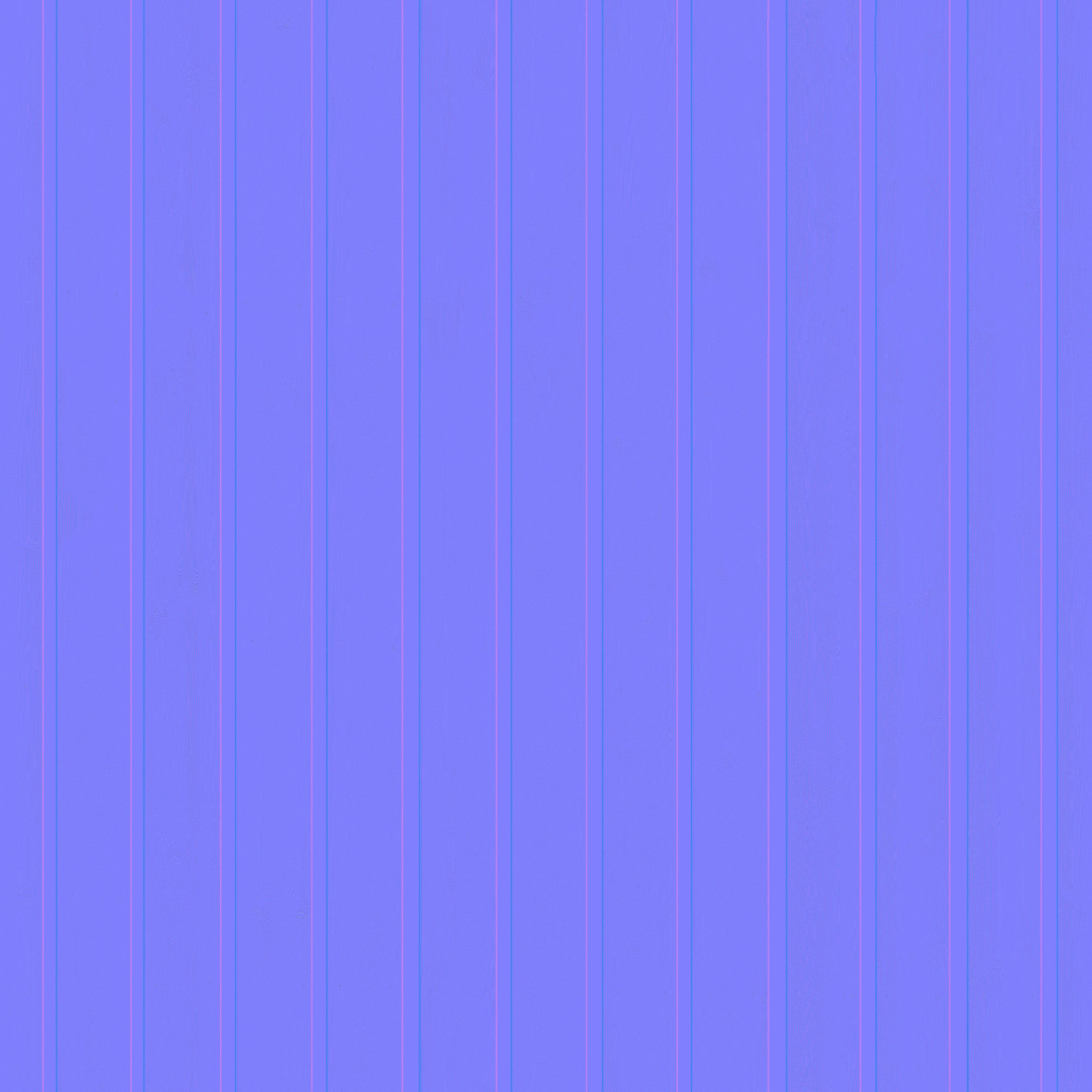} };
        \node[draw,thick,inner sep=0pt,right=1pt of D64] (ref) { \includegraphics[width=1.8cm]{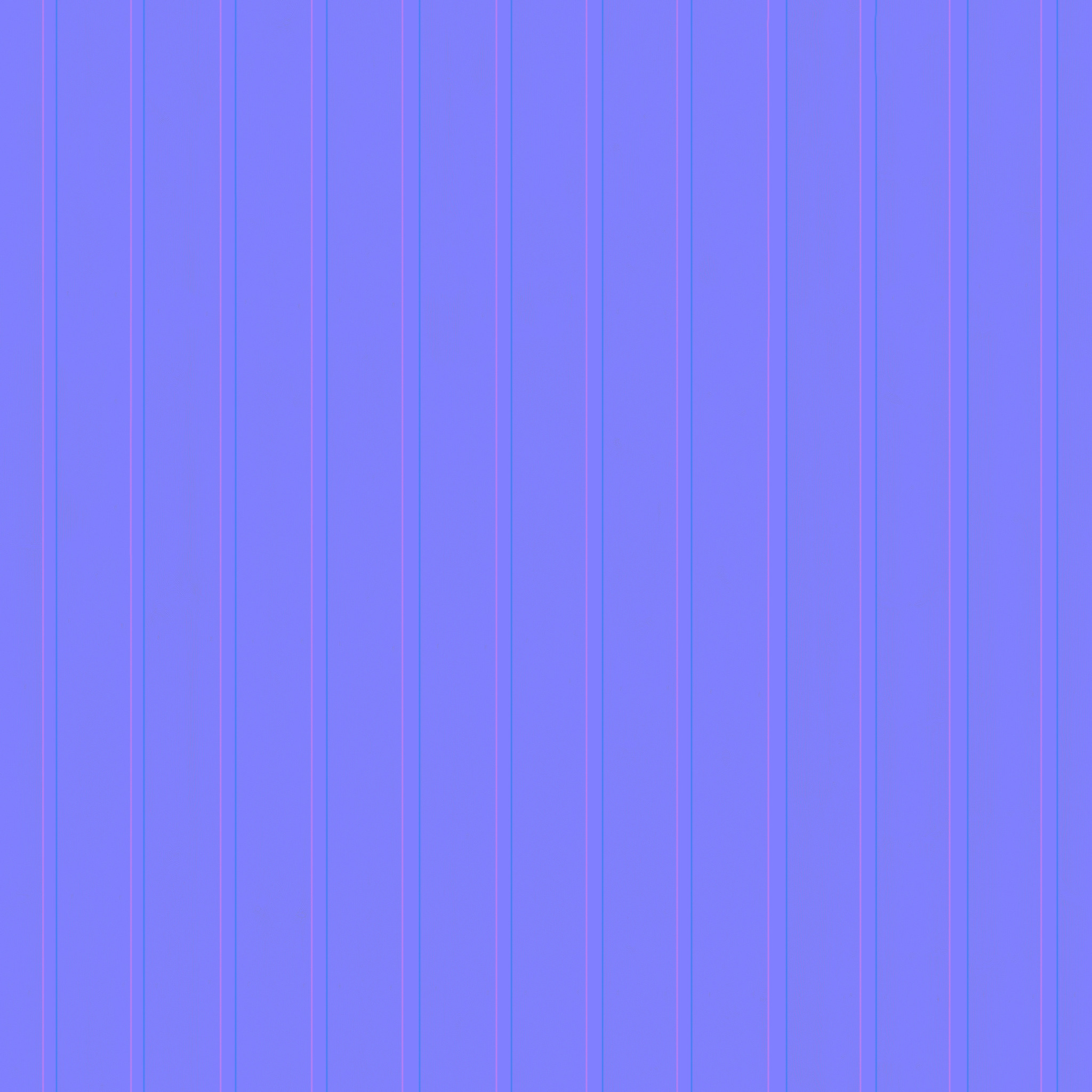} };
        \node[rotate=90] at ($(D16.west)+(-0.14,0.0)$) {\textbf{Normal}};

        

    \end{tikzpicture}
    \caption{\textbf{Multiple Heads.} We use the same inference of our Hypernetwork to produce multiple MLP weights and biases for the same latents. For this texture set, we generate $D_h = 16, 32$ and $64$.
        \label{fig:results_multiple_heads}
    }
\end{figure}
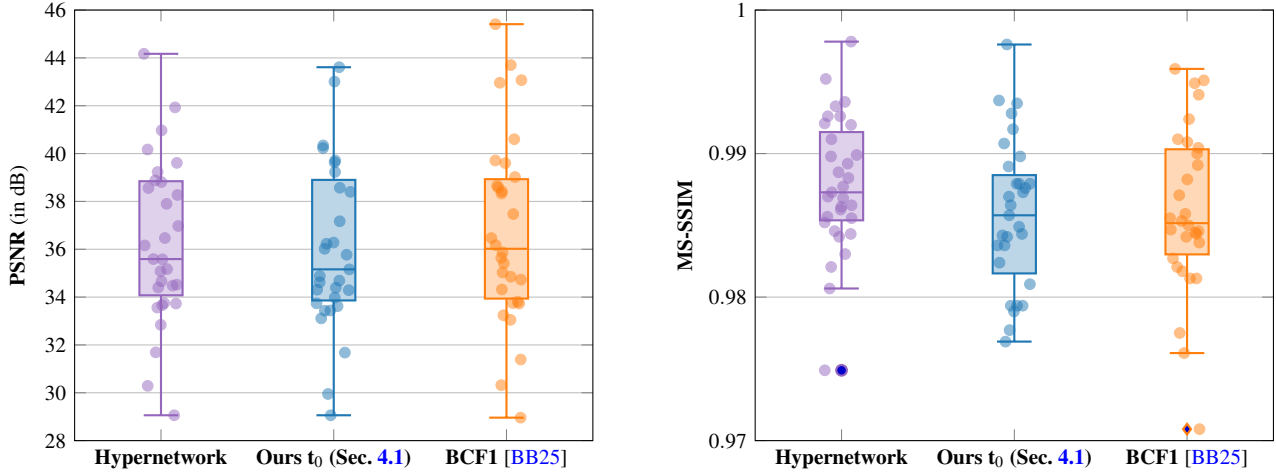
\begin{figure*}[t]
    \vspace{-0.0cm}
\definecolor{tab_blue}{rgb}{0.12156862745098039,0.4666666666666667,0.7058823529411765}
\definecolor{tab_orange}{rgb}{1.0,0.4980392156862745,0.054901960784313725}
\definecolor{tab_green}{rgb}{0.17254901960784313,0.6274509803921569,0.17254901960784313}
\definecolor{tab_red}{rgb}{0.8392156862745098,0.15294117647058825,0.1568627450980392}
\definecolor{tab_purple}{rgb}{0.5803921568627451,0.403921568627451,0.7411764705882353}
\definecolor{tab_brown}{rgb}{0.5490196078431373,0.33725490196078434,0.29411764705882354}
\definecolor{tab_pink}{rgb}{0.8901960784313725,0.4666666666666667,0.7607843137254902}
\definecolor{tab_gray}{rgb}{0.4980392156862745,0.4980392156862745,0.4980392156862745}
\definecolor{tab_olive}{rgb}{0.7372549019607844,0.7411764705882353,0.13333333333333333}
\definecolor{tab_cyan}{rgb}{0.09019607843137255,0.7450980392156863,0.8117647058823529}          
\begin{tikzpicture}[font=\small]

    \begin{axis}[
        boxplot/draw direction=y, 
        ytick distance=2, 
        xmin=0.5, xmax=3.5,
        xticklabels={
            \textbf{Hypernetwork},
            \textbf{Ours $\mathbf{t}_0$~(Sec.~\ref{sec:tokens_to_mlps})},
            \textbf{BCF1}~\cite{belcour2025tsnc},
        },
        ymin=28.0, ymax=46.0,
        ymajorgrids, 
        xtick={1,2,3}, 
        ylabel={\textbf{PSNR} (in dB)},
    ]

\addplot+ [boxplot, thick, solid, tab_purple, fill=tab_purple!30!white,boxplot/box extend=0.25,] table [row sep=\\,y index=0] {
34.52 \\33.75 \\35.08 \\41.93 \\35.58 \\35.59 \\36.16 \\36.47 \\34.48 \\34.40 \\39.23 \\35.17 \\40.17 \\38.81 \\32.84 \\38.27 \\33.56 \\31.69 \\34.66 \\39.61 \\47.16 \\30.29 \\44.17 \\40.97 \\36.97 \\33.73 \\37.90 \\38.56 \\33.66 \\29.06 \\38.88 \\};
\addplot [tab_purple,only marks, mark=*, opacity=0.5] coordinates {
	(1+ 0.1*rand, 34.52)
	(1+ 0.1*rand, 33.75)
	(1+ 0.1*rand, 35.08)
	(1+ 0.1*rand, 41.93)
	(1+ 0.1*rand, 35.58)
	(1+ 0.1*rand, 35.59)
	(1+ 0.1*rand, 36.16)
	(1+ 0.1*rand, 36.47)
	(1+ 0.1*rand, 34.48)
	(1+ 0.1*rand, 34.40)
	(1+ 0.1*rand, 39.23)
	(1+ 0.1*rand, 35.17)
	(1+ 0.1*rand, 40.17)
	(1+ 0.1*rand, 38.81)
	(1+ 0.1*rand, 32.84)
	(1+ 0.1*rand, 38.27)
	(1+ 0.1*rand, 33.56)
	(1+ 0.1*rand, 31.69)
	(1+ 0.1*rand, 34.66)
	(1+ 0.1*rand, 39.61)
	(1+ 0.1*rand, 47.16)
	(1+ 0.1*rand, 30.29)
	(1+ 0.1*rand, 44.17)
	(1+ 0.1*rand, 40.97)
	(1+ 0.1*rand, 36.97)
	(1+ 0.1*rand, 33.73)
	(1+ 0.1*rand, 37.90)
	(1+ 0.1*rand, 38.56)
	(1+ 0.1*rand, 33.66)
	(1+ 0.1*rand, 29.06)
	(1+ 0.1*rand, 38.88)
};
\addplot+ [boxplot, thick, solid, tab_blue, fill=tab_blue!30!white,boxplot/box extend=0.25,] table [row sep=\\,y index=0] {
34.31 \\33.44 \\34.88 \\43.01 \\34.62 \\36.28 \\36.02 \\35.16 \\34.29 \\34.69 \\40.23 \\34.39 \\39.23 \\38.40 \\33.43 \\36.23 \\33.74 \\31.68 \\33.11 \\39.71 \\47.27 \\29.95 \\43.61 \\39.63 \\37.17 \\33.98 \\35.77 \\40.34 \\33.62 \\29.06 \\38.57 \\};
\addplot [tab_blue,only marks, mark=*, opacity=0.5] coordinates {
	(2+ 0.1*rand, 34.31)
	(2+ 0.1*rand, 33.44)
	(2+ 0.1*rand, 34.88)
	(2+ 0.1*rand, 43.01)
	(2+ 0.1*rand, 34.62)
	(2+ 0.1*rand, 36.28)
	(2+ 0.1*rand, 36.02)
	(2+ 0.1*rand, 35.16)
	(2+ 0.1*rand, 34.29)
	(2+ 0.1*rand, 34.69)
	(2+ 0.1*rand, 40.23)
	(2+ 0.1*rand, 34.39)
	(2+ 0.1*rand, 39.23)
	(2+ 0.1*rand, 38.40)
	(2+ 0.1*rand, 33.43)
	(2+ 0.1*rand, 36.23)
	(2+ 0.1*rand, 33.74)
	(2+ 0.1*rand, 31.68)
	(2+ 0.1*rand, 33.11)
	(2+ 0.1*rand, 39.71)
	(2+ 0.1*rand, 47.27)
	(2+ 0.1*rand, 29.95)
	(2+ 0.1*rand, 43.61)
	(2+ 0.1*rand, 39.63)
	(2+ 0.1*rand, 37.17)
	(2+ 0.1*rand, 33.98)
	(2+ 0.1*rand, 35.77)
	(2+ 0.1*rand, 40.34)
	(2+ 0.1*rand, 33.62)
	(2+ 0.1*rand, 29.06)
	(2+ 0.1*rand, 38.57)
};
\addplot+ [boxplot, thick, solid, tab_orange, fill=tab_orange!30!white,boxplot/box extend=0.25,] table [row sep=\\,y index=0] {
34.32 \\33.81 \\34.85 \\43.07 \\35.87 \\38.33 \\36.47 \\37.47 \\35.03 \\35.41 \\40.60 \\34.73 \\38.58 \\39.60 \\33.75 \\35.64 \\33.24 \\31.39 \\36.17 \\39.71 \\30.32 \\45.41 \\43.70 \\39.02 \\33.73 \\38.42 \\42.96 \\33.05 \\28.96 \\38.66 \\};
\addplot [tab_orange,only marks, mark=*, opacity=0.5] coordinates {
	(3+ 0.1*rand, 34.32)
	(3+ 0.1*rand, 33.81)
	(3+ 0.1*rand, 34.85)
	(3+ 0.1*rand, 43.07)
	(3+ 0.1*rand, 35.87)
	(3+ 0.1*rand, 38.33)
	(3+ 0.1*rand, 36.47)
	(3+ 0.1*rand, 37.47)
	(3+ 0.1*rand, 35.03)
	(3+ 0.1*rand, 35.41)
	(3+ 0.1*rand, 40.60)
	(3+ 0.1*rand, 34.73)
	(3+ 0.1*rand, 38.58)
	(3+ 0.1*rand, 39.60)
	(3+ 0.1*rand, 33.75)
	(3+ 0.1*rand, 35.64)
	(3+ 0.1*rand, 33.24)
	(3+ 0.1*rand, 31.39)
	(3+ 0.1*rand, 36.17)
	(3+ 0.1*rand, 39.71)
	(3+ 0.1*rand, 30.32)
	(3+ 0.1*rand, 45.41)
	(3+ 0.1*rand, 43.70)
	(3+ 0.1*rand, 39.02)
	(3+ 0.1*rand, 33.73)
	(3+ 0.1*rand, 38.42)
	(3+ 0.1*rand, 42.96)
	(3+ 0.1*rand, 33.05)
	(3+ 0.1*rand, 28.96)
	(3+ 0.1*rand, 38.66)
};
    \end{axis}
    \begin{scope}[xshift=9cm]
    \begin{axis}[
        boxplot/draw direction=y, 
        ytick distance=0.01, 
        xmin=0.5, xmax=3.5,
        xticklabels={
            \textbf{Hypernetwork},
            \textbf{Ours $\mathbf{t}_0$~(Sec.~\ref{sec:tokens_to_mlps})},
            \textbf{BCF1}~\cite{belcour2025tsnc},
        },
        ymin=0.97, ymax=1.0,
        ymajorgrids, 
        xtick={1,2,3}, 
        ylabel={\textbf{MS-SSIM}},
    ]

\addplot+ [boxplot, thick, solid, tab_purple, fill=tab_purple!30!white,boxplot/box extend=0.25,] table [row sep=\\,y index=0] {
0.9852 \\0.9861 \\0.9887 \\0.9936 \\0.9855 \\0.9870 \\0.9863 \\0.9846 \\0.9806 \\0.9864 \\0.9926 \\0.9821 \\0.9933 \\0.9910 \\0.9844 \\0.9893 \\0.9883 \\0.9869 \\0.9842 \\0.9926 \\0.9978 \\0.9749 \\0.9952 \\0.9920 \\0.9899 \\0.9877 \\0.9856 \\0.9898 \\0.9873 \\0.9830 \\0.9921 \\};
\addplot [tab_purple,only marks, mark=*, opacity=0.5] coordinates {
	(1+ 0.1*rand, 0.9852)
	(1+ 0.1*rand, 0.9861)
	(1+ 0.1*rand, 0.9887)
	(1+ 0.1*rand, 0.9936)
	(1+ 0.1*rand, 0.9855)
	(1+ 0.1*rand, 0.9870)
	(1+ 0.1*rand, 0.9863)
	(1+ 0.1*rand, 0.9846)
	(1+ 0.1*rand, 0.9806)
	(1+ 0.1*rand, 0.9864)
	(1+ 0.1*rand, 0.9926)
	(1+ 0.1*rand, 0.9821)
	(1+ 0.1*rand, 0.9933)
	(1+ 0.1*rand, 0.9910)
	(1+ 0.1*rand, 0.9844)
	(1+ 0.1*rand, 0.9893)
	(1+ 0.1*rand, 0.9883)
	(1+ 0.1*rand, 0.9869)
	(1+ 0.1*rand, 0.9842)
	(1+ 0.1*rand, 0.9926)
	(1+ 0.1*rand, 0.9978)
	(1+ 0.1*rand, 0.9749)
	(1+ 0.1*rand, 0.9952)
	(1+ 0.1*rand, 0.9920)
	(1+ 0.1*rand, 0.9899)
	(1+ 0.1*rand, 0.9877)
	(1+ 0.1*rand, 0.9856)
	(1+ 0.1*rand, 0.9898)
	(1+ 0.1*rand, 0.9873)
	(1+ 0.1*rand, 0.9830)
	(1+ 0.1*rand, 0.9921)
};

\addplot+ [boxplot, thick, solid, tab_blue, fill=tab_blue!30!white,boxplot/box extend=0.25,] table [row sep=\\,y index=0] {
0.9844 \\0.9824 \\0.9857 \\0.9935 \\0.9794 \\0.9873 \\0.9842 \\0.9809 \\0.9769 \\0.9843 \\0.9928 \\0.9794 \\0.9907 \\0.9879 \\0.9836 \\0.9876 \\0.9870 \\0.9836 \\0.9790 \\0.9917 \\0.9976 \\0.9678 \\0.9937 \\0.9879 \\0.9879 \\0.9864 \\0.9777 \\0.9891 \\0.9849 \\0.9794 \\0.9898 \\};
\addplot [tab_blue,only marks, mark=*, opacity=0.5] coordinates {
	(2+ 0.1*rand, 0.9844)
	(2+ 0.1*rand, 0.9824)
	(2+ 0.1*rand, 0.9857)
	(2+ 0.1*rand, 0.9935)
	(2+ 0.1*rand, 0.9794)
	(2+ 0.1*rand, 0.9873)
	(2+ 0.1*rand, 0.9842)
	(2+ 0.1*rand, 0.9809)
	(2+ 0.1*rand, 0.9769)
	(2+ 0.1*rand, 0.9843)
	(2+ 0.1*rand, 0.9928)
	(2+ 0.1*rand, 0.9794)
	(2+ 0.1*rand, 0.9907)
	(2+ 0.1*rand, 0.9879)
	(2+ 0.1*rand, 0.9836)
	(2+ 0.1*rand, 0.9876)
	(2+ 0.1*rand, 0.9870)
	(2+ 0.1*rand, 0.9836)
	(2+ 0.1*rand, 0.9790)
	(2+ 0.1*rand, 0.9917)
	(2+ 0.1*rand, 0.9976)
	(2+ 0.1*rand, 0.9678)
	(2+ 0.1*rand, 0.9937)
	(2+ 0.1*rand, 0.9879)
	(2+ 0.1*rand, 0.9879)
	(2+ 0.1*rand, 0.9864)
	(2+ 0.1*rand, 0.9777)
	(2+ 0.1*rand, 0.9891)
	(2+ 0.1*rand, 0.9849)
	(2+ 0.1*rand, 0.9794)
	(2+ 0.1*rand, 0.9898)
};
\addplot+ [boxplot, thick, solid, tab_orange, fill=tab_orange!30!white,boxplot/box extend=0.25,] table [row sep=\\,y index=0] {
0.9818 \\0.9845 \\0.9846 \\0.9941 \\0.9853 \\0.9892 \\0.9842 \\0.9855 \\0.9821 \\0.9858 \\0.9924 \\0.9775 \\0.9904 \\0.9882 \\0.9827 \\0.9847 \\0.9838 \\0.9813 \\0.9850 \\0.9910 \\0.9708 \\0.9959 \\0.9949 \\0.9908 \\0.9844 \\0.9871 \\0.9951 \\0.9813 \\0.9761 \\0.9900 \\};
\addplot [tab_orange,only marks, mark=*, opacity=0.5] coordinates {
	(3+ 0.1*rand, 0.9818)
	(3+ 0.1*rand, 0.9845)
	(3+ 0.1*rand, 0.9846)
	(3+ 0.1*rand, 0.9941)
	(3+ 0.1*rand, 0.9853)
	(3+ 0.1*rand, 0.9892)
	(3+ 0.1*rand, 0.9842)
	(3+ 0.1*rand, 0.9855)
	(3+ 0.1*rand, 0.9821)
	(3+ 0.1*rand, 0.9858)
	(3+ 0.1*rand, 0.9924)
	(3+ 0.1*rand, 0.9775)
	(3+ 0.1*rand, 0.9904)
	(3+ 0.1*rand, 0.9882)
	(3+ 0.1*rand, 0.9827)
	(3+ 0.1*rand, 0.9847)
	(3+ 0.1*rand, 0.9838)
	(3+ 0.1*rand, 0.9813)
	(3+ 0.1*rand, 0.9850)
	(3+ 0.1*rand, 0.9910)
	(3+ 0.1*rand, 0.9708)
	(3+ 0.1*rand, 0.9959)
	(3+ 0.1*rand, 0.9949)
	(3+ 0.1*rand, 0.9908)
	(3+ 0.1*rand, 0.9844)
	(3+ 0.1*rand, 0.9871)
	(3+ 0.1*rand, 0.9951)
	(3+ 0.1*rand, 0.9813)
	(3+ 0.1*rand, 0.9761)
	(3+ 0.1*rand, 0.9900)
};
    \end{axis}
    \end{scope}

\end{tikzpicture}

    \vspace{-5pt}
\caption{\textbf{Measured statistics.} 
We report the mean \textbf{PSNR} (left plot) and \textbf{MS-SSIM} for random texture sets from Polyhaven (both Texture and Assets). Each dot corresponds to a single texture set and the bar plots display the statistics. For the PSNR, our method  (in {purple}) matches the overall statistics of the BCF1 optimizer of Belcour and Benyoub~\cite{belcour2025tsnc} (in {orange}) but has a worse average PSNR (over all texture sets). However, our method performs better when looking at the MS-SSIM. It is not surprising, as we include a perceptual loss (DISTS) to train our Hypernetwork. We also report the statistics of using the same token $\tokent_0$ for every texture set in blue. As expected, this method has lower quality, but still provides good results.
\label{fig:results_metrics_polyhaven}
\vspace{-0.50cm}}
\end{figure*}

\paragraph*{Multiple heads.} In Figure~\ref{fig:results_multiple_heads}, we show that the Hypernetwork is capable of outputting MLPs of different hidden dimensions (here $D_h = 16, 32$ and $64$) but for the same latent textures. This is interesting to provide different quality/performance tradeoffs without increasing the storage space.
\section{Limitations \& Conclusion}

\paragraph*{Color consistency.} We observed that neural texture compression favors frequency consistency over color consistency. This can lead to global color shift compared to the reference. This is likely due to the hidden dimension of the MLP being too small to encode the full color space of a texture set. Both our Hypernetwork and the BCF1 method of Belcour and Benyoub~\cite{belcour2025tsnc} are affected by this issue.

We have shown that using a Hypernetwork to compress texture sets is a viable solution. Furthermore, this opens new possibilities for the use of small neural networks for jointly solving different problems (such as texture compression and upscaling). While we use BCF1, it could be retargeted for a different neural texture compression format (such as NTC~\cite{vaidyanathan2023ntc}). A future research direction could even be to provide a generic compressor that can target different compression formats from the same internal latent, drawing inspiration from meta-learning~\cite{finn2017maml}.

\section*{Acknowledgement}
The author thanks Alban Fichet for technical support and proof-reading,
Jonathan Dupuy for his firefighter help, and Marissa Dubois for discussion and
proof-reading. The author acknowlege using Creative Common assets from Loïc
Norgeot (fruits scene), Robert Lindner (Kabuto scene), and @DailyArt (violin
scene) in this paper.

\bibliographystyle{sty/eg-alpha-doi} 
\bibliography{egsr2026}       



\end{document}